\documentclass[aps,twocolumn,nofootinbib,preprintnumbers,superscriptaddress]{revtex4}
\usepackage{mathrsfs}

\usepackage{amsfonts}
\usepackage{amsmath}

\usepackage{amssymb}

\RequirePackage[dvips]{graphicx}
\RequirePackage{CJKutf8}
\usepackage[colorlinks=true,  % not ture - STD
            citecolor=red,% magenta , cyan  % refs cite
            urlcolor= blue,                     %
            filecolor=black,
            linktocpage=true,  % Contents
            linkcolor=blue,%{[rgb]{0.0,0.618,0.0}},  % Eqs
             ]{hyperref}

\usepackage[dvips]{color}
\usepackage[normalem]{ulem}
\usepackage{epsfig}
\usepackage{yfonts}
\usepackage{bm}

\newcommand{\beq}{\begin{equation}}
\newcommand{\eeq}{\end{equation}}
\newcommand{\bq}{\begin{equation}}
\newcommand{\eq}{\end{equation}}
\newcommand{\ba}{\begin{array}}
\newcommand{\ea}{\end{array}}
\newcommand{\beqa}{\begin{eqnarray}}
\newcommand{\eeqa}{\end{eqnarray}}
\def\bc{\begin{center}}
\def\ec{\end{center}}
\def\bnum{\begin{enumerate} }
\def\enum{\end{enumerate}}
\def\nn{\nonumber}
\def\ii{\!\!\!\!\!\!}  % \! can compress 1/6 space for left moving its right part
\def\3i{\!\!\!}
\def\2i{\!\!}

\def\ea{{e_a}}

\def\ec{{e_c}}

\def\log{\ln}

\def\hg{ \hat{g} }   % \hat{g} is used for bulk metric

\def\nn{\nonumber}

\def\[{\left[}
\def\]{\right]}
\def\({\left(}
\def\){\right)}

\def\>{\rightarrow}

\def\Diracslash#1{\not{\hbox{\kern-4pt $#1$}}}

\def\Dslash{\not{\hbox{\kern-4pt $D$}}}
\def\pslash{\not{\hbox{\kern-4pt $p$}}}
\def\qslash{\not{\hbox{\kern-4pt $q$}}}

\def\lv{\not{\hbox{\kern-4pt $L$}}}
\def\lsim{\mathrel{\raise.3ex\hbox{$<$\kern-.75em\lower1ex\hbox{$\sim$}}}}
\def\gsim{\mathrel{\raise.3ex\hbox{$>$\kern-.75em\lower1ex\hbox{$\sim$}}}}
\def\ifmath#1{\relax\ifmmode #1\else $#1$\fi}

\evensidemargin=\oddsidemargin

\def\a{ \hat{H} }%{\alpha}

\numberwithin{equation}{section}

\begin{document}
\begin{CJK}{UTF8}{gbsn}

\begin{titlepage}
\begin{flushright}
\end{flushright}

\begin{center}
 \vspace*{10mm}

{\LARGE\bf
Topological Charged Black Holes in Generalized Ho\v{r}ava-Lifshitz Gravity
%{ \cblue with $z=d$ UV Fixed Point }
}\\
\medskip
\bigskip\vspace{0.6cm}

{\large {Tian-Jun Li$^{\star}$,} \, {Yong-Hui Qi$^{\dag,\star,\ast}$,} \, {Yue-Liang Wu$^{\star,\ddag}$,}\, {Yun-Long Zhang$^{\star}$} }
\\[7mm]
{\it
$^{\dag}$Center for High Energy Physics, Peking University, Beijing 100871, People's Republic of China\\
$^{\star}$State Key Laboratory of Theoretical Physics, Kavli Institute for Theoretical Physics China, \\
Institute of Theoretical Physics, Chinese Academy of Sciences, Beijing 100190, People's Republic of China\\
$^{\ast}$Key Laboratory of Particle Astrophysics, \\
Institute of High Energy Physics, Chinese Academy of Sciences, Beijing 100190, People's Republic of China\\
$^{\ddag }$University of Chinese Academy of Sciences, People's Republic of China
}
\vspace*{0.3cm} \\
 {\tt tli@itp.ac.cn,~qiyh@itp.ac.cn,~ylwu@itp.ac.cn,~zhangyl@itp.ac.cn}
\bigskip\bigskip\bigskip

{
\centerline{\large\bf Abstract}
\begin{quote}
As a candidate of quantum gravity in ultrahigh energy, the $(3+1)$-dimensional Ho\v{r}ava-Lifshitz (HL) gravity with critical exponent $z\ne 1$, indicates anisotropy between time and space at short distance.
In the paper, we investigate the most general $z=d$ Ho\v{r}ava-Lifshitz gravity in arbitrary spatial dimension $d$,
with a generic dynamical Ricci flow parameter $\lambda$ and a detailed balance violation parameter $\epsilon$. In arbitrary dimensional generalized HL$_{d+1}$ gravity with $z\ge d$ at long distance, we study the topological neutral black hole solutions with general $\lambda$ in $z=d$ HL$_{d+1}$, as well as the topological charged black holes with $\lambda=1$ in $z=d$ HL$_{d+1}$. The HL gravity in the Lagrangian formulation is adopted, while in the Hamiltonian formulation, it reduces to Dirac$-$De Witt's canonical gravity with $\lambda=1$. In particular, the topological charged black holes in $z=5$ HL$_6$, $z=4$ HL$_5$, $z=3,4$ HL$_4$ and $z=2$ HL$_3$ with $\lambda=1$ are solved. Their asymptotical behaviors near the infinite boundary and near the horizon are explored respectively. We also study the behavior of the topological black holes in the $(d+1)$-dimensional HL gravity with $U(1)$ gauge field in the zero temperature limit and finite temperature limit, respectively. Thermodynamics of the topological charged black holes with $\lambda=1$, including temperature, entropy, heat capacity, and free energy are evaluated.
\bigskip \\
PACS numbers: 04.60.-m, 04.70.Dy, 11.10.Ef, 04.50.Gh
\end{quote}}
\end{center}
\end{titlepage}

\tableofcontents

\section{Introduction}
\label{sec:intro}

The main problem of traditional quantum gravity is that the gravitational coupling constant is dimensionful, with a negative dimension in mass units $[G_N]=-2$. As a result, at the quantum loop level, the theory is by definition not a renormalizable theory~\cite{'tHooft:1974bx}. In Refs.~\cite{Horava:2008jf,Horava:2008ih,Horava:2009uw,Horava:2009if,Horava:2010zj,Horava:2011gd}, Ho\v{r}ava proposed a quantum field theory of gravity with dynamical critical exponent $z$ ($z\ne1$ signifies that the scalings of time and space are anisotropy at short distance, and thus the relativistic invariance is broken at ultrahigh energy scale), which measures the anisotropy between space and time, although the homogeneous of pure spatial space are still assumed. The theory describes an interacting nonrelativistic renormalizable gravity at short distance. Assuming that the theory is in the $(d+1)$-dimensional space-time(where $d$ denotes the dimension of space), then the ultraviolet (UV) theory with $z=d$ is power-counting renormalizable at the UV energy scale. The theory flows naturally to the relativistic one with critical exponent $z=1$ at long wave limit, with Lorentz invariance presenting as an accidental symmetry restored at large distance. Meanwhile the theory becomes the classical Einstein's general relativity(GR) in the infrared(IR) limit. In the literatures, this theory is called Ho\v{r}ava-Lifshitz gravity as a candidate of quantum gravity. Although there still exit some problems, the higher derivative terms can improve the renormalizability of the theory, without the usual attendant problem of ghosts.

In this paper, we begin with the topological black holes in $(d+1)$-dimensional Ho\v{r}ava-Lifshitz gravity with the critical exponent $z=d$ in ultrahigh energy, which indicates anisotropy between time and space at short distance, while the pure spatial space is still assumed homogeneous with different topological structure, depicted with topological index, e.g., $k=0,\pm 1$, corresponding to Ricci flat, sphere and hyperbolic spatial hyersurface, respectively. The gravity is generalized to include an arbitrary given critical exponent $z=d$, a generic Ricci flow parameter $\lambda$, as well as a detailed balance violation parameter $\epsilon$. The motivation of the generalization to arbitrary dimensions originates from the fact that the dimensions of the space-time are essential to critical exponent, which describes the critical phenomena behavior of physical quantities(or Landau's order parameters) near continuous phase transitions~\cite{Wilson:1971dc}. For example, the critical exponents of the ferromagnetic transition in Ising model, namely, $\alpha$, $\beta$, $\gamma$, $\delta$, $\nu$, etc., depend on the dimensions of the space-time. This can be understood through the renormalization group paradigm~\cite{Wilson:1971bg-dh,Wilson:1973jj}, e.g., in perturbative approach, by using the dimensional regularization~\cite{'tHooft:1972fi,'tHooft:1973mm} in momentum space-time, to calculate Feynman loop diagrams, which indicate that the quantum loop effects in arbitrary dimensions, e.g., $d=2$ and $d=3$ have significantly distinguishable physical consequence. To be brief, we investigate a generalized HL$_{d+1}$ gravity with Lifshitz points $z\ge d$ at both short distance and long distance, where the high curvature terms are relevant and irrelevant respectively. In the pure spatial sphere symmetric case with Ricci scalar index $k=1$, the topological neutral black holes just reduce to be anti-de Sitter (AdS) or de Sitter (dS) Schwarzschild black hole, with negative (with $\lambda>1/d$) or positive (with $\lambda<1/d$) effective cosmological constant $\Lambda$, respectively. The corresponding topological charged black holes are solved in the relativistic limit with Lifshitz point $z=1$.
In the complete violation of the detailed balance condition with $\epsilon=1$, the topological charged black holes just reduce to be Reissner-Nordstr\"{o}m (RN) AdS topological black holes due to the classical Einstein-Maxwell dynamics at large distance. The HL gravity in the Lagrangian formulation is adopted, while the dynamic Ricci flow equation for the Riemannian manifold with $\lambda=1$ is emphasized, which reduces to the Dirac$-$De Witt's canonical gravity~\cite{Dirac:1958sc,DeWitt:1967} in Hamiltonian formulation. In particular, by using the Hamiltonian approach, we have solved the topological charged black holes in HL gravity with $\lambda=1$ in a $U(1)$ gauge field. The topological charged black holes in HL$_{d+1}$ with $z=d$, including $z=5,4,3,2$ cases are solved. For the intuitive purpose, the topological charged black holes in HL$_4$ with $z=4$ and in HL$_3$ with $z=2$ cases are explored in more detail. Their asymptotic behavior in the infinite boundary and near horizon behavior are explored respectively, both of which deduce scale invariances on the black branes. Moreover, we study the behaviors of the topological black holes in the HL$_4$ gravity with $U(1)$ gauge field in zero temperature and finite temperature limit, respectively at and near extremal black holes horizon. Thermodynamics of topological charged black holes with $\lambda=1$, including temperature, entropy, heat capacity, and free energy are discussed as well.

This paper is organized as follows. In Sec. \ref{sec:HL}, we briefly introduce the essential elements of HL gravity. In Sec. \ref{sec:(d+1)HL-z=d}, first we investigate the most general HL gravity at $z=d$ UV fixed point and solve the topological black holes in HL$_{d+1}$ gravity. Second, we study the $z=5$ HL gravity and solve the topological black holes in HL$_{d+1}$ gravity. Then we investigate the HL$_{d+1}$ gravity with both critical exponent $z=3$ and $z=4$ at a UV fixed point, with the detailed balance violation parameter $\epsilon$, but with Ricci flow parameter $\lambda=1$. In particular, the solutions in HL$_3$ gravity with critical exponent $z=2$ at the UV fixed point with $\lambda=1$ are obtained, too. This metric essentially has significantly different physics from the HL$_4$ gravity case. While in this paper, we mainly focus on HL$_4$ gravity with critical exponent $z=3$ and $z=4$ at the UV fixed point.
In Sec. \ref{sec:CFT-HL}, we explore both infinite boundary and near horizon behaviors of the topological charged black holes in HL gravity. In the infinite boundary limit, we can recover the pure AdS$_{d+1}$ metric,
with constant negative Ricci scalar curvature and isomorphic group\footnote{In the paper, we adapt the $\{-,+,+,\ldots  \}$ signature.} $SO(2,d)$, which is the same as the $d$-dimensional conformal symmetry group of $d$-dimensional Minkowski space ${\mathbb R}^{1,d-1}$, while in the near extremal horizon limit, we obtain AdS$_2$ metric along $(r,t)$ plane with IR conformal symmetry, which is perpendicular to the pure spatial hypersurface of topological charged black holes with topological index $k=0,\pm 1$.
By using Hawking temperature at or near extremal black holes horizon, one would expect to have corresponding scale invariance embedding in IR $(r,t)$ boundary at both zero and finite temperature, respectively.
Some basic physical thermodynamic quantities of the topological charged black holes in HL$_4$ are calculated and summarized in Section \ref{sec:thermodynamics}.
The conclusion is given in Sec \ref{sec:concl}. In Appendix \ref{app:HL}, we briefly review the field theory strucuture of Ho\v{r}ava-Lifshitz gravity, including both %path integral
Lagrangian and Hamiltonian formulation. The characteristic physical consequence of the HL gravity in various dimensional space-time, in particular HL$_3$ with $z=2$, HL$_{4}$ with $z = 3,4$, HL$_5$ with $z=4$ and HL$_6$ with $z=5$ are briefly summarized in Appendix~\ref{app:HL(d+1)}. The notation of the differential form for the most general Lovelock gravity in $d$-dimensional spatial time is introduced in Appendix~\ref{app:diff-form-Lovelock}. The useful results on general coordinate invariant curvature terms in the whole space-time are put in Appendix~\ref{app:causal-singulariy}.

In this paper, we use the small letter $i,j,=1,2,...,d$ to denote the index of pure spatial coordinate
$x^{i}$, the greek symbols $\mu,\nu,= 0,1,...,d$, to denote the index
of the ordinary space-time coordinate $x^{\mu}\sim (\tau,x^{i})$.

\section{Ho\v{r}ava-Lifshitz Gravity}
\label{sec:HL}

\subsection{Symmetry of space-time}

In the HL gravity, time is a special dimension in space-time, and a codimension-$q$($q=1$) foliation preserving diffeomorphisms is imposed on the space-time manifold ${\mathcal M}$. The levels of the foliation are the hypersurfaces of constant time, and all leaves of foliation are topologically equivalent to a $d$-dimensional manifold with fixed time with intrinsic metric $g_{ij}$ and extrinsic curvature $K_{ij}$ induced along the leaves. The functions that take constant values on each leaf are called projectable functions. Intuitively, the foliation diffeomorphism can be viewed as a gauge invariance; thus, $N_i$ can be interpreted as gauge fields associated with the space-time-dependent spatial diffeomorphism and $N$ as the time-dependent time reparametrizations. The generators of their infinitesimal transformations are respectively in nonrelativistic limit,
\beqa
\delta x^i = \zeta^i(t,x^j) + o(1/c^2), \quad \delta t = c\,f(t) + o(1/c),
\eeqa
under which, one says that HL gravity manifests $d$-dimensional spatial general covariance and time-reparametrization invariance. The dynamical field consists of $d$-dimensional spatial metric $g_{ij}$(a tensor), shift variable $N^i$(a vector) and lapse function $N$(a scalar), which transforms, respectively, as
\beqa
\delta g_{ij} \! &=& \! \partial_i \zeta^k g_{jk} + \partial_j \zeta^k g_{ik} + \zeta^k \partial_k g_{ij} + c f\dot{g}_{ij}, \nn\\
\delta N_i \! &=& \! \partial_i \zeta^j N_j + \zeta^j \partial_j N_i + \dot\zeta^j g_{ij} + \dot{f}N_i + f\dot{N}_i, ~~ \\
\delta N \! &=& \! \zeta^j \partial_j N + \dot{f} N + f \dot{N}, \nn
\eeqa
where they are all functions of $t$ and $x^i$. The lapse function $N$ is viewed as a gauge field for time reparametrizations, and is effectively restricted to depend only on time $t$, but not the spatial coordinates $x^i$ ~\cite{Horava:2009uw}.

\subsection{Scaling dimension of space-time}

The anisotropic scaling properties of the space and time are respectively depicted by the dynamical critical exponent(or Lifshitz index) $z$ at the fixed point,
\beqa
x^i \to \ell \, x^i, \quad t\to \ell^z \, t.
\eeqa
Thus the theory does not have the full diffeomorphism invariance of GR but only a subset, or local Galilean invariance, while when $z=1$, it is the standard relativistic scale invariance.
The space and time are dimensionful in the units of mass(spatial momentum or inverse spatial length),
\beqa
[x^i]=-1, \quad [t] =-z.   \label{Eq:scaling_x_t}
\eeqa
 As a result, the scaling dimension of the speed of light at the fixed point is
\beqa
[c] = z-1.
\eeqa
At $z=1$, the speed of light is dimensionless and the theory just reduces to the usual general relativity in IR limit at long distance. While in the UV limit, the HL gravity switches to other $z\ne 1$ to make the theory renormalizable.
The dynamical field of HL gravity theory consists of spatial metric $g_{ij}$, a spacial vector $N_i$, and a spatial scalar $N$, which are respectively the spatial sector of metric, the shift variable and the lapse function in $(d+1)$ split of the $(d+1)$-dimensional relativistic space-time metric. For a generic metric, it can be decomposed in the $(d+1)$-dimensional Arnowitt-Deser-Misner(ADM) formalism,
\beqa
ds^2 = - N^2 dt^2 + g_{ij}(dx^i + N^i dt)(dx^j + N^j dt), \label{Eq:ds2-ADM}
\eeqa
Generally speaking, one can start from the covariant metric $g_{\mu\nu}$ associated with the time $t$. The light speed c is restored so that the time coordinate $x_0 = ct$, and then one can expand the metric in the nonrelativistic limit $c\rightarrow\infty$

%Generally speaking, one can start from a usual relativistic metric $g_{\mu\nu}$ in the usual ADM $(d+1)$ decomposition with $c$ restored that $x^0=c\,t$ and expanded the metric in the nonrelativistic limit $c\to \infty$.

The metric and its inverse can be written explicitly as below
\beqa
 \hg_{\mu\nu} &=& \left(
                   \begin{array}{cc}
                       -N^2 + N_i N^i /c^2 & N_i/c\\
                       N_i/c & g_{ij}\\
                   \end{array}
                    \right), \\
 \hg^{\mu\nu} &=& \left(
                   \begin{array}{cc}
                       -1/N^2  & N_i/N^2\\
                       N_i/N^2 & g_{ij} - N^i N^j/N^2\\
                   \end{array}
                    \right).
\eeqa

The definition of extrinsic curvature(second fundamental form) of the leaves at constant time moment in spatial sector is
\beqa
K_{ij}=\frac{1}{2N}(\dot{g}_{ij}-\nabla_i N_j - \nabla_j N_i), \label{Eq:Kij}
\eeqa
where the covariant derivatives are defined with respect to the spatial metric $g_{ij}$. Then according to the scaling of space-time defined in Eq.(\ref{Eq:scaling_x_t}),  one can obtain the classical scaling dimensions of the dynamical fields as below
\beqa
[g_{ij}] = 0, \quad [N_i]= z-1, \quad [N]=0,   \label{Eq:scaling_gij_Ni_N}
\eeqa
since the scaling dimension of extrinsic curvature inversely proportional to that to time coordinate, $[K_{ij}]=[\dot{g}_{ij}]=[\partial_t g_{ij}]=z=[\nabla_i N_j]=1+[N_j]$.

\section{Topological Black Holes in Generalized Ho\v{r}ava-Lifshitz Gravity}
\label{sec:(d+1)HL-z=d}

\subsection{Topological black holes in generalized HL$_{d+1}$ gravity}

The easiest way to obtain the black hole solutions of the HL$_{d+1}$ gravity is through making a metric ansatz, and then substituting the metric ansatz into the action to solve the metric.

Let us consider a general topological black hole solution with arbitrary constant scalar curvature horizon~\cite{Cai:1998vy}. The metric ansatz is
\beqa
ds^2 = - \tilde{N}^2(r)f(r)c^2 dt^2 + \frac{dr^2}{f(r)}+ r^2 d\Omega_{d-1,k}^2, \label{Eq:ds2-topological-HL}
\eeqa
where $d\Omega_{d-1}^2$ denote the line element of a $d-1$-dimensional manifold with constant scalar curvature $(d-1)k$ and area $\Omega_{d-1,k}^2$, where $k=0,\pm 1$ indicate different topology of the spatial hypersurface,
\beqa
d\Omega_{d-1,k}^2 = \gamma_{mn} dx^m dx^n,
\eeqa
where $\gamma_{mn}$ is the metric of the hypersurface.
%\beqa
%d\Omega_{d-1,k}^2 = \sum_{i=1}^{d-1}\prod_{j=1}^{i-1}\Theta^2_{j,k_j} d\theta_i^2,
%\eeqa
%and the most intuitive flag topological space-time could be parameterized through $\Theta_{j,k_j} \equiv \sin\theta_j, \sinh\theta_j, \theta_j$ for $k_j= +1, -1, 0$ respectively, where $\theta_{i}\in[0,\pi]$ for $i=1,\ldots d-2$, and $\theta_{d-1}\in[0,2\pi]$.
For example, when $d=3$, for $k=\pm 1$ case we have made an static, spherically/hyperbolic symmetric metric ansatz that
\beqa
k=1:  && d\Omega_{2,1}^2 = d\theta^2 + \sin^2\theta \, d\phi^2,  \label{Eq:ds2_d=3-k=1} \\
k=-1: && d\Omega_{2,-1}^2 = d\theta^2 + \sinh^2\theta \, d\phi^2, \label{Eq:ds2_d=3-k=-1}
\eeqa
for $k=0$ case, we have a static plane with translational symmetric black brane ansatz
\beqa
 k=0: \quad d\Omega_{2,0}^2 = d\theta^2 + \theta^2 \, d\phi^2 \equiv dx^2 +dy^2, \label{Eq:ds2_d=3-k=0}
\eeqa
where we have transformed the metric in polar coordinates into one in Cartesian coordinates, by introducing
\beqa
(x, y) \equiv \theta(\cos\phi, \sin\phi).
\eeqa

\subsection{$z=d$ HL$_{d+1}$ gravity}

For a general theory of gravity in dimensions $d \ge 5$, by imposing the requirement of ($1$)general coordinate invariance, ($2$)the field equation for the metric to be second order, and the Lagrangian to be a most general local $d$-form up to a closed differential form($d{\mathcal L}=0$) with fully spatial rotational $SO(d-1)$ invariant symmetry in tangent space, we consider a spatial potential action given by a sum of dimensionally continued topological invariant characteristic classes of lower dimensions, i.e., extended Euler densities~\cite{Banados:1993ur}. In the differential form, it can be expressed as
\beqa
W_d &=& \int_{{\mathcal M}}  \sum_{p=0}^{[d/2]} \alpha_p \epsilon_{a_1 \ldots a_d} R^{a_1 a_2}\wedge \ldots  \wedge R^{a_{2p-1} a_{2p}}  \nn\\
&& \wedge   e^{a_{2p+1}}\wedge \ldots \wedge e^{a_d}. \label{Eq:Wd}
\eeqa
By observing Eqs.(\ref{Eq:Wd-d-form}), (\ref{Eq:Wd-d-form_1}), (\ref{Eq:Wd-d-form_2}), and (\ref{Eq:Wd-d-form_3})in Appendix ~\ref{app:diff-form-Lovelock}, the above differential form of potential action is equivalent to Lovelock gravity~\cite{Lovelock:1971yv}
\beqa
W_d =\int d^d x \sqrt{g} \sum_{p}^{[d/2]} (d-2p)! \alpha_p  W_{(p)}, \label{Eq:W_d_(d+1)D}
\eeqa
where $\alpha_p$ is an arbitrary constant, namely, Lovelock coefficients with scaling dimension $[\alpha_p] = d - 2p$, ${\mathcal R}^p$ is the Euler density of a $2p$-dimensional manifold,
\beqa
W_{(p)}={\mathcal R}^p \equiv \frac{1}{2^p}\delta^{i_1 \ldots i_{2p}}_{j_1 \ldots j_{2p}}\prod_{k=1}^p R^{j_{2k-1} j_{2k}}_{~~ i_{2k-1} i_{2k}},  \label{Eq:Wp}
\eeqa
where $[W_{(p)}]=[{\mathcal R}^p]=2p$, $R^{IJ}_{~~KL}$ is the Rieman tensor, and the generalized Kronecker $\delta$ symbol is defined as the totally antisymmetric product of Kronecker deltas,
\beqa
\ii\ii
\delta^{i_1 \ldots i_{2p}}_{j_1,\ldots j_{2p}}
&=& \begin{vmatrix}
\delta^{i_1}_{j_1}    & \cdots   & \delta^{i_1}_{j_{2p}}\\
\vdots  &   \ddots     & \vdots  \\
\delta^{i_{2p}}_{j_1}    & \cdots   & \delta^{i_{2p}}_{j_{2p}}\\
\end{vmatrix}  \\
&=&  (2p)! \,\delta^{i_1\ldots i_{2p}}_{[j_1 \ldots j_{2p}]} = (2p)! \,\delta^{[ i_1 \ldots i_{2p}]}_{ j_1,\ldots j_{2p}} \nn\\
&=& \varepsilon^{i_1\ldots i_{2p}}\varepsilon_{ j_1,\ldots j_{2p}}, \nn
\eeqa
In the second to last equality, the antisymmetrization has been adopted. While in the last quality, the Levi-Civita symbols are related to Kronecker symbols through $\varepsilon_{j_{1} \ldots j_{2p}}=\delta_{j_{1} ... j_{2p}}^{{1} \ldots {2p}}$ and $\varepsilon^{i_{1} \ldots i_{2p}}=\delta^{i_{1} \ldots i_{2p}}_{{1} \ldots {2p}}$. Therefore,
\beqa
{\mathcal R}^p \!=\! \frac{1}{2^p}\varepsilon^{i_1 i_2 \ldots i_{2p-1}i_{2p}}\varepsilon_{j_1 j_2\ldots j_{2p-1} j_{2p}}
R^{j_1 j_2}_{~~i_1 i_2}\ldots R^{j_{2p-1} j_{2p}}_{~~i_{2p-1} i_{2p}}. \nn
\eeqa
Each term ${\mathcal R}^p$ corresponds to the dimensional extension of the Euler density in $2p$ dimensions, and $p<(d+1)/2$ for a given $d$ dimension. The up limit $n=d/2$ for even spatial dimensions and $n=(d-1)/2$ for odd spatial dimensions, or $n \equiv [d/2]$, e.g., $n=0$ for $d=0,1$; $n=1$ for $d=2,3$; $n=2$ for $d=4,5$; $n=3$ for $d=6,7$; etc. For $p=0$, $W_{(0)}={\mathcal R}^0=1$ and $\alpha_0$ is just the cosmological constant
\beqa
E_{ij}^{(0)} = -\frac{1}{2}g_{ik}\delta^{k}_{j} = - \frac{1}{2} g_{ij}, \label{Eq:Eij_W0}
\eeqa
with trace
\beqa
E^{(0)} = -\frac{d}{2}. \label{Eq:trace_Eij_W0}
\eeqa
For $p=1$, $W_{(1)}={\mathcal R}$ is
\beqa
{\mathcal R} = \frac{1}{2} \delta^{i_1 i_2}_{j_1 j_2}R^{j_1 j_2}_{~~i_1 i_2} = \frac{1}{2} (\delta^{i_1}_{j_1}\delta^{i_2}_{j_2}- \delta^{i_1}_{j_2}\delta^{i_2}_{j_1})R^{j_1 j_2}_{~~ i_1 i_2} = R, \nn
\eeqa
which just gives the usual Einstein Hilbert term,
\beqa
E_{ij}^{(1)} = - \frac{1}{2^2} g_{ik} \delta^{k i_1 i_2}_{j j_1 j_2} R^{j_1j_2}_{~~i_1 i_2}= R_{ij} - \frac{1}{2}g_{ij}R. \label{Eq:Eij_W1}
\eeqa
with trace
\beqa
E^{(1)} = \bigg(1-\frac{d}{2}\bigg) R. \label{Eq:trace_Eij_W1}
\eeqa
For $p=2$, $W_{(2)}={\mathcal R}^2$ is
\beqa
{\mathcal R}^2 = \frac{1}{4} \delta^{i_1 i_2 i_3 i_4}_{j_1 j_2 j_3 j_3}  R^{j_1 j_2}_{~~ i_1 i_2}  R^{j_3 j_4}_{~~ i_3 i_4}
= {\mathcal L}_{GB}.
\eeqa
In this case, ${\mathcal R}^2$ is precisely the quadratic Gauss-Bonnet term, which is the dimensionally extended version of the four-dimensional Euler density. The corresponding tensor $E_{ij}$ according to the detailed balance condition defined in Eq.(\ref{Eq:Eij}) becomes
\beqa
E_{ij}^{(2)} &=& - \frac{1}{2^3}g_{ik}\delta^{k i_1 i_2 i_3}_{j j_1 j_2 j_3}R^{j_1 j_2}_{~~i_1 i_2}R^{j_3 j_4}_{~~ i_3 i_4} \nn\\
&=& 2( R_{i}^{klm}R_{jklm} - 2R_{ikjl}R^{kl} -2 R_{ik}R^{k}_{~j} + R R_{ij}) \nn\\
&& - \frac{1}{2}g_{ij}(R^2- 4 R^{kl}R_{kl} + R^{klmn}R_{klmn}), \label{Eq:Eij_W2}
\eeqa
with trace
\beqa
E^{(2)} = \bigg(2 - \frac{d}{2} \bigg){\mathcal L}_{GB}, \label{Eq:trace_Eij_W2}
\eeqa
which are consistent with Eqs.(\ref{Eq:Jij-GB}) and (\ref{Eq:J-GB}).

For example, HL$_6$ gravity with $z=5$ UV fixed point($d=5$,$p=2$), can be generated through the potential action $W_5$
\beqa
W_5 = \int d^5 x \sqrt{g}(5!\alpha_0 + 3!\alpha_1 R + \alpha_2 {\mathcal R}_{GB} ), \nn
\eeqa
with the scaling mass dimension $[\alpha_0]=d=5$, $[\alpha_1]=d-2=3$, $[\alpha_2]=d-4=1$.
In order to match with Eqs.(\ref{Eq:W_1-6D}) and (\ref{Eq:W_5-6D_2}), one can choose parameters as below
\beqa
\alpha_0 = -\frac{2\Lambda_W}{5!\kappa_W^2}, \quad \alpha_1 = \frac{1}{3! \kappa_W^2}, \quad \alpha_2 = \gamma. \label{Eq:alpha0-alpha1-alpha2}
\eeqa
Moreover, HL$_7$ gravity with $z=6$ UV fixed point can be generated through a spatial potential $W_6$,
\beqa
W_6 = \int d^6x \sqrt{g}(6!\alpha_0 + 4!\alpha_1 R + 2!\alpha_2 {\mathcal L}_{GB}), \nn
\eeqa
with the scaling mass dimension $[\alpha_0]=d=6$, $[\alpha_1]=d-2=4$ and $[\alpha_2]=d-4=2$.

For HL$_8$ gravity with $z=7$ UV fixed point can be generated through a spatial potential $W_7$,
\beqa
W_7 = \int d^7x \sqrt{g}(7!\alpha_0 + 5!\alpha_1 R + 3!\alpha_2 {\mathcal L}_{GB} + \alpha_3 W_{(3)}), \nn
\eeqa
with the scaling mass dimension $[\alpha_0]=d=7$, $[\alpha_1]=d-2=5$, $[\alpha_2]=d-4=3$ and $[\alpha_3]=d-6=1$.

The additional term $W_{(3)}={\mathcal R}^3$ is
\beqa
{\mathcal R}^3 = \frac{1}{2^3}\delta^{i_1\ldots i_6}_{j_1 \ldots j_6} R^{j_1 j_2}_{~~i_1 i_2} R^{ j_3 j_4}_{~~i_3 i_4} R^{ j_5 j_6}_{~~i_5 i_6}.
\eeqa
where $W_{(3)}$ consists of cubic term of Ricci scalar, i.e., $W_{(3)}$ is at an order of ${\mathcal O}(R^3)$ as expected. This is obvious by re-expressing $W_{(3)}$ more explicitly as below:
\beqa
W_{(3)} &=& R^3 - 12 R R^{ij}R_{ij} + 16 R^{ij}R_{jk}R^{k}_{~i} + 3 R R^{ijkl}R_{ijkl} \nn\\
&& + 24 R^{ijkl}R_{ik}R_{jl} + 24 R^{ijkl}R_{ijkm}R^{m}_{~l} \nn\\
&& + 2 R^{ijkl}R_{ijmn}R^{mn}_{~~kl} + 8 R^{ij}_{~~kn}R^{kl}_{~~im}R^{mn}_{~~jl}.
\eeqa
The corresponding $E_{ij}$ becomes
\beqa
\ii
E_{ij}^{(3)} = - \frac{1}{2^4}g_{ik}\delta^{k i_1 i_2 i_3 i_4 i_5 i_6}_{j j_1 j_2 j_3 j_4 j_5 j_6}R^{j_1 j_2}_{~~ i_1 i_2}R^{j_3 j_4}_{~~ i_3 i_4} R^{j_5 j_6}_{~~ i_5 i_6}.  \label{Eq:Eij_W3}
\eeqa
In a similar procedure, HL$_{d+1}$ gravity with $z=d$ UV fixed point can be generated through the spatial potential $W_{(d)}$ defined in Eq.(\ref{Eq:W_d_(d+1)D}) and the corresponding $E_{ij}$ turns out to be
\beqa
E_{ij}^{(p)} = - \frac{1}{2^{p+1}} g_{il} \delta^{l i_1 \ldots i_{2p}}_{j j_1 \ldots j_{2p}}\prod_{k=1}^p R^{j_{2k-1} j_{2k}}_{~~ i_{2k-1} i_{2k}}.  \label{Eq:Eij_Wp}
\eeqa
Consequently, the potential density in Eq.(\ref{Eq:SV}) becomes
\beqa
\ii
{\mathcal V}[g]
\2i &=& \2i
\frac{\kappa^2}{8}\sum_{p,q=0}^{[d/2]}(d-2p)!\alpha_p (d-2q)!\alpha_q \nn\\
&& \times (E^{(p)ij}E_{ij}^{(q)}-\tilde\lambda E^{(p)}E^{(q)}), \label{Eq:SV_1_dD-2}
\eeqa
where $[\alpha_p]=d-2p$, $[\alpha_q]=d-2q$, $[\kappa^2]=z-d$, and $\tilde\lambda$ is defined in Eq.(\ref{Eq:Wheeler-DeWitt-inverse}).
The Ricci flow equation in Eq.(\ref{Eq:gij-Ricci-Flow-1}) becomes
\beqa
\dot{g}_{ij} \3i &=& \3i 2 \nabla_{(i} N_{j)} - \frac{N\kappa^2}{2}\sum_{p} (d-2p)! \alpha_p(E_{ij}^{(p)}-\tilde\lambda g_{ij}E^{(p)} ) \nn\\
& \!-\! &  N  \frac{\kappa^2}{2\kappa_W^2}\bigg(R_{ij}-\frac{2\lambda-1}{2(d\lambda-1)}R\,g_{ij} - \frac{1}{d\lambda-1}\Lambda_W g_{ij}\bigg). \nn
\eeqa

In this case, the scaling dimension of $[{\mathcal V}[g]]=z+d\le 2d$; therefore, the action can be modified by relevant operators with dimension less or equal than $2d$. For different relevant operators with scaling dimensions
\beqa
\ii [E^{(p)}]=2p, ~ [E^{(q)}]= 2q, ~ [E^{(p)}E^{(q)}]=2(p+q),
\eeqa
the terms with the highest curvature due to $W_d$ will improve the UV behavior of the HL$_{d+1}$ gravity at short distance, since the operators with highest scaling dimensions in the potential dominate in the UV limit, i.e., terms inherited from $W_d$; the theory exhibits $z=d$ Lifshitz fixed points. While in the IR limit, the operators with lower dimensions in the potential density, i.e., Ricci scalar $R$ and those with cosmological constant $\Lambda_W$ will still dominate at long distance.

Combining Eqs.(\ref{Eq:S_K+V}) and (\ref{Eq:Eij_Wp}) together, the total action can be written as
\beqa
S \3i &=& \3i \int dt \int d^d x \sqrt{g}  N \bigg[  \frac{2}{\kappa^2}(K_{ij}K^{ij}-\lambda K^2)  \nn\\
\3i &-& \3i  \frac{\kappa^2}{8\kappa_W^4}\bigg(R^{ij}R_{ij} \!-\! \frac{1 \!-\! d/4 \!-\! \lambda }{1 \!-\! d\lambda}R^2 \!-\! \frac{(d \!-\! 2)\Lambda_W R \!-\! d\Lambda_W^2}{1 \!-\! d\lambda} \bigg) \nn\\
\3i &-& \3i \frac{\kappa^2}{4\kappa_W^2} \! \sum_{p=2}^{[d/2]} \! \alpha_p  (d-2p)! \bigg(E^{(p)ij}\big(R_{ij}-\frac{1}{2} R g_{ij} + \Lambda_W g_{ij}\big)  \nn\\
\3i && \3i     - \tilde\lambda E^{(p)}\Big(\frac{2-d}{2} R + d\Lambda_W \Big) \bigg) -\frac{\kappa^2}{8}\sum_{p,q=2}^{[d/2]}\alpha_p (d-2p)!\nn\\
\3i && \3i \times  \alpha_q (d-2q)! (E^{(p)ij}E_{ij}^{(q)}-\tilde\lambda E^{(p)}E^{(q)}) \bigg].
\label{Eq:S-(d+1)D}
\eeqa
The scaling dimension of the parameters for the theory becomes
\beqa
\ii && [\kappa]=\frac{z-d}{2}, ~ [\lambda]=0, ~ [\kappa_W] = -\frac{d-2}{2}, ~ [E^{(p)}] = 2p, \nn \\
\ii && [dtd^dx] = -z-d, ~ [{\mathcal L}] = z+d, ~ [\alpha_p]=d-2p,  \nn
\eeqa
from which one obtains the scaling dimension of the couplings of kinetic terms and potential terms with renormalizability,
\beqa
\ii [ \kappa^{-2} ] = d-z \le 0, ~  [ \kappa^2 \alpha_p \alpha_q ] = z+d - 2(p+q) \le 0. \nn
\eeqa
Or, equivalently the renormalizable conditions for kinetic term and potential term are respectively,
\beqa
z\ge d, \quad  2(p+q) \ge z+d .
\eeqa
The two inequalities together imply the renormalizability condition for the full Lagrangian,
\beqa
2(p+q) = [E^{(p)}E^{(q)}] \ge [{\mathcal L}] = z+d \ge 2d.  \label{Eq:RE-operators}
\eeqa
By observing the scaling dimensions of the relevant operators, the power-counting renormalizability of $d+1$-dimensional HL gravity theory necessarily requires the addition of higher order spatial curvature terms with at
least $2d$-th order derivative.

In the UV limit, the operators with higher scaling dimensions in the potential density dominate at short distance, i.e., terms inherited from $W_{(p)}$ with $p=[d/2]$; the theory exhibits the $z=d$ Lifshitz fixed point. While in the IR limit, the operators with lower dimensions in the potential density will dominate at long distance, i.e. the terms proportional to Ricci scalar and those with cosmological constant.

To match with the leading order terms in the IR limit as shown in Eq.(\ref{Eq:S-HL(d+1)-z=3}), we have used the notation defined as shown in Eq.(\ref{Eq:alpha0-alpha1-alpha2}), and thus Eqs.(\ref{Eq:Eij_W0}) and (\ref{Eq:Eij_W1}), which gives
\beqa
E_{ij}^{(1)} - 2\Lambda_W E_{ij}^{(0)} &=& R_{ij}-\frac{1}{2} R g_{ij} + \Lambda_W g_{ij}, \nn\\
E^{(1)} - 2\Lambda_W E^{(0)} &=& \frac{2-d}{2} R + d\Lambda_W.
\eeqa

For the generic potential in Eq.(\ref{Eq:W_d_(d+1)D}) with arbitrary parameters, the metric solution to the black hole of the action in Eq.(\ref{Eq:S-(d+1)D}) is determined by solving for the real roots of a polynomial equation. For arbitrary $\alpha_p$s, on one hand, it is not easy to extract physics from the solution, e.g., there are negative energy solutions with horizons and positive energy solutions with naked singularities, resulting in physics difficulties. Therefore, we will consider a particular choice of the Lovelock coefficients $\alpha_p$ in Eq.(\ref{Eq:Wp}), requiring that the theory posses a unique cosmological constant~\cite{Banados:1993ur,Troncoso:1999pk},
\beqa
\alpha_p =
\alpha_0 (2\delta)^p \left\{ \begin{aligned}
&  \frac{d}{d-2p} \left(\begin{array}{c}
                   n-1 \\
                   p
                 \end{array}\right)  ,  \quad d= 2n-1 \\
&  \left(\begin{array}{c}
                   n \\
                   p
                 \end{array}\right),  \quad d=2n \\
\end{aligned} \right.  \label{Eq:alphap}
\eeqa
where $\alpha_0$ is the zeroth order Lovelock coefficient and $\delta$ is a constant to be determined.
In order to match with the lowest order, e.g., Eqs.(\ref{Eq:W_1-6D}) and (\ref{Eq:W_5-6D_2}), one can choose parameters as below
\beqa
\alpha_0 = -\frac{2\Lambda_W}{d!\kappa_W^2}, \quad \alpha_1 = \frac{1}{(d-2)! \kappa_W^2}. \label{Eq:alpha0-alpha1_d}
%\quad \alpha_2 = \frac{\gamma}{(d-4)!}.
\eeqa

In the following, let us consider the most general action of the generalized HL$_{d+1}$ gravity,
\beqa
\ii\ii
S  \! = \! \int \! dt \int d^d x \sqrt{g}  N \bigg(  \frac{2}{\kappa^2}(K_{ij}K^{ij} \!-\! \lambda K^2)  \!-\! {\mathcal V}[g] \bigg),
\label{Eq:S-(d+1)D_2}
\eeqa
where ${\mathcal V}[g]$ is given in Eq.(\ref{Eq:SV_1_dD-2}).

By direct substituting the metric in Eq.(\ref{Eq:ds2-topological-HL}) back into the action in Eq.(\ref{Eq:S-(d+1)D_2}), one obtains
\beqa
\3i
S \3i &=& \3i  \int dt \int dr  \tilde{N} \Omega_{d-1,k}  r^{d-1}  {\mathcal L},  \label{Eq:S-L-(d+1)D}\\
{\mathcal L} \3i &=& \3i - \! \frac{\kappa^2}{8}  \sum_{p,q=0}^{[d/2]} (d\!-\!2p)!\alpha_p (d\!-\!2q)! \alpha_q( E^{(p)ij}E_{ij}^{(q)} \!-\! \tilde\lambda E^{(p)}E^{(q)} ),  \ii \nn
\eeqa
where $\sqrt{g}N = \sqrt{g_{rr}(r^2)^{d-1}\gamma} \sqrt{g_{tt}}\tilde{N} = \tilde{N}\sqrt{\gamma}r^{d-1}$, and $\int d^dx = \int dx^{d-1} dr$, $\int \sqrt{\gamma} dx^{d-1} = \Omega_{d-1,k}$. In addition, according to Eq.(\ref{Eq:Eij_Wp}), we have
\beqa
\ii
E^{(p)} &=& - \frac{1}{2^{p+1}} \delta^j_i \delta^{i i_1 \ldots i_{2p}}_{j j_1 \ldots j_{2p}}\prod_{k=1}^p R^{j_{2k-1} j_{2k}}_{~~ i_{2k-1} i_{2k}} \nn\\
&=& - \frac{1}{2} \frac{s!}{(s-2p)!}  r^{-s} \bigg[  r^{s+1} \bigg( \frac{g(r)}{r^2} \bigg)^p \bigg]^\prime, \quad\label{Eq:Eij(p)-E(p)}
\eeqa
where $s \equiv \delta^m_n \delta^n_m = d-1$, $ t \equiv \delta^i_j \delta^j_i = d$ are defined as the range of the indices for spatial space $\{x^m\}$ and $\{r, x^m\}$ respectively. Note that $E^{(p)} \equiv -\frac{(t-2p)}{2} {\mathcal R}^{(p)}$.
In the above derivation, we have used the identities between the antisymmetric Kronecker deltas symbols,
\beqa
\3i
\delta^{i_1 \ldots i_p}_{j_1 \ldots j_p}\delta^{j_1}_{i_1}\delta^{j_2}_{i_2} \ldots \delta^{j_m}_{i_m} \3i & = &\3i \frac{(s - p +m)!}{(s-p)!} \delta^{i_{m+1}\ldots i_p}_{j_{m+1}\ldots j_p} , \nn\\
\3i
\delta^{i_1 \ldots i_{2p}}_{j_1 \ldots j_{2p}} \delta^{j_1 j_2}_{i_1 i_2} \ldots \delta^{j_{2m-1} j_{2m}}_{i_{2m-1} i_{2m}} \3i &=& \3i \frac{2^m (s - 2p + 2m)!}{(s - 2p)!}\delta^{i_{2m+1}\ldots i_{2p}}_{j_{2m+1}\ldots j_{2p}} , \nn \\
\3i
\delta^{i i_1 \ldots i_{2p}}_{j j_1 \ldots j_{2p}} \delta^{j_1 j_2}_{i_1 i_2} \ldots \delta^{j_{2m\!-\!1} j_{2m}}_{i_{2m\!-\!1} i_{2m}} \3i &=& \3i \frac{2^m (s \!-\! (2p\!+\!1) \!+\! 2m)!}{(s \!-\! (2p\!+\!1))!}\delta^{i i_{2m\!+\!1}\ldots i_{2p}}_{j j_{2m\!+\!1}\ldots j_{2p}} . \nn
\eeqa
The above results in Eq.(\ref{Eq:Eij(p)-E(p)}) are correct, by using Eq.(\ref{Eq:R-Rij2-Rijkl2}), which reproduces same results for the trace $E^{(p)}$ at leading order in Eqs.(\ref{Eq:trace_Eij_W0}), (\ref{Eq:trace_Eij_W1}) and (\ref{Eq:trace_Eij_W2}), respectively.
For the field equation of motion $E^{i(p)}_j$, the nonvanishing component\footnote{All other $E^{m(p)}_r=E^{r(p)}_n=0$.} turns out to be $E^{m(p)}_n$ and $E^{r(p)}_r$,
\beqa
E^{r(p)}_r &=& - \frac{1}{2^{p+1}} \delta^{r i_1 \ldots i_{2p}}_{r j_1 \ldots j_{2p}}\prod_{k=1}^p R^{j_{2k-1} j_{2k}}_{~~ i_{2k-1} i_{2k}} \nn\\
&=& -\frac{1}{2}\frac{s!}{(s-2p)!}\bigg( \frac{g(r)}{r^2} \bigg)^p,  \label{Eq:Err(p)}
\eeqa
where all of index $i_k,j_k$ with $k=1,\ldots 2p$ does not include $r$ component, otherwise it is vanishing due to the totally antisymmetric properties of the generalized Kronecker deltas. And
\beqa
\ii
E_{n}^{m(p)} &=& - \frac{1}{2^{p+1}} \delta^{m i_1 \ldots i_{2p}}_{n j_1 \ldots j_{2p}}\prod_{k=1}^p R^{j_{2k-1} j_{2k}}_{~~ i_{2k-1} i_{2k}} \nn\\
&=& -\frac{1}{2} \frac{(s-1)!}{(s-2p)!}  r^{-(s-1)} \bigg[ r^{s} \bigg( \frac{g(r)}{r^2} \bigg)^p \bigg]^\prime \delta^m_n.  \qquad \label{Eq:Emn(p)}
\eeqa
It is straightforward to check the consistency of the results in Eqs.(\ref{Eq:Eij(p)-E(p)}), (\ref{Eq:Emn(p)}), and (\ref{Eq:Err(p)}) by observing that
\beqa
E^{(p)} = E^{m(p)}_{n}\delta^n_m + E^{r(p)}_r.
\eeqa
Thus, the Lagrangian density in Eq.(\ref{Eq:S-L-(d+1)D}) becomes
\beqa
\2i
{\mathcal L} \3i  & =  & \3i  -\! \frac{\kappa^2}{8}  \sum_{p,q=0}^{[d/2]} (d\!-\!2p)!\alpha_p (d\!-\! 2q)! \alpha_q \nn\\
\3i && \3i \times [ (E^{m(p)}_n E^{n(q)}_m +  E^{r(p)}_r E^{r(q)}_r) \!-\!  \tilde\lambda (E^{m(p)}_{n}\delta^n_m + E^{r(p)}_r) ], \nn
\eeqa
where we have used $E^{(p)ij} E^{(q)}_{ij} =  E^{m(p)}_n E^{n(q)}_m +  E^{r(p)}_r E^{r(q)}_r$.
By substituting back the Lovelock coefficients as shown in Eq.(\ref{Eq:alphap}), for both odd $d=2n-1$ and even $d=2n$ dimensions respectively, and by using the matching parameters in Eq.(\ref{Eq:alpha0-alpha1_d}), the Lagrangian can be simplified as shown below:
\begin{widetext}
\beqa
{\mathcal L}
&=& - \frac{\kappa^2}{8} \frac{\Lambda_W^2}{\kappa_W^4} \Bigg[ \frac{1}{d-1}   \bigg[ r^{-(d-2)}\Big[ r^{d-1} G_{n-1}(r) \Big]^\prime \bigg]^2   +   G_{n-1}^2(r)  - \tilde\lambda   \bigg[  r^{-(d-1)}\Big[ r^{d} G_{n-1}(r) \Big]^\prime  \bigg]^2 \Bigg]  \nn\\
&=& \frac{\kappa^2}{8} \frac{\Lambda_W^2}{\kappa_W^4}\frac{d}{d\lambda-1} \bigg[   G_{n-1}^2(r) + 2\frac{r}{d} G_{n-1}(r)G_{n-1}^\prime(r) -  \frac{\lambda-1}{d(d-1)} r^2[G_{n-1}^\prime(r)]^2  \bigg].
\eeqa
where $G_n(r) \equiv  \Big(1 + 2\delta {g(r)}/{r^2} \Big)^{n}$.
Therefore, the action in Eq.(\ref{Eq:S-L-(d+1)D}) can be simplified as shown below:
\beqa
S = -\frac{\Omega_{d-1,k} c^3}{16\pi G_N} \int dt \int dr \tilde{N}(r) \frac{d\Lambda_W}{(d-2)}  r^{d-1} \bigg(   G_{n-1}^2(r) + 2\frac{r}{d} G_{n-1}(r)G_{n-1}^\prime(r) -  \frac{\lambda-1}{d(d-1)} r^2[G_{n-1}^\prime(r)]^2  \bigg) ,  \label{Eq:S-HL(d+1)D-n}
\eeqa
where we have used the matching condition in Eq.(\ref{Eq:NormGravity}).
By doing variation upon the action with respect to the $\tilde{N}$ and $g(r)$, respectively, we obtain the most general field equations of motion, i.e., $\delta S/\delta \tilde{N}(r)=0$ and $\delta S/\delta g(r)=0$, respectively, gives
\beqa
&& G_{n-1}^2(r) + 2\frac{r}{d} G_{n-1}(r)G_{n-1}^\prime(r) -  \frac{\lambda-1}{d(d-1)} r^2[G_{n-1}^\prime(r)]^2 =0 , \label{Eq:EOMs-G(n-1)} \\
&& \frac{r^d}{d}\bigg[  \bigg( -G_{n-1}(r) + r \frac{\lambda-1}{d-1}G_{n-1}^\prime(r)   \bigg) \tilde{N}^\prime(r) + (\lambda-1)\bigg( \frac{d+1}{d-1} G_{n-1}^\prime(r) + \frac{r}{d-1} G_{n-1}^{\prime\prime}(r) \bigg) \tilde{N}(r)  \bigg] \frac{\delta G_{n-1}^\prime(r)}{\delta g} = 0. \nn
\eeqa
\end{widetext}

For the relativistic case with $\lambda=1$, the field equations of motion are
\beqa
G_{n-1}^2(r) + 2\frac{r}{d} G_{n-1}(r)G_{n-1}^\prime(r)=0, \quad \tilde{N}^\prime(r) = 0, \nn
\eeqa
which give solutions
\beqa
G_{n-1}(r) = C_G r^{-\frac{d}{2}}, \quad \text{and} \quad \tilde{N}(r) = \text{const.},
\eeqa
with $C_G=0$ or $C_G\ne 0$. Without losing generality, by choosing the constant to be the norm, the solutions become
\beqa
g(r) = - \frac{r^2}{2\delta}(1- C_G r^{-\frac{d}{2(n-1)}}), \quad \tilde{N}(r)=1.
\eeqa
Therefore, according to Eq.(\ref{Eq:gr-fr}), we obtain
\beqa
f(r) = k + \frac{1}{2\delta}r^2  - C_F r^{2-\frac{d}{2(n-1)}}, \label{Eq:fr-HL(d+1)-delta}
\eeqa
with $C_F=0$ or $C_F\ne 0$. The first solution is nothing but the asymptotic AdS solution; thus, the constant $\delta$ can be matched to the AdS case as
\beqa
\frac{1}{2\delta} \equiv - \frac{2\Lambda_W}{(d-1)(d-2)} = \frac{2}{\ell^2},  \label{Eq:delta-LambdaW}
\eeqa
where we have used Eq.(\ref{Eq:Lambda_W-ell}), while the second solution in Eq.(\ref{Eq:fr-HL(d+1)-delta}) with $C_F\ne 0$ is
\beqa
f(r) = \left\{ \begin{aligned}
&  k - \frac{2\Lambda_W}{(d-1)(d-2)} r^2  - C_F r^{\frac{2n-3}{2n-2}} , ~  d=2n-1 \\
&  k - \frac{2\Lambda_W}{(d-1)(d-2)} r^2  - C_F r^{\frac{n-2}{n-1}}  ,  ~  d=2n \\
\end{aligned} \right.
\eeqa
For the field equations of motion given in Eq.(\ref{Eq:EOMs-G(n-1)}), the most general solution turns out to be
\beqa
G_{n-1}(r) \2i &=& \2i  C_G r^{\lambda_G^\pm}, ~ \lambda_G^\pm \equiv \frac{d\!-\!1\pm \sqrt{(d\!-\!1)(d\lambda \!-\!1)}}{\lambda\!-\!1}, \label{Eq:Ntr-HL(d+1)}\\
\tilde{N}(r) \2i &=& \2i C_N r^{\lambda_N^\pm}, ~ \lambda_N^\pm \equiv -\frac{(d\!-\!2)\!+\!d\lambda \! \pm \! 2\sqrt{(d\!-\!1)(d\lambda\!-\!1)}}{\lambda\!-\!1},  \nn
\eeqa
and it is worthy of notice that
\beqa
\lambda_N^\pm + 2 \lambda_G^\pm = -d,  \label{Eq:lambdaN-lambdaG}
\eeqa
thus $\lambda_N^\pm$ are not independent of $\lambda_G^\pm$ respectively.

The exponent in some special limit reduces to be
\beqa
\lim_{\lambda\to 1^\pm}\lambda_{N}^+ &=& \mp\infty, \quad \lim_{\lambda\to 1^\pm}\lambda_{N}^- = 0,  \nn\\
\lim_{\lambda\to {1}/{d}}\lambda_{N}^\pm &=& d, ~ \lim_{\lambda\to \pm \infty}\lambda_{N}^\pm = -d.
\eeqa
For both odd $d=2n-1$ and even $d=2n$ dimensions respectively,
\beqa
f(r) &=& k - \frac{2\Lambda_W}{(d-1)(d-2)}r^2  - C_F r^{\lambda_F^\pm}, \label{Eq:fr-HL(d+1)-n}\\
\lambda_F^\pm  &=& \frac{2\lambda(n-1)+d-(2n-1)\pm \sqrt{(d-1)(d\lambda - 1 )}}{(\lambda-1)(n-1)}.  \nn
\eeqa
where $\lambda_F^\pm  \equiv  2 + {\lambda_G^\pm}/{(n-1)}$ and according to Eq.(\ref{Eq:lambdaN-lambdaG}), one obtains
\beqa
\lambda_N + 2(n-1)\lambda_F = 4(n-1)-d.  \label{Eq:lambdaN-lambdaF}
\eeqa
The corresponding exponent has the behavior
\beqa
  \lim_{\lambda\to 1^\pm}\lambda_F^+ &=& \pm\infty, ~ \lim_{\lambda\to 1^\pm}\lambda_F^- = 2 - \frac{d}{2(n-1)}, \nn\\
  \lim_{\lambda\to {1}/{d}}\lambda_F^\pm &=& 2-\frac{d}{n-1}, ~  \lim_{\lambda\to \pm \infty}\lambda_F^\pm = +2.
\eeqa
In summary, the general solution of the topological black holes in HL$_{d+1}$ gravity is given in Eq.(\ref{Eq:ds2-topological-HL}) with
\beqa
\ii\ii
\tilde{N}^2(r)  \2i &=& \2i C_N^2 r^{2\lambda^\pm_N},\nn\\
f(r) \2i &=& \2i k - \frac{2\Lambda_W}{(d-1)(d-2)} r^2 - C_F r^{\lambda_F^\pm}. \label{Eq:ds2-HL(d+1)}
\eeqa
it is worthy of notice the difference between the metric of Lifshitz gravity in Refs.~\cite{Kachru:2008yh,Griffin:2012qx} and the metric of topological neutral black holes in HL gravity with $k=0$ case. To be more concrete,

\begin{enumerate}

\item $\lambda=1$,
\beqa
ds^2 &=&  - f(r) dt^2 + f(r)^{-1}dr^2 + r^2 d\Omega_{d-1,k}^2.  \nn\\
f(r) &=& k - \frac{2\Lambda_W}{(d-1)(d-2)} r^2 - C_F r^{2-\frac{d}{2(n-1)}}, \label{Eq:ds2-topological-HL(d+1)-lambda=1}
\eeqa
For odd $d=2n-1$ and even $d=2n$ dimensions, one has
\beqa
f(r)
=  \left\{ \begin{aligned}
&  k \!-\! \frac{2\Lambda_W}{(d-1)(d-2)} r^2 \!-\! C_F r^{\frac{2n-3}{2n-2}} , ~ d=2n-1 \\
&  k \!-\! \frac{2\Lambda_W}{(d-1)(d-2)} r^2 \!-\! C_F r^{\frac{n-2}{n-1}}   , ~ d=2n \\
\end{aligned} \right.
\eeqa

\item $\lambda=\frac{1}{d}$,
\beqa
ds^2  &=&  - r^{2d} f(r) dt^2 + f(r)^{-1}dr^2 + r^2 d\Omega_{d-1,k}^2; \nn\\
f(r) &=& k - \frac{2\Lambda_W}{(d-1)(d-2)} r^2 - \frac{C_F}{r^{\frac{d}{n-1}-2}} \label{Eq:ds2-topological-HL(d+1)-lambda=1/3}
\eeqa
where $C_N$ is normalized to be unit and $C_F$ is irrelevant in the infinite boundary. The solution has an IR singularity at $r=0$ if $C_F\ne 0$ in the large distance where $C_F$ is relevant. For odd $d=2n-1$ and even $d=2n$ dimensions, one has
\beqa
f(r)
=  \left\{ \begin{aligned}
&  k \!-\! \frac{2\Lambda_W}{(d-1)(d-2)} r^2 \!-\! C_Fr^{n-1} ,  ~ d=2n-1 \\
&  k \!-\! \frac{2\Lambda_W}{(d-1)(d-2)} r^2 \!-\! C_Fr^{\frac{n-1}{2}} ,  ~ d=2n \\
\end{aligned} \right.
\eeqa

\item $\lambda=\pm \infty$,
\beqa
ds^2 &=& - r^{-2d} f(r) dt^2 +  f(r)^{-1} \ii  dr^2 + r^2 d\Omega_{d-1,k}^2; \nn\\
f(r) &=& k  -  \frac{2\Lambda_W}{(d-1)(d-2)} r^2 - C_F r^2, \label{Eq:ds2-topological-HL(d+1)-lambda=infty}
\eeqa
where $C_N$ is normalized to be unit and $C_F$ can be absorbed into the cosmological constant.

\end{enumerate}

Note for $p=0,1$ case, the summation up to $p=1$; thus, $n-1=1$, so the solution in Eq.(\ref{Eq:fr-HL(d+1)-n}) recovers
\beqa
\lambda_F^\pm = \frac{2\lambda+(d-3)\pm \sqrt{(d-1)(d\lambda - 1 )}}{\lambda-1},
\eeqa
where $\lambda_F^\pm \equiv 2 + \lambda_G^\pm$. It just recovers the results as shown in Eq.(\ref{Eq:Fr}). This can be understood, since by considering the lowest curvature terms $W_{(p)}$ defined in Eq.(\ref{Eq:Wp}) for the $p=0,1$ case, i.e., with the summation over to $p=1$, thus $n-1=1$, the action in Eq.(\ref{Eq:S-HL(d+1)D-n}) becomes
\begin{widetext}
\beqa
S = -\frac{\Omega_{d-1,k} c^3}{16\pi G_N} \int dt \int dr \tilde{N}(r) \frac{d\Lambda_W}{(d-2)}  r^{d-1} \bigg[   G_{1}^2(r) + 2\frac{r}{d} G_{1}(r)G_{1}^\prime(r) -  \frac{\lambda-1}{d(d-1)} r^2[G_{1}^\prime(r)]^2  \bigg] . %\nn\\
\eeqa
By substituting $\delta$ defined in Eq.(\ref{Eq:delta-LambdaW}) back, one obtains
\beqa
S &=&  \frac{\Omega_{d-1,k} c^3}{16\pi G_N} \int dt \int dr \tilde{N}(r)   r^{d-1} \bigg[ -\frac{d\Lambda_W}{d-2}d +\frac{(d-1)(d-2)}{\Lambda_W}\bigg( \frac{\Lambda_W g^\prime(r)}{(d-2)r} + \frac{\lambda-1}{4r^2}[g^\prime(r)]^2 \nn\\
&& \qquad \qquad - \frac{(d-3)+2\lambda}{2r^3}g(r)g^\prime(r) - \frac{(d-2)(d-3)+2(1-2\lambda)}{4r^4}g(r)^2    \bigg)    \bigg],
\eeqa
\end{widetext}
which just recovers the result $L_0+L_1$ as defined in Eqs.(\ref{Eq:L0}) and (\ref{Eq:L1}) separately.

For $p=0,1,2$ case, the summation up to $p=2$($n-1=2$), as we have shown before Eq.(\ref{Eq:Eij_W2}), with the Gauss-Bonnet term. In this case, the solution becomes
\beqa
\lambda_F^\pm  &=& \frac{4\lambda + d - 5 \pm \sqrt{(d-1)(d\lambda - 1 )}}{2(\lambda-1)}\overset{\lambda=1}{=} 2 - \frac{d}{4}, \nn
\eeqa
where $\lambda_F^\pm \equiv 2 + {\lambda_G^\pm}/{2}$.
Therefore, in the relativistic limit $\lambda=1$,
\beqa
f(r) = k \!-\! \frac{2\Lambda_W}{(d-1)(d-2)} r^2 \bigg( 1 \!-\! \frac{C_G}{r^{\frac{d}{4}}} \bigg),
\eeqa
for charged black holes with additional Maxwell action in Eq.(\ref{Eq:S-Maxwell-Einstein-RSS}), considering the relativistic case with $\lambda=1$. By varying the action $S$ with respect to $\phi(r)$, $p(r)$ and $\tilde{N}(r)$ respectively, one has the equations of motion
\beqa
&&  (r^{d-1} p)^\prime = 0, \quad  -r^{d-1} (\tilde{N}p+\phi^\prime) = 0, \nn\\
&& -\frac{d\Lambda_W}{(d-2)}  r^{d-1} \bigg(   G_{n-1}^2(r) + 2\frac{r}{d} G_{n-1}(r)G_{n-1}^\prime(r) \nn\\
&& -  \frac{\lambda-1}{d(d-1)} r^2[G_{n-1}^\prime(r)]^2  \bigg)  - \frac{1}{2}r^{d-1} p^2 = 0. \label{Eq:HL(d+1)-Nt-p-phi-z=1}
\eeqa
Finally, one obtains
\beqa
G_{n-1}(r) = r^{-\frac{d}{2}}\sqrt{C_G + \frac{q_0^2 }{2\Lambda_W r^{d-2}}},
\eeqa
where $C_G$ is an integral constant, and
\begin{widetext}
\beqa
f(r) = k - \frac{2\Lambda_W r^2}{(d-1)(d-2)} - r^{2-\frac{d}{2(n-1)}} \bigg(c_0 + \frac{ 2\Lambda_W }{ [(d-1)(d-2)]^2  }\frac{q_0^2}{r^{d-2}} \bigg)^{\frac{1}{2(n-1)}}.
\eeqa
\end{widetext}

\subsection{$z=5$ HL$_{d+1}$ gravity}

In $(d+1)$ dimensions, the coupling $\kappa$ will be dimensionless if $z=d$; thus, the $z=5$ HL$_{d+1}$ gravity will be a power-counting renormalizable UV theory at $z=5$ Lifshitz points at short distance in $d \le 5$. For $z=5$ HL$_{d+1}$ gravity, except for the leading order spatial potential,
\beqa
W_1 = \frac{1}{\kappa^2_W}\int d^d x \sqrt{g}(R-2\Lambda_W), \label{Eq:W_1-D}
\eeqa
there will be a new term present in spatial potential since $d = 5$, i.e., Gauss-Bonnet term\cite{Boulware:1985wk,Cai:2001dz},
\beqa
{\mathcal L}_{GB} =  R^2 - 4R_{ij}R^{ij} +  R_{ijkl}R^{ijkl}, \label{Eq:GB}
\eeqa
which is nontrivial and is present since $d = 5$ dimensions. It first appears in $(4+1)$ dimensions but reduces to be a topological invariant surface term describing the Euler number of the spatial surface in $(3+1)$ dimensions and less, i.e., $\chi_{E} = \int d^4x \sqrt{g}{\mathcal L}_{GB}$. Consequently, this term contributes nothing to the field equation of motion in $d\le 4$; thus, naively it can be viewed as absent in $d\le 4$.
\beqa
\ii
W_5 \!= \!  \int d^d x \sqrt{g}   \gamma {\mathcal L}_{GB},  \label{Eq:W_5-6D_2}
\eeqa
where $[\gamma]=1$. The action in Eq.(\ref{Eq:W_1-D}) together with that in Eq.(\ref{Eq:W_5-6D_2}) consists the spatial potential generating the $z=5$ HL$_{d+1}$ gravity since $d = 5$. For the purpose of generalization, the homogeneous spatial $W_5$ potential, which consists of the most general quadratic curvature terms with three independent
\footnote{ Since $C^{ijkl}C_{ijkl}$ is not independent of $R^2$, $R^{ij}R_{ij}$ and ${\mathcal L}_{GB}$ according to Eq.(\ref{Eq:Weyl-square-d-GB}). We do not consider the topological Pontryagin density term $\tilde{R}R$, which is totally depicted by the Weyl tensor, since
\beqa
\tilde{R}R=\tilde{C}C, \, \tilde{C}^{ijkl}=\frac{1}{2}\epsilon^{ijmn}C^{kl}_{~~mn}, \, \tilde{R}^{ijkl}=\frac{1}{2}\epsilon^{ijmn}R^{kl}_{~~mn}. \nn
\eeqa
which violates the parity of the Lagrangian~\cite{Cai:2009ar}.
}
dimensionless couplings $\alpha$, $\beta$, $\gamma$,

%\footnote{That is, since $C^{ijkl}C_{ijkl}$ is not independent of $R^2$, $R^{ij}R_{ij}$ and ${\mathcal L}_{GB}$ according to Eqs.(\ref{Eq:Weyl-anomaly-d}) and (\ref{Eq:GB}). We do not consider the topological Pontryagin density term $\tilde{R}R$, which is totally depicted by the Weyl tensor, since the term
%\beqa
%\tilde{R} R=\tilde{C}C, \, \tilde{C}^{ijkl}=\frac{1}{2}\epsilon^{ijmn}C^{kl}_{mn}, \, \tilde{R}^{ijkl}=\frac{1}{2}\epsilon^{ijmn}R^{kl}_{~~mn}.
%\eeqa
%violates the parity of the Lagrangian~\cite{Cai:2009ar}.
%},
\beqa
\ii
W_5 \!&=&\!  \int d^d x \sqrt{g} (   \alpha R^{ij}R_{ij} + \beta R^2 + \gamma {\mathcal L}_{GB} ) ,  \label{Eq:W_5-D}
\eeqa
where $[\alpha]=[\beta]=[\gamma]=d-4$. The action in Eq.(\ref{Eq:W_5-D}) is still renormalizable since the scaling dimension of the four parameters $\alpha$, $\beta$, $\gamma$ is not dimensionless.

According to the detailed balance condition in Eq.(\ref{Eq:Eij}), one obtains $E_{ij} = (\alpha-4\gamma) E_{ij}(R^{ij}R_{ij}) + (\beta+\gamma) E_{ij}(R^2)  + \gamma E_{ij}(R^{ijkl}R_{ijkl})$, with
\beqa
&& E_{ij}(R^2) = 2 [ R R_{ij} + (g_{ij}\nabla^2-\nabla_i\nabla_j) R] - \frac{1}{2}g_{ij}R^2, \nn\\
&& E_{ij}(R^{ij}R_{ij}) = 2 R^{kl}\bigg( R_{ikjl} - \frac{1}{4}g_{ij}R_{kl}\bigg) + \nabla^2 G_{ij} \nn\\
&& \qquad \qquad\qquad + (g_{ij}\nabla^2 -\nabla_i \nabla_j)R, \nn\\
&& E_{ij}(R^{ijkl}R_{ijkl}) = 2 R_{i lmn }R_j^{~ lmn} + 4 R^{kl} R_{ikjl} - 4R_{ik}R^{k}_{~j} \nn\\
&& \quad + 4\nabla^2 G_{ij} + 2 (g_{ij} \nabla^2-\nabla_i \nabla_j) R - \frac{1}{2}g_{ij}R^{klmn}R_{klmn}, \nn
\eeqa
where $G_{ij} \equiv  R_{ij}- g_{ij} R/2 $.
Therefore, one obtains $J_{ij}$ and its trace
\beqa
E_{ij}(W_5) &=&  (\alpha + 2\beta )(g_{ij}\nabla^2 -\nabla_i \nabla_j)R + \alpha \nabla^2G_{ij} \nn\\
&& + 2(\alpha-2\gamma)  R^{kl}( R_{ikjl} - \frac{1}{4}g_{ij}R_{kl}) \nn\\
&& + 2(\beta+\gamma) R(R_{ij}  - \frac{1}{4}g_{ij}R)  \nn\\
&& + 2\gamma [ R_{i lmn }R_j^{~ lmn}  - 2 R^{k}_{~i}R_{kj} \nn\\
&& + \frac{1}{2}g_{ij}(R^{kl}R_{kl} - \frac{1}{2}R^{klmn}R_{klmn})] , \label{Eq:Jij} \\
E(W_5) & =&   2\bigg(1 - \frac{d}{4} \bigg)(\alpha R^{kl}R_{kl} + \beta R^2 + \gamma{\mathcal L}_{GB}) \nn\\
&& + 2\bigg(\frac{d}{4}\alpha + (d-1)\beta \bigg)\nabla^2 R , \label{Eq:J}
\eeqa
where $E \equiv  g^{ij}E_{ij}$.

Since $R^{ij}R_{ij}$ and $R^2$ will be considered in $z=4$ HL$_{d+1}$ gravity since $d=4$ dimensions in the next section, in the following we will mainly consider the term since $d=5$ dimensions, i.e., $\gamma\ne 0$ is imposed and Eq.(\ref{Eq:W_5-6D}) becomes Eq.(\ref{Eq:W_5-6D_2}).

\subsubsection{Gauss-Bonnet term:$\gamma\ne 0$, $\alpha=\beta=0$}

When $\gamma\ne 0$, and $\alpha=\beta=0$, according to Eqs.(\ref{Eq:Jij}) and (\ref{Eq:J}), for Eq.(\ref{Eq:W_5-6D_2}), one obtains
\beqa
E_{ij} = \gamma J_{ij}, \quad [J_{ij}]= [E_{ij}] - (d-4) =4,
%E^{(2)ij} = -\gamma J^{ij}, \quad [E^{(2)ij}]= [J^{ij}] +1 =5,
\eeqa
where $E_{ij}\equiv E_{ij}(W_5)$ and $E \equiv E(W_5)$ as below,
\beqa
J_{ij} &=& 2 [ R_{i lmn }R_j^{~ lmn} - 2 R^{kl} R_{ikjl}  - 2 R^{k}_{~i}R_{kj} + R R_{ij} \nn\\
&& - \frac{1}{4}g_{ij}(R^2 - 4 R^{ij}R_{ij} + R^{ijkl}R_{ijkl})], \label{Eq:Jij-GB} \\
J &\equiv &  g^{ij}J_{ij} = 2 \bigg(1 - \frac{d}{4} \bigg) {\mathcal L}_{GB}, \label{Eq:J-GB}
\eeqa
which are consistent with those results in Eq.(\ref{Eq:Eij_W2}), namely, $J_{ij}=E_{ij}^{(2)}$ and $J=E^{(2)}$.
The $z=5$ HL$_{d+1}$ gravity action is
\beqa
S \3i &=& \3i \int dt \int d^d x \sqrt{g}  N \bigg[  \frac{2}{\kappa^2}(K_{ij}K^{ij}-\lambda K^2)  \nn\\
\3i &-& \3i  \frac{\kappa^2}{8\kappa_W^4}\bigg(R^{ij}R_{ij} \!-\! \frac{1 \!-\! d/4 \!-\! \lambda }{1 \!-\! d\lambda}R^2 \!-\! \frac{(d \!-\! 2)\Lambda_W R \!-\! d\Lambda_W^2}{1 \!-\! d\lambda} \bigg) \nn\\
\3i &-& \3i \frac{\kappa^2}{4\kappa_W^2} \gamma \bigg( E^{(2)ij}R_{ij} \!-\! \frac{1\!-\!2\lambda}{2(1\!-\!d\lambda)}E^{(2)}R \!+\! \frac{\Lambda_W}{1\!-\!d\lambda} E^{(2)} \bigg) \nn\\
\3i &-& \3i \frac{\kappa^2}{8}\gamma^2  (E^{(2)ij}E_{ij}^{(2)}-\tilde\lambda E^{(2)}E^{(2)}) \bigg], \label{Eq:S-HL(d+1)-z=5}
\eeqa
which is a special case of Eq.(\ref{Eq:S-(d+1)D}) with $\alpha_2 (d-4)! = \gamma $ since $d=5$.

Under the metric ansatz in Eq.(\ref{Eq:ds2-topological-HL}), $E^{ij(2)}$ and $E^{(2)}$ are given by Eqs.(\ref{Eq:Eij(p)-E(p)}), (\ref{Eq:Err(p)}), and (\ref{Eq:Emn(p)}),
\beqa
E^{(p)} &=& - \frac{1}{2} \frac{(d-1)!}{(d-5)!}  r^{-(d-1)} \bigg[  r^{d} \bigg( \frac{g(r)}{r^2} \bigg)^2 \bigg]^\prime, \nn\\
E^{r(2)}_r &=& -\frac{1}{2}\frac{(d-1)!}{(d-5)!}\bigg( \frac{g(r)}{r^2} \bigg)^2,   \nn \\
\ii
E_{n}^{m(2)}
&=& -\frac{1}{2} \frac{(d-2)!}{(d-5)!}  r^{-(d-2)} \bigg[ r^{d-1} \bigg( \frac{g(r)}{r^2} \bigg)^2 \bigg]^\prime \delta^m_n.  \nn
\eeqa

The equations of motions can be obtained by substituting Eq.(\ref{Eq:R-Rij2-Rijkl2}),
and we find that there are four branch solutions to HL$_{d+1}$ at $z=5$ UV fixed point,
\begin{widetext}
\beqa
g(r) \! = \!  -\frac{r^2}{2(d-3)(d-4)\gamma\kappa_W^2} \pm r^{2-\frac{d}{4}}\sqrt{C_G  + \frac{1}{4(d-3)^2(d-4)^2\gamma^2\kappa_W^4}\bigg( 1 + \frac{8(d-3)(d-4)\Lambda_W}{(d-1)(d-2)}\gamma\kappa_W^2  \bigg)r^{\frac{d}{2}} } ,
\eeqa
with $C_G=0$ or $C_G\ne 0$. While there are only two physical solutions, assuming that the solution should recover those in the limit $\gamma\to 0$,
\beqa
f(r) \!=\! k \!+\! \frac{r^2}{2(d-3)(d-4)\gamma\kappa_W^2}\bigg[ 1 \!-\!  \sqrt{ \bigg( 1 \!+\! \frac{8(d-3)(d-4)\Lambda_W}{(d-1)(d-2)}\gamma\kappa_W^2 \bigg) \!+\! \frac{C_F}{r^{\frac{d}{2}}}  } \bigg] \approx k \!-\! \frac{2\Lambda_W}{(d-1)(d-2)}r^2 + {\mathcal O}(\gamma),
\eeqa
with $C_F=0$ or $C_F=[2(d-3)(d-4)\gamma\kappa_W^2]^2C_G\ne 0$. In particular, for $d=3$ and $d=4$, one obtains HL$_4$ and HL$_5$ at $z=5$, respectively,
\beqa
\text{HL}_4:  f(r) = k - \Lambda_W r^2 \pm C \sqrt{r}; \quad \text{HL}_5: f(r) = k - \frac{\Lambda_W}{3}r^2 \pm C,
\eeqa
with $C=0$ or $C \ne 0$.
For charged black holes of HL$_{d+1}$ gravity with $z=5$ UV fixed point, we obtain
\beqa
f(r) \!=\! k + \frac{r^2}{2(d-3)(d-4)\gamma \kappa_W^2} \bigg[  1 \!-\! \sqrt{ \bigg( 1 \!+\! \frac{8(d-3)(d-4) \Lambda_W \gamma\kappa_W^2 }{(d-1)(d-2) } \bigg) \!+\!  \frac{\gamma \kappa_W^2}{r^{\frac{d}{2}}}  \sqrt{   \frac{C_F^2}{\gamma^2 \kappa_W^4}  \!+\!  \frac{ 32(d-3)^2(d-4)^2 \Lambda_W }{(d-1)^2(d-2)^2}\frac{q_0^2  }{r^{d-2}}      }   } \bigg] , \nn
\eeqa
with $C_F\ne 0$. For $d=5$, one obtains the charged black holes of HL$_6$ at $z=5$, one obtains
\beqa
f(r) \!=\! k + \frac{r^2}{4\gamma \kappa_W^2} \bigg(  1 \!-\! \sqrt{  1 \!+\! \frac{4 \Lambda_W \gamma\kappa_W^2 }{ 3 }  \!+\!  \frac{\gamma \kappa_W^2}{r^{\frac{5}{2}}}  \sqrt{   \frac{C_F^2}{\gamma^2 \kappa_W^4}  \!+\!  \frac{8\Lambda_W }{9}\frac{q_0^2  }{r^{3}}      }   } \bigg). \nn
\eeqa
\end{widetext}

\subsubsection{Conformal term:$\gamma\ne 0$, $\alpha\ne 0$ and $\beta\ne 0$ }

When $ \gamma\ne 0$, but $\alpha\ne 0$ and $\beta\ne 0$ too , the action in Eq.(\ref{Eq:W_5-D}) becomes
\beqa
\ii\ii
W_5 \2i=\2i \int d^dx \sqrt{g} \gamma \bigg[ \frac{\alpha}{\gamma}\bigg(  R^{ij}R_{ij} + \frac{\beta}{\alpha} R^2 \bigg) +  {\mathcal L}_{GB} )\bigg] . \label{Eq:W_5-D_GB}
\eeqa
Let us consider a special case for Eq.(\ref{Eq:W_5-D}). When
\beqa
\beta = -\frac{d}{4(d-1)}\alpha, \label{Eq:beta-over-alpha}
\eeqa
according to Eq.(\ref{Eq:J}), one obtains the trace of the field equation of motion,
\beqa
E(W_5) = 2\bigg(1 \!-\! \frac{d}{4} \bigg) \bigg[ \alpha \bigg( R^{kl}R_{kl} \!-\! \frac{d}{4(d-1)} R^2 \bigg) \!+\! \gamma {\mathcal L}_{GB} \bigg], \nn
\eeqa
since the Laplacian term of Ricci scalar in Eq.(\ref{Eq:J}) vanishes.
In addition to Eq.(\ref{Eq:beta-over-alpha}), by choosing
\beqa
\frac{\alpha}{\gamma} = \frac{4(d-3)}{d-2}, \label{Eq:gamma-over-alpha}
\eeqa
the action Eq.(\ref{Eq:W_5-D_GB}) becomes that of conformal gravity,
\beqa
W_5 = \int d^dx \sqrt{g} \gamma C^{ijkl}C_{ijkl}, \label{Eq:W_5-Weyl-square}
\eeqa
where $C_{ijkl}$ are Weyl tensor defined as
\beqa
C_{ijkl} &\equiv & R_{ijkl} - \frac{2}{d-2}\big(g_{i[k}R_{l]j}-g_{j[k}R_{l]i}\big) \nn\\
&+&\frac{2}{(d-1)(d-2)}g_{i[k}g_{l]j}R,  \label{Eq:Weyl-d}
\eeqa
which describes completely the traceless part of the Riemann tensor.
In three dimensions, the Weyl tensor vanishes identically. Therefore, the term in Eq.(\ref{Eq:Weyl-d}) is present since $d = 4$. In $ d \ge 4$, the vanishing of $C_{ijkl}$ implies an equivalent condition of conformal flatness of a Riemann metric.

By using the definition of Weyl tensor in Eq.(\ref{Eq:Weyl-d}), one obtains the identity of the square of Weyl tensor, i.e.,
Weyl anomaly $C^{ijkl}C_{ijkl}$ as below,
\beqa
\ii
R_{ijkl}R^{ijkl} \!-\! \frac{4}{d-2}R_{ij}R^{ij} \!+\! \frac{2}{(d-1)(d-2)}R^2, \label{Eq:Weyl-anomaly-d}
\eeqa
which can be expressed as
\beqa
\ii
{\mathcal L}_{GB} + \frac{4(d-3)}{d-2}R_{ij}R^{ij} - \frac{d(d-3)}{(d-1)(d-2)}R^2,\label{Eq:Weyl-square-d-GB}
\eeqa
where ${\mathcal L}_{GB}$ is the Gauss-Bonnet term defined in Eq.(\ref{Eq:GB}). Alternatively, by comparing Weyl anomaly in Eq.(\ref{Eq:Weyl-square-d-GB}) with the action in Eq.(\ref{Eq:W_5-D_GB}), one recovers the ratio chosen in Eqs.(\ref{Eq:beta-over-alpha}) and (\ref{Eq:gamma-over-alpha}).

Therefore, the dynamics controlled by the action in Eq.(\ref{Eq:W_5-Weyl-square}) together with Eq.(\ref{Eq:W_1-D}) consists of the critical gravity in Riemman manifold with pure spatial metric $g$ since $d=5$ dimensions.

The trace of the field equation of motion become
%one just obtains Eq.(\ref{Eq:Weyl-square-d-GB})
\beqa
E(W_5) = \gamma E^{(2)} = 2\bigg( 1 - \frac{d}{4}\bigg) \gamma C^{ijkl}C_{ijkl}. \label{Eq:traceJ-d-critical}
\eeqa
In this case, one finds that, the equations of motion becomes
\beqa
E^m_{~n} &=& L^m_{~n} + \gamma E^{(2)m}_{~n}  =0 , \nn\\
E^r_{~r} &=& L^r_{~r} + \gamma E^{(2)r}_{~r}  =0 , \nn\\
E &=& L + \gamma E^{(2)} = 0,
\eeqa
where $L$s are given in Eqs.(\ref{Eq:Lmn}), (\ref{Eq:Lrr}), and (\ref{Eq:L-MG}); $J$s are given in Eq.(\ref{Eq:Jij-GB}) and Eq.(\ref{Eq:J-GB}).
Therefore, the equations of motion are completely determined by the leading order potential due to Eq.(\ref{Eq:W}).

It turns out that there are two branches for the solution to the topological neutral black hole in HL$_{d+1}$ inheriting from $d$-dimensional critical gravity
\begin{widetext}
\beqa
g(r) = \frac{2\Lambda_W}{(d-1)(d-2)}r^2 + C_G r^{2-\frac{d}{2}},
\eeqa
with $C_G=0$ or $C_G\ne 0$. The corresponding topological charged black hole with $\lambda=1$ becomes
\beqa
g(r) = \frac{2\Lambda_W}{(d-1)(d-2)}r^2  \pm r^{2-\frac{d}{2}}\sqrt{C_G+\frac{2\Lambda_W }{(d-1)^2(d-2)^2}\frac{q_0^2}{r^{d-2}}},
\eeqa
where only the branch with plus sign is physical, which reduces to the neutral black holes when the charge is absent. Therefore, the topological charged black holes of HL$_{d+1}$ gravity with conformal term, are given by
\beqa
f(r) = k - \frac{2\Lambda_W}{(d-1)(d-2)}r^2  - r^{2-\frac{d}{2}}\sqrt{C_G+\frac{2\Lambda_W }{(d-1)^2(d-2)^2}\frac{q_0^2}{r^{d-2}}},
\eeqa
with $C_G=0$ or $C_G\ne 0$.
\end{widetext}

\subsection{$z=4$ HL$_{d+1}$ gravity}

In this section, Let us consider the $z=4$ HL$_{d+1}$ gravity since $d = 4$. As stated before, the Gauss-Bonnet term is the topological surface term in $d \le 4$. As a result, it becomes a total derivative in the Lagrangian, so that it is absent in the equations of motion. This can be studied as a special case in the most general potential in Eq.(\ref{Eq:W_5-6D}) with $\gamma=0$,
\beqa
W_5 = \int d^dx \sqrt{g} \alpha \bigg( R^{ij}R_{ij} + \frac{\beta}{\alpha}  R^2 \bigg) . \label{Eq:W_5-D_NMG}
\eeqa
The corresponding field equations of motion are straightforward according to Eqs.(\ref{Eq:Jij})and (\ref{Eq:J}) respectively,
%\begin{widetext}
\beqa
L_{ij}(W_d) \3i &=& \3i (\alpha \!+\! 2\beta )(g_{ij}\nabla^2 \!-\! \nabla_i \nabla_j)R \!+\! \alpha \nabla^2G_{ij} \label{Eq:Lij-MG}\\
\3i &+& \3i  2\alpha  R^{kl}\bigg( R_{ikjl} \!-\! \frac{1}{4}g_{ij}R_{kl}\bigg)  \!+\! 2\beta R\bigg(R_{ij}  \!-\! \frac{1}{4}g_{ij}R\bigg), \ii \nn \\
L(W_d) \3i & = & \3i  2\bigg(\frac{d}{4}\alpha + (d-1)\beta \bigg)\nabla^2 R \nn\\
\3i &+& \3i 2\bigg(1 - \frac{d}{4} \bigg)(\alpha R^{kl}R_{kl} + \beta R^2), \label{Eq:L-MG}
\eeqa
where $L \equiv  g^{ij}L_{ij}$.

The topological black holes cannot be analytically solved unless one chooses a special parameter
\beqa
\frac{\beta}{\alpha} = -\frac{d}{4(d-1)} =
\left\{ \begin{aligned}
& -\frac{3}{8}, \quad d=3 \\
& -\frac{1}{3}, \quad d=4. %\\
%& -\frac{5}{16}, \quad d=5
\end{aligned} \right.
\eeqa
In this set of parameters, the diagonal terms of Laplacian of Ricci scalar as shown in Eq.(\ref{Eq:L-MG}) are vanishing, e.g. in the Lagrangian of z=4 HL gravity, the term $[g^{\prime\prime\prime}(r)]^2 \in L^2$ will be absent. Therefore, the corresponding algebra equation of the field equation of motion is a polynomial with the largest order no more than $4$.
The potential action in Eq.(\ref{Eq:W_5-D_NMG}) together with Eq.(\ref{Eq:W_1-D}) is nothing but the new massive gravity action in $d \ge 3$. By choosing the parameter
\beqa
\alpha\equiv \frac{1}{M},  \label{Eq:alpha-1/M}
\eeqa
for $d=3,4$ separately, the potential action in Eq.(\ref{Eq:W_5-D_NMG}) recovers that for $z=4$ HL$_4$ gravity in Eq.(\ref{Eq:W_3-3D}), and that for $z=4$ HL$_5$ gravity in Eq.(\ref{Eq:W_4-5D-2}) in Appendix ~\ref{app:HL(4+1)}. In the case, according to Eq.(\ref{Eq:Lij-MG}), one obtains the field equations of motion $E_{ij}\equiv E_{ij}(W_5)$ and $E \equiv E(W_5)$,
\beqa
E_{ij} = \alpha L_{ij}, \quad [L_{ij}]= [E_{ij}] - (d-4) =4,
\eeqa
where
\beqa
L^m_{~n} &=&  \frac{d-2}{2(d-1)}   \bigg(\nabla^2 R + \frac{g(r)}{r} R^\prime \bigg) \delta^m_n  +  \nabla^2  G^m_{~n} \nn\\
&+&  2 ( R^{pq} R^m_{~pnq}  +  R^{rr}R^m_{~rnr}) - \frac{1}{2} R^{kl}R_{kl}\delta^m_{~n}  \nn\\
&-&   \frac{d}{2(d-1)} R \bigg(R^m_{~n}  - \frac{1}{4}\delta^m_{~n}R \bigg) ,  \label{Eq:Lmn}  \\
L^r_{~r} &=&  \frac{d-2}{2(d-1)} \bigg[  \nabla^2  R  + \bigg(g(r)R^{\prime\prime} + \frac{1}{2}g^\prime(r)R^\prime\bigg)\bigg] \nn\\
&+&   \nabla^2 G^r_{~r} + 2  R^{mn}R^r_{~mrn} - \frac{1}{2} \alpha R^{kl}R_{kl}  \nn\\
&-&  \frac{d}{2(d-1)} R \bigg(R^r_{~r}  - \frac{1}{4}R  \bigg) , \label{Eq:Lrr}\\
L &=& 2\bigg(1 - \frac{d}{4} \bigg)\bigg(R^{kl}R_{kl}- \frac{d}{4(d-1)}R^2\bigg) , \label{Eq:L}
\eeqa
where we have used $G^m_{~n} = R^m_{~n} - R \delta^m_{n}/2$, and $G^r_{~r} = R^r_{~r} - R/2$.

The full action of theory in Eq.(\ref{Eq:L0-L1-4D-z3}) without the cotton tensor can be generalized to be that in $(d+1)$ dimensions as shown in Eq.(\ref{Eq:S-HL(d+1)-z=3}). The action of HL$_{d+1}$ gravity with $z=4$ UV fixed point can be expressed more explicitly as below
\beqa
S \3i &=& \3i \int \! dt \! \int \! d^dx  \sqrt{g} N  \bigg[ {\mathcal L }_K  \!+\!   \frac{\kappa^2}{8\kappa_W^4}\bigg( \frac{(d-2) R \!-\! d\Lambda_W^2}{1\!-\!d\lambda} \nn\\
\3i &+& \3i \frac{\big( 1 \!-\!\frac{d}{4} \!-\! \lambda \big)}{1\!-\!d\lambda}R^2  \!-\! R^{ij}R_{ij} \bigg) \!-\! \frac{\kappa^2 \alpha^2}{8}(L^{ij}L_{ij} \!-\! \tilde\lambda L^2) \qquad \qquad \3i \nn\\
\3i &-& \3i \frac{\kappa^2\alpha}{4\kappa_W^2}\bigg(L_{ij}R^{ij} \!-\! \frac{1\!-\!2\lambda}{2(1\!-\!d\lambda)}LR \!+\! \frac{1}{1\!-\!d\lambda}\Lambda_W L \bigg) \bigg], \label{Eq:S-HL(d+1)-z=4}
\eeqa
where ${\mathcal L}_K\equiv {2}(K_{ij}K^{ij} - \lambda K^2)/\kappa^2$. In the above, we have used the subtraction of metric in pure spatial time, e.g., $g^{ij}g_{ij}=d$, so that
\beqa
 && {\mathcal G}_{ijkl}\bigg[R^{kl}+\bigg(\Lambda_W-\frac{1}{2}R \bigg)g^{kl}\bigg] \nn\\
 && = R_{ij} +\bigg[ \bigg( \Lambda_W - \frac{1}{2}R \bigg)(1-d\tilde\lambda) - \tilde\lambda R \bigg] g_{ij}, \qquad \quad
\eeqa
where $\tilde\lambda$ is defined in Eq.(\ref{Eq:Wheeler-DeWitt-inverse}), and the identity
\beqa
&& T^{ij}{\mathcal G}_{ijkl}\bigg(R^{kl}-\frac{1}{2}Rg^{kl}+\Lambda_W g^{kl}\bigg) \nn\\
&& = T_{kl}R^{kl} - \frac{1-2\lambda}{2(1-d\lambda)}TR + \frac{1}{1-d\lambda}\Lambda_W T ,  \label{Eq:Tij-gRij-Rij-d} \qquad \quad
\eeqa
where $T$ is a second order tensor. By choosing the parameter in Eq.(\ref{Eq:alpha-1/M}) for $d=3,4$ separately, the action in Eq.(\ref{Eq:S-HL(d+1)-z=4}) recovers that for $z=4$ HL$_4$ gravity in Eq.(\ref{Eq:S-HL(3+1)-z=4}), and that for $z=4$ HL$_5$ gravity in Eq.(\ref{Eq:S-HL(4+1)-z=4}) in Appendix ~\ref{app:HL(4+1)}.

With the metric assumed in Eq.(\ref{Eq:ds2-topological-HL}), by substituting Eqs.(\ref{Eq:Lmn}), (\ref{Eq:Lrr}), and (\ref{Eq:L}) back into the action, by using the identity $\delta^{mp}_{nq}\delta^q_p = (s-1) \delta^m_n$, where $s\equiv g_{mn}g^{mn} = d-1$, and by using Eqs.(\ref{Eq:Rijkl-Rrjrl}), (\ref{Eq:Rmn-Rrr}), and (\ref{Eq:R-Rij2-Rijkl2}), the action can be expressed as a function of $g(r)$. Finally, the field equations of motion can be solved. There are four branches for the black hole solution to HL$_{d+1}$ at $z=4$ UV fixed point,
\begin{widetext}
\beqa
g(r) = \frac{2r^2}{(d-2)(d-4)\alpha\kappa_W^2}\bigg[1 \pm \sqrt{ \bigg( 1- \frac{2(d-4)\alpha\kappa_W^2\Lambda_W}{d-1} \bigg) + \frac{[(d-2)(d-4)]^2\alpha^2\kappa_W^4}{4r^{\frac{d}{2}}} C_G} \bigg],
\eeqa
with $C_G=0$ or $C_G\ne 0$. While there are only two physical solutions, assuming that the solution should recover those in the limit $\gamma\to 0$,
\beqa
f(r) = k - \frac{2r^2}{(d-2)(d-4)\alpha\kappa_W^2}\bigg[1 -  \sqrt{ \bigg( 1- \frac{2(d-4)\alpha\kappa_W^2\Lambda_W}{d-1} \bigg) + \frac{C_F}{r^{\frac{d}{2}}}  } \bigg] \approx k - \frac{2\Lambda_W}{(d-1)(d-2)}r^2 + {\mathcal O}(\alpha),
\eeqa
with $C_F=0$ or $C_F=[(d-2)(d-4)\alpha\kappa_W^2]^2C_G/4$.
For $d=3$ one obtains HL$_{4}$ at $z=4$,
\beqa
f(r) = k + \frac{2r^2}{\alpha\kappa_W^2}\bigg( 1 -  \sqrt{ 1 + \alpha\kappa_W^2\Lambda_W + \frac{\alpha^2\kappa_W^4}{4r^{\frac{3}{2}}} C_G} \bigg),
\eeqa
with $C_G=0$ or $C_G\ne 0$. For $d=4$, one obtains HL$_5$ at $z=4$,
\beqa
f(r) = k - \frac{\Lambda_W}{3}r^2 \mp C,
\eeqa
with $C=0$ or $C \ne 0$. The results recover the $z=4$ HL gravity~\cite{Cai:2009ar} at $d=3,4$ by choosing $\alpha=M^{-1}$.

For topological charged black holes of HL$_{d+1}$ gravity with $z=4$ UV fixed point, we obtain
\beqa
\ii\ii
f(r) = k - \frac{2r^2}{(d-2)(d-4)\alpha \kappa_W^2} \bigg[  1 - \sqrt{ \bigg( 1 - \frac{2(d-4)  \alpha\kappa_W^2 \Lambda_W}{d-1 } \bigg) + \frac{\alpha \kappa_W^2}{r^{\frac{d}{2}}}  \sqrt{ \frac{C_F^2}{\alpha^2 \kappa_W^4} + \frac{ 2 (d-4)^2  \Lambda_W}{(d-1)^2  } \frac{  q_0^2  }{r^{d-2} }  }  } \bigg] , ~~
\eeqa
with $C_F\ne 0$. For $d=3$, one obtains HL$_4$ at $z=4$,
\beqa
\ii\ii
f(r) = k + \frac{2r^2}{\alpha \kappa_W^2} \bigg(  1 - \sqrt{ 1 +  \alpha\kappa_W^2 \Lambda_W + \frac{\alpha \kappa_W^2}{r^{\frac{3}{2}}}  \sqrt{ \frac{C_F^2}{\alpha^2 \kappa_W^4} + \frac{ \Lambda_W}{ 2  } \frac{  q_0^2  }{r }  }  } \bigg). ~~
\eeqa
\end{widetext}
We will study in more detail the topological charged black holes in Sec. \ref{sec:HL(3+1)-z=4}, where without losing generality, we have chosen the parameter as shown in Eq.(\ref{Eq:alpha-1/M}). In this case, the above solution to the black holes metric in HL$_4$ at $z=4$, just recovers that in Eq.(\ref{Eq:HL-charged-z=4}).

\subsection{$z=3$ HL$_{d+1}$ gravity}
\label{sec:(d+1)HL-z=3}

For $z = 3$ HL$_{d+1}$ gravity in $d \le 3$, the most general spatial potential $W$ is the $d$-dimensional spatial isotropic Einstein-Hilbert potential\footnote{For $z=3$ HL$_4$ gravity special case, the Chern-Simons gravity action will be present; for more detail, refer to Appendix.~\ref{app:(3+1)HL-z=3}.} in Eq.(\ref{Eq:W_1-D}),
\beqa
\ii
W_1 = \frac{1}{\kappa^2_W}\int d^dx \sqrt{g}(R-2\Lambda_W),  \label{Eq:W}
\eeqa
with scaling dimension
\beqa
 [\kappa_W] = \frac{2-d}{2}.  \label{Eq:scaling-kappa_W}
\eeqa
it is worthy of notice that for $d=3$ case, the scaling dimension of the coupling $\kappa_W$ becomes negative. This implies that Eq.(\ref{Eq:W}) is not a complete potential action for the UV theory, but merely a low-energy effective field theory for $d=3$. It will be broken down at an energy scale set by the dimensionful couplings $\kappa_W$, unless including a new potential action term as shown in Eq.(\ref{Eq:W_2}), namely, the three-dimensional Chern-Simons gravity(CS$_3$) action. For the general case, one finds that in the background metric in Eq.(\ref{Eq:ds2-topological-HL}), the field equations of motion are due to the CS$_3$, i.e., the Cotton tensor is vanishing. Therefore, in the following, one just needs to consider the action $W_1$ in Eq.(\ref{Eq:W}), but leave $d=3$ case discussed in Appendix.~\ref{app:(3+1)HL-z=3}.

According to Eq.(\ref{Eq:Eij}), we obtain $E^{ij}$ and its scaling dimension
\beqa
\ii
E^{ij} \2i=\2i -\frac{1}{\kappa_W^2}\bigg(R^{ij} \!-\! \frac{1}{2}Rg^{ij} \!+\! \Lambda_W g^{ij}\bigg), \label{Eq:Eij_2} ~
[E^{ij}] = d ,   \quad \label{Eq:scaling-Eij}
\eeqa
where we have used
\beqa
\frac{\delta R}{\delta g_{ij}} &=&  \frac{\delta (g_{kl}R^{kl})}{\delta g_{ij}} = -R^{ij} + g_{kl} \frac{\delta R_{kl}}{\delta g_{ij}}, \nn \\
\delta \sqrt{g} &=& \frac{1}{2}\sqrt{g} g^{ij}\delta g_{ij} = - \frac{1}{2}\sqrt{g} g_{ij}\delta g^{ij}.
\eeqa
In this case, the Ho\v{r}ava-Lifshitz gravity theory in $(d+1)$ dimensions takes the potential density form as below
\beqa
{\mathcal V}[g]
&=& \frac{\kappa^2}{8\kappa_W^4}\bigg(R^{ij}R_{ij} - \frac{1 - d/4 -\lambda }{1 - d\lambda}R^2 \nn\\
&-& \frac{(d-2)\Lambda_W R - d\Lambda_W^2}{1-d\lambda} \bigg). \label{Eq:SV_1}
\eeqa
Note that when $d > 2$, lower dimension operator $\Lambda_W$ in the $W$ potential will induce two new terms, proportional to $\Lambda_W R$ and $\Lambda_W^2$, which are combined as the third term in the above expression.

According to Eqs.(\ref{Eq:Eij_2}) and (\ref{Eq:gij-Ricci-Flow-1}), we have a generalized covariantized Ricci flow equation with parameter $\lambda$,
\beqa
\dot{g}_{ij} &=& 2 \nabla_{(i} N_{j)} - N  \frac{\kappa^2}{2\kappa_W^2}\bigg( R_{ij} \nn\\
&-&\frac{2\lambda-1}{2(d\lambda-1)}R\,g_{ij}-\frac{1}{d\lambda-1}\Lambda_W g_{ij}\bigg). ~ \label{Eq:gij-Ricci-Flow-2}
\eeqa
The flow takes account of a spatial diffeomorphism(possibly time dependent) and a time reparametrization. The naive Ricci flow equation is obtained by choosing a popular gauge $N=1$, $N_i=0$, and $\lambda=1$; we have
\beqa
\ii\ii
\dot{g}_{ij} \!=\! -  \frac{\kappa^2}{2\kappa_W^2}\bigg( R_{ij} \! - \! \frac{1}{2(d-1)}R\,g_{ij} \!-\! \frac{\Lambda_W}{d-1} g_{ij}\bigg), \label{Eq:gij-Ricci-Flow-GR}
\eeqa
where $R$ is the average(mean) of the scalar curvature, and the normalized equation preserves the volume of the metric. The time evolution equation of the metric $g_{ij}$ is called the geometric evolution equation with normalized Ricci flow for a Riemannian manifold in math literature. The normalization coefficient in front of the right hand side of the equation is not important since it can be changed to any nonzero real number by rescaling time $t$, while the minus sign in front is relevant since it ensures that the Ricci flow is well-defined for sufficiently small positive times, which means $g_{ij}$ can run forwards in time but not usually backwards. The physical meaning of the equation is that the Ricci flow tends to expand negatively curved regions of the manifold, and contract positively curved regions of the manifold.

The full action of theory is
\beqa
S \3i &=&\int dt(L_0+L_1), \label{Eq:S-HL(d+1)-z=3}\\
L_0 \3i &\equiv & \3i \int d^dx \sqrt{g} N \bigg[ {\mathcal L}_K \!+\! \frac{\kappa^2}{8\kappa_W^4}\bigg( \frac{(d-2)\Lambda_W}{1-d\lambda}R \!-\! \frac{d\Lambda_W^2}{1-d\lambda} \bigg) \bigg], \nn\\
L_1 \3i &\equiv & \3i \int d^dx \sqrt{g} N \frac{\kappa^2}{8\kappa_W^4} \bigg( \frac{\big( 1 -\frac{d}{4} - \lambda \big)}{1-d\lambda}R^2  - R^{ij}R_{ij} \bigg), \nn
\eeqa
where ${\mathcal L}_K \equiv {2}(K_{ij}K^{ij} - \lambda K^2)/\kappa^2$. The first two terms in the ${\mathcal L}_0$ are kinetic density, while the residues are potential density. The third term in ${\mathcal L}_0$ consists of two parts: one is the spatial Ricci scalar term $\Lambda_W R$ and the other is the spatial volume term $\Lambda_W^2$, which will dominate the dynamics of theory at the long distance. The lower dimensional operator $\Lambda_W$ in the $W$ potential, will induce terms proportional to $\Lambda_W R$ and $\Lambda_W^2$ from arbitrary $z$ and flow to $z=1$ in the infrared, which is the fixed point of the usual Einstein's gravity.
it is worthy of notice that
\beqa
\ii &&\ii  [dtd^dx]=-z-d, \quad [R^2] = 4, \quad  \nn \\
\ii &&\ii \Big[ \frac{\kappa^2}{\kappa_W^4} \Big] = (z-d) - 2(2-d) = z + d - 4.  \label{Eq:kappa2-kappaW4}
\eeqa
Consequently, $ [{\mathcal V}[g] ]  = z +d = [dt d^dx]^{-1} $; thus, the Lagrangian density is dimensionless.

By substituting the metric ansatz in Eq.(\ref{Eq:ds2-topological-HL}) into the Lagrangian of the bulk theory in Eq.(\ref{Eq:S-HL(d+1)-z=3}), and by using the matching conditions in Eqs.(\ref{Eq:G_N-Lambda}) and (\ref{Eq:c/G_N}), the action can be re-expressed explicitly as
\beqa
L_0
\3i &=& \3i \int dr \tilde{N} \frac{  \Omega_{d-1,k} c^3}{16\pi G_N}  r^{d-1} \bigg( - \frac{d}{d-2} \Lambda_W   \nn\\
\3i &-& \3i  (d-1)\frac{[(d-2)(f(r)-k) + r f^\prime(r)]}{r^2}  \bigg), \label{Eq:L0} \\
L_1
\3i &=& \3i \int dr \tilde{N} \frac{ \Omega_{d-1,k} c^3}{16\pi G_N} r^{d-1} \frac{1}{\Lambda_W}  (d-1)(d-2) \nn\\
\3i &\times& \3i \bigg( \frac{\lambda-1}{4r^2}f^\prime(r)^2 - \frac{2\lambda+(d-3)}{2r^3} (f(r)-k)f^\prime(r)  \nn\\
\3i&+& \3i  \frac{4\lambda-[2+(d-2)(d-3)]}{4r^4}(f(r)-k)^2  \bigg), ~~\label{Eq:L1}
\eeqa
where we have used Eqs.(\ref{Eq:ds2-ADM}) and (\ref{Eq:ds2-topological-HL}) so that $\sqrt{g}N = (\sqrt{g_{rr}}\sqrt{\gamma}) \sqrt{g_{xx}^{d-1}}  \sqrt{g_{tt}} = \sqrt{\gamma}\tilde{N}  r^{d-1}  $, since $g_{tt}=\tilde{N}^2g^{rr}$, $\Omega_{d-1,k}=\int dx^{d-1}\sqrt{\gamma}$, and
\beqa
R \3i &=& \3i - \frac{d-1}{r^2}[(d-2)(f(r)-k)+r f^\prime(r)], \nn\\
R_{ij}R^{ij} \3i &=& \3i \frac{(d\!-\!1)^2}{4r^2}f^\prime(r)^2 \!+\! \frac{d\!-\!1}{4r^4}[2(d\!-\!2)(f(r)\!-\!k)\!+\!rf^\prime(r)]^2. \nn
\eeqa
These can be obtained through Eq.(\ref{Eq:R-Rij2-Rijkl2}), by remembering that $g(r)\equiv k-f(r)$. As a special case, the IR behavior of the HL$_4$ gravity in the background of the metric can be obtained from Eq.(\ref{Eq:L0}) by setting $d=3$,
\beqa
L_0
&=& \frac{ \Omega_{2,k} c^3}{16\pi G_N }\int dr \tilde{N}(r)\big[- 3 \Lambda_W r^2  \nn\\
&-&2[(f(r)-k) + r \,f^\prime(r) ] \big],  \label{Eq:L0-HL}
\eeqa
which recovers the $L_0$ in Eq.(\ref{Eq:L0-L1-4D-z3}).

\subsubsection{IR relevant dynamics: $L_0$}

By doing variation of Eq.(\ref{Eq:L0}) with respect to $f(r)$ and $\tilde{N}(r)$, respectively,\footnote{
Note that the derivation of $f^\prime$ with respect to $f$, can be simplified by exchanging the variation and differential as $\delta_f f^\prime(r)=\delta_f \overrightarrow\partial_r f = \overrightarrow\partial_r (\delta_f f) = - \overleftarrow\partial_r$, with the minus sign originates from the partial integral operation.}, one obtains the equations of motion
\beqa
0 \3i &=& \3i (d-1)(d-2) r^{d-2}\tilde{N}^\prime,\nn\\
0 \3i &=& \3i -d \Lambda_W r^2 - (d-1)(d-2)[(d-2)(f(r)-k)+r f^\prime(r)], \nn
\eeqa
where for $d=1,2$ case, the $\tilde{N}$ can be an arbitrary function; thus, one can choose them to be a constant for convenience.
The solution turns out to be
\beqa
\ii\ii
\tilde{N}(r) \!=\! \text{const},  f(r) \!=\! k \!-\! \frac{\Lambda_W}{(d-1)(d-2)} r^2 \!-\! \frac{c_1}{r^{d-2}}, \label{Eq:fr-IR}
\eeqa
where $c_1$ is an integral constant with length dimension $d-2$ that can be normalized to be a radial radius $c_1 \equiv \ell^{d-2}$. Note that the const $\tilde{N}(r)$ can be absorbed into the time coordinate. For convenience, one can make the choice for the constants.
\beqa
 \tilde{N}(r) =  1, \quad f(r) = k \!-\! \frac{2\Lambda}{d(d-1)}r^2 \!-\! \frac{\ell^{d-2}}{r^{d-2}},
\eeqa
where $\Lambda$ will be given in Eq.(\ref{Eq:G_N-Lambda}) for $d\ge 3$ case.
Consequently, the exact solution of the topological metric in Eq.(\ref{Eq:ds2-topological-HL}) in the IR limit of HL gravity turns out to be
\beqa
\ii\ii\ii\ii\ii
ds^2 &=&  - f(r) dt^2 + f(r)^{-1} dr^2 + r^2 d\Omega_{d-1,k}^2. \nn\\
f(r) &=&  k - \frac{2\Lambda}{d(d-1)}r^2 - \frac{\ell^{d-2}}{r^{d-2}}. \label{Eq:ds2-topological-Einstein}
\eeqa
This is nothing but the topological black holes in $(d+1)$-dimensional Einstein's general relativity with $\Lambda_W \ne 0$.

In $(3+1)$ dimensions, for the spherically symmetric case $k=1$ in Eq.(\ref{Eq:ds2_d=3-k=1}), by using the effective cosmological constant in Eq.(\ref{Eq:G_N-Lambda})($\Lambda_W = 2\Lambda/3$), we obtain the metric as the solution with respect to the Lagrangian ${\mathcal L}_0$, which recovers that in Ref.~\cite{Lu:2009em}.
\beqa
ds^2 &=& - f(r) dt^2 + f(r)^{-1} dr^2 + r^2(d\theta^2 + \sin^2 \theta d\phi^2), \nn\\
f(r) &=& 1 - \frac{\Lambda}{3}r^2 - \frac{\ell}{r}, \label{Eq:L0_A|dS_Schwarzschild}
\eeqa
which is an AdS(for $\Lambda<0$ if $\lambda>1/3$) Schwarzschild black hole or dS(for $\Lambda>0$ if $\lambda<1/3$) Schwarzschild black hole. This is consistent with the result in Ref.~\cite{Lu:2009em}. For AdS case with negative cosmological constant, by using Eq.(\ref{Eq:G_N-Lambda}), let us make the notation that
\beqa
\ii\ii
 \Lambda \!\equiv\!  - \frac{d(d\!-\!1)}{2\ell^2} < 0,  \Rightarrow  \Lambda_W %= \frac{2(d-2)}{d}\Lambda
\!=\! - \frac{(d\!-\!1)(d\!-\!2)}{\ell^2}<0. \quad \label{Eq:Lambda_W-ell}
\eeqa
In this case, Eq.(\ref{Eq:ds2-topological-Einstein}) becomes
\beqa
ds^2 &=& - f(r) dt^2 + f(r)^{-1} dr^2 + r^2 d\Omega_{d-1,k}^2, \nn\\
f(r) &=& k - \frac{r^2}{\ell^2} - \frac{\ell^{d-2}}{r^{d-2}},
\eeqa
i.e., for $(3+1)$ dimensions, we have $\Lambda=-3/\ell^2$, $\Lambda_W = - {2}/{\ell^2}$.
Therefore, the Lagrangian density ${\mathcal L}_0$ that dominates at IR still recover the AdS$_4$ black hole with negative cosmological constant in the usual Hilbert-Einstein action with $\lambda=1$, while if the UV sector ${\mathcal L}_1$ is taken into account, the IR physics of the black hole will be modified as we will see later on.

\subsubsection{UV relevant dynamics: $L_0+L_1$}

The full Lagrangian $  L _0+  L _1$ in the background of topological metric in Eq.(\ref{Eq:ds2-topological-HL}) can be calculated explicitly with Eqs.(\ref{Eq:L0})-(\ref{Eq:L1}). By observing the asymptotic behavior of the IR relevant solution in Eq.(\ref{Eq:fr-IR}),
\beqa
f(r) \overset{r\to \infty}{\approx} k - \frac{\Lambda_W}{(d-1)(d-2)}r^2 + \ldots \, .
\eeqa
One can define a new function $F(r)$,
\beqa
 F(r) \equiv k - \frac{2\Lambda_W}{(d-1)(d-2)} r^2 - f(r). \label{Eq:Fr-fr}
 %\quad F(r) \overset{d=3}{\equiv} k - \Lambda_W r^2 - f(r),
\eeqa
According to Eqs.(\ref{Eq:L0}) and (\ref{Eq:L1}), the full bulk Lagrangian $L_0+L_1$ takes the form,
\beqa
L_0 + L_1 \3i &=& \3i \frac{ \Omega_{d-1,k} c^3}{16\pi G_N }\int dr \tilde{N}(r)\frac{(d-1)(d-2)}{2\Lambda_W}r^{d-3} \times \nn\\
\3i && \3i \bigg( \frac{\lambda-1}{2}F^\prime(r)^2  - \frac{2\lambda +(d-3)}{r} F(r)F^\prime(r) + \nn\\
\3i && \3i \frac{2(2\lambda-1)-(d-2)(d-3)}{2r^2}F(r)^2\bigg),  \label{Eq:L01_lambda}
\eeqa
which recovers the relativistic case by setting $\lambda=1$,
\beqa
\ii\ii
\frac{ \Omega_{d-1,k} c^3}{16\pi G_N } \!\int \! dr \tilde{N}(r) \bigg( \!-\! \frac{(d-1)^2(d-2)}{4\Lambda_W}\frac{F(r)^2}{r^{4-d}}  \bigg)^\prime. ~\label{Eq:L0-L1-lambda=1}
\eeqa
By doing variation with respect to $F(r)$ and $\tilde{N}(r)$ respectively, the equations of motion are obtained
\beqa
0 \3i &=& \3i \tilde{N}(r)(\lambda-1)\bigg(\frac{2(d-2)}{r^2}F(r) -\frac{(d-3)}{r}F^\prime(r)-F^{\prime\prime}(r) \bigg) \nn\\
\3i &+& \3i \tilde{N}^\prime(r)\bigg( \frac{2\lambda +(d-3)}{r}F(r) - (\lambda-1)F^\prime(r) \bigg),\label{Eq:EOM_Nt}  \\
0 \3i &=& \3i \frac{\lambda-1}{2}F^\prime(r)^2  - \frac{2\lambda +(d-3)}{r}F(r)F^\prime(r) \nn\\
\3i &+& \3i \frac{2(2\lambda-1)-(d-2)(d-3)}{2r^2}F(r)^2, \label{Eq:EOM_Fr}
\eeqa
which give solutions
\beqa
 F(r) &=& C_F r^{\lambda^\pm_F} , \label{Eq:Fr}\\
 \lambda_F^\pm &\equiv & \frac{2\lambda+(d-3)\pm\sqrt{(d-1)(d\lambda-1)}}{\lambda-1}, \nn \\
 \tilde{N}(r) &=& C_N r^{\lambda^\pm_N},  \label{Eq:Nr}\\
 \lambda_N^\pm & \equiv & -\frac{(d-2)+d\lambda \pm 2\sqrt{(d\lambda-1)(d-1)}}{\lambda-1},\nn
\eeqa
where $C_{F,N}$ are both integration constants and it is worthy of notice that $\lambda_N^\pm$ and $\lambda_F^\pm$ are not independent, respectively;
\beqa
\lambda_N^\pm = 4 - d -  2 \lambda_F^\pm,
\eeqa
which is just a special case of Eq.(\ref{Eq:lambdaN-lambdaF}) with $n=2$.
%When $C_F=0$, namely, $F(r)=0$.
In the case that $C_F=0$, $\tilde{N}(r)$ is not restricted by the equations of motion, and thus is an arbitrary function,
\beqa
F(r)=0, \quad \tilde{N}(r) = \text{arbitrary}.
\eeqa
In this case, the metric solution in Eq.(\ref{Eq:ds2-topological-HL}) becomes
\beqa
ds^2 \3i &=& \3i - \tilde{N}(r) f(r) c^2 dt^2 + f(r) dr^2 + r^2 d\Omega^2_{d-1,k}. \nn \\
f(r) &=& k - \frac{2\Lambda_W}{(d-1)(d-2)} r^2.
\eeqa
The exponent coefficients in some special limit reduce to be
\beqa
 \lim_{\lambda\to {1}/{d}}\lambda_F^+ &=& 2-d , ~ \lim_{\lambda\to 1^\pm}\lambda_F^+ = \pm\infty, ~  \lim_{\lambda\to \pm \infty}\lambda_F^+ = +2; \nn\\
 \lim_{\lambda\to {1}/{d}}\lambda_F^- &=& 2-d , ~ \lim_{\lambda\to 1^\pm}\lambda_F^- = 2 - \frac{d}{2}, ~  \lim_{\lambda\to \pm \infty}\lambda_F^-= +2. \nn
\eeqa
For the negative branch, $\lambda_F^-$ is an increasing function of $\lambda$,
and is always less than $2$ for positive $\lambda\in [1/d,+\infty)$.
It is always less than $2$ for positive $\lambda$, thus, the $r^2$ term in $F(r)$ always dominates at large distance $r \to \infty$.

In other words, the negative branch of the function $F(r)$ in HL gravity affects only the IR behavior of the topological black holes, while for the positive branch, there is a singularity at $\lambda=1$. In summary, the value range of exponent of function $F(r)$ is
\beqa
\ii \lambda_F^+ \in [2-d,-\infty]\cup[+\infty,+2], ~ \lambda_F^- \in [2-d,+2]. \nn
\eeqa
i.e, $\lambda_F^+(\lambda=1/3)=-1,-4,-4-\sqrt{6}\approx -6.449$ for $d=3,4,5$ respectively, while $\lambda_F^-(\lambda=1/3)=-1,-1,-4+\sqrt{6}\approx -1.551$ for $d=3,4,5$. Thus $\lambda_F^+(\lambda=1/3)$ is a monotonically decreasing function of spatial dimensions $d$, while $\lambda_F^-(\lambda=1/3)$ is a decreasing function of $d$, and achieves its maximal value $-1$ at both $d=3$ and $d=4$.
\beqa
 \lim_{\lambda\to {1}/{d}}\lambda_{N}^+ &=& d, \quad \lim_{\lambda\to 1^\pm}\lambda_{N}^+ = \mp\infty, \quad  \lim_{\lambda\to \pm \infty}\lambda_{N}^+ = -d; \nn\\
 \lim_{\lambda\to {1}/{d}}\lambda_{N}^- &=& d, \quad \lim_{\lambda\to 1^pm}\lambda_{N}^- = 0, \quad  \lim_{\lambda\to \pm \infty}\lambda_{N}^- = -d. \nn
\eeqa
i.e, $\lambda_N^+(\lambda=1/3)=3,8,7+2\sqrt{6}\approx 11.899$ for $d=3,4,5$ respectively, while for $d=4$, $\lambda_N^-(\lambda=1/3)=3,2,7-2\sqrt{6}\approx 2.101$ for $d=3,4,5$. Thus, $\lambda_N^+(\lambda=1/3)$ is a monotonically increasing function of spatial dimensions, while $\lambda_N^-(\lambda=1/3)$ achieves its minimal value $2$ at $d=3$. Therefore, the value range of exponents of function $F(r)$ is
\beqa
\lambda_{N}^+ \in [d,+\infty]\cup[-\infty,-d], \quad \lambda_{N}^- \in [d,-d].
\eeqa
In summary, the general solution of the topological black holes in HL gravity is given in Eq.(\ref{Eq:ds2-topological-HL}) with
\beqa
\tilde{N}^2(r)\! &=& \! C_N^2 r^{2\lambda^\pm_N}, \nn\\
f(r)\! &=& \! k \!-\! \frac{2\Lambda_W}{(d-1)(d-2)} r^2 \!-\! C_F r^{\lambda_F^\pm}. \label{Eq:ds2-HL-generic-lambda}
\eeqa
It is worthy of notice the difference between the metric of Lifshitz gravity in Refs.~\cite{Kachru:2008yh,Griffin:2012qx} and the metric of topological neutral black holes in z=d Ho\v{a}va-Lifshitz(HL) gravity with $k=0$ case at long distance in the IR limit. It turns out that the metric of Lifshitz gravity, i.e., $Lif_{d+1}$ with $z\ne 1$, is a special vacuum solution of $z=2$ HL gravity with negative cosmological constant, by including a low energy effective Lagrangian at long distance in the IR limit.
To be more concrete,
\begin{enumerate}
\item $\lambda=\pm \infty$,
\beqa
ds^2 & = & - r^{-2d} f(r) dt^2 + f(r)^{-1} dr^2 + r^2 d\Omega_{d-1,k}^2, \nn\\
f(r)&=& k - \frac{2\Lambda_W}{(d-1)(d-2)} r^2 - C_F r^2, \label{Eq:ds2-topological-HL-lambda=infty}
\eeqa
where $C_N$ is normalized to be a unit and $C_F$ can be absorbed into the cosmological constant.
\item $\lambda=\frac{1}{d}$,
\beqa
ds^2  &=& - r^{2d} f(r) dt^2 + f(r)^{-1}dr^2 + r^2 d\Omega_{d-1,k}^2,  \nn \\
f(r)  &=& k - \frac{2\Lambda_W}{(d-1)(d-2)} r^2 - \frac{C_F}{r^{d-2}}, \label{Eq:ds2-topological-HL-lambda=1/3}
\eeqa
where $C_N$ is normalized to be unit and $C_F$ is irrelevant in the infinite boundary. The solution has an IR singularity at $r=0$ if $C_F\ne 0$ in the large distance where $C_F$ is relevant.
\item $\lambda=1$,
\beqa
ds^2  &=& - f(r) dt^2 +  f(r)^{-1}dr^2 + r^2 d\Omega_{d-1,k}^2, \nn \\
f(r)  &=& k - \frac{2\Lambda_W}{(d-1)(d-2)} r^2 - C_F r^{2-\frac{d}{2}}. \label{Eq:ds2-topological-HL-lambda=1}
\eeqa
The stable solution for $\lambda=1$($\Lambda_W<0$) is asymptotically AdS$_{d+1}$ at large distance and it has an IR singularity at ($r=0$) if $C_F\ne 0$, which could be covered by a black holes horizon at $r=r_+$, where $r_+$ is the largest root of the equation $f(r)=0$. Note that for $k=1$ case, the solution does not recover the usual AdS$_{d+1}$ Schwarzschild black hole, namely, the solution in Eq.(\ref{Eq:ds2-topological-HL-lambda=1}) suggests that the solution to the field equations of motion of the HL gravity with Lagrangian $L_0+L_1$ in the IR limit does not repeat the usual AdS Schwarzschild black holes in the general relativity at large distance, as shown in Eq.(\ref{Eq:ds2-topological-Einstein}).

\end{enumerate}

\subsection{Topological charged black holes in generalized HL$_{d+1}$ gravity with generic $\epsilon$}
The electric charge can be incorporated by coupling the electromagnetic field to the gravitational sector of the action\cite{Banados:1993ur,Cai:1998vy,Borzou:2011bp,Janiszewski:2014ewa}.
In the section, we consider the charged generalization of the topological black holes. The Hamiltonian action of the Maxwell field[or $U(1)$ gauge field] in a curved space-time is
\beqa
S^{EM} \3i &=& \3i \int \! dt d^{d}x {\mathcal L}^{EM} + S_{B}^{EM}, \label{Eq:S-Maxwell-Einstein} \\
{\mathcal L}^{EM} \3i &=& \3i p^i \dot{A}_i - \frac{1}{2}N\bigg(  \frac{\alpha}{\sqrt{g}} p^i p_i + \frac{\sqrt{g}}{2\alpha} F_{ij}F^{ij}  \bigg) + \phi\, \partial_i p^i, \nn
\eeqa
where $N$ is the lapse function, $g$ is the determinant of the induced metric of the ADM decomposition of space-time. $p^i$ is the momentum conjugate to the spatial components of the Maxwell field $A_i$ or the vector potential, and meanwhile the scalar potential is $\phi=A_t$. $S_B^{EM}$ is a surface term that depends on the boundary conditions. The constant $\alpha$ is a parameter, which may be conveniently taken to be equal to the area of the hypersurface $\Omega_{d-1,k}$, i.e., a $(d-1)$-unit sphere for $k=1$ case. We are only interested in the solutions without magnetic charge, radial, static Maxwell field, namely,
\beqa
\ii\ii
F_{ij}=0, \quad p^i = (0,p^r, 0, \ldots, 0), \quad \dot{A}_i =0 = \dot{p}^i,
\eeqa
where $i =1, 2, \ldots d-1$, and the conjugate momentum $[p^i]=[L^{-1}]=1$. After imposing the above conditions, the action is reduced to be one for the Coulomb field taking the form $\bar{S}^{EM}\equiv S^{EM} - S_B^{EM}$,
\beqa
\ii\ii\ii
\bar{S}^{EM} \2i\3i &=& \3i \frac{\Omega_{d-1,k}}{\alpha} \! \int \! dt dr \bigg( \!-\! \frac{1}{2}\tilde{N}r^{d-1} p^2 \!+\! \phi (r^{d-1}p)^\prime \bigg), ~~ \label{Eq:S-Maxwell-Einstein-RSS}
\eeqa
where $\tilde{N}(r)\equiv N\sqrt{g_{rr}}$ and $p$ is the rescaled radial component of $p^i$. $\gamma$ is the determinate of the $(d-1)$-dimensional Einstein space $d\Omega^2_{d-1}=\gamma_{mn}dx^m dx^n$, namely, $\gamma=\det (\gamma_{mn})$. It is obvious that $\sqrt{g}=\sqrt{\gamma}\sqrt{g_{rr}}$ and $\Omega_{d-1,k}=\int d^{d-1}x \sqrt{\gamma}$. The prime is the derivative with respect to the radial coordinate. In this above derivation, we have used that
\beqa
&& \dot{p_i}=0, \quad p^r = \frac{\sqrt{\gamma}}{\alpha} r^{d-1} p, \quad \partial_i p^i = \partial_r p^r  = \frac{\sqrt{\gamma}}{\alpha}(pr^{d-1})^\prime, \nn \\
&& \frac{\alpha}{\sqrt{g}} p^i p_i = \frac{\alpha}{\sqrt{g}}g_{ij}p^i p^j = \alpha \frac{g_{rr}(p^r)^2 }{\sqrt{g_{rr}\gamma}} = \frac{\sqrt{g}}{\alpha}p^2 r^{2(d-1)}. \nn
\eeqa

The Hamiltonian action of Maxwell field in the zero magnetic charge, radial static and spherically symmetric background in Eq.(\ref{Eq:S-Maxwell-Einstein-RSS}) is general. By considering the metric ansatz in Eq.(\ref{Eq:ds2-topological-HL}), we find that $N=\sqrt{g_{tt}}=\tilde{N}\sqrt{f(r)}=\tilde{N}\sqrt{g^{rr}}$.

In the following, let us consider the $\lambda=1$ case\footnote{Note, in the following, we will mainly focus on the $\lambda=1$ case, which is more physical interesting in the sense that its Hamiltonian density reduce to be the Wheeler Dewitt equation in the canonical quantization of the gravity, i.e., ${\mathcal G}_{ijkl}=(g_{ik}g_{jl}+g_{il}g_{jk}-g_{ij}g_{kl})/2$.}, in addition, to investigate the physical consequence for the deviation from the detailed balance condition; we assume a small derivation from the detailed balance condition, measured by the detail balance violation parameter $\epsilon\in [0,1)$,
\beqa
S_\epsilon = \int dt L_\epsilon, \quad L_\epsilon \equiv L_0 + (1-\epsilon^2)L_1,
\eeqa
where $L_0$ accounts for the IR relevant Lagrangian, $\epsilon^2\ne0$ is a small variation from zero, the Lagrangian reduces to be the IR relevant one when $\epsilon=1$, but reduces to be the one with the detailed balance condition with $\epsilon=0$. After substituting the metric ansatz in Eq.(\ref{Eq:ds2-topological-HL}), according to Eqs.(\ref{Eq:L0}) and (\ref{Eq:L1}), one obtains the full modified Lagrangian in $(d+1)$ dimensions,
\beqa
 L_\epsilon \3i &=& \3i \frac{\Omega_{d-1,k}c^3}{16\pi G_N}\int dr \tilde{N}(r) U_d(r,\lambda) , \label{Eq:L_epsilon_lambda} \\
 U_{d}(r,\lambda) \3i &=& \3i\frac{(d-1)(d-2)}{2\Lambda_W}r^{d-3}\bigg[\epsilon^2 \bigg( \frac{2d\Lambda_W^2}{(d-1)(d-2)^2}r^2 \nn\\
&+& 2\Lambda_W F(r) + \frac{2\Lambda_W}{d-2} r F^\prime(r) \bigg) + (1-\epsilon^2)  \nn\\
&\times & \bigg( \frac{\lambda-1}{2}F^\prime(r)^2  - \frac{2\lambda +(d-3)}{r} F(r)F^\prime(r) \nn\\
&+& \frac{2(2\lambda-1)-(d-2)(d-3)}{2r^2}F(r)^2 \bigg) \bigg], \nn
\eeqa
where we have used the matching condition according to Eq.(\ref{Eq:c/G_N}),
\beqa
\frac{\kappa^2}{8\kappa_W^4} = -\frac{c^3}{16\pi G_N}\frac{d\lambda -1}{(d-2)\Lambda_W}. \label{Eq:NormGravity}
\eeqa
When $\epsilon=0$, one obtains the full bulk Lagrangian $L_0+L_1$ as shown in Eq.(\ref{Eq:L01_lambda}).

It is worthy of emphasizing that the $d=2$ case should be handled separately; by using Eq.(\ref{Eq:LambdaW-Lambda}), Eq.(\ref{Eq:L_epsilon_lambda}) can be re-expressed as
\beqa
&& U_{d}(r,\lambda) \! = \! \frac{d(d\!-\!1)}{4\Lambda}r^{d-3}\bigg[\epsilon^2 \bigg( \frac{8\Lambda^2}{d(d\!-\!1)}r^2 \! +\!  \frac{4(d\!-\!2)\Lambda}{d} F(r) \nn\\
\3i && \3i +  \frac{4\Lambda}{d} r F^\prime(r) \bigg) \!+\! (1\!-\!\epsilon^2)\bigg( \frac{2(2\lambda-1)-(d-2)(d-3)}{2r^2}F(r)^2  \nn\\
\3i && \3i +  \frac{\lambda-1}{2}F^\prime(r)^2  - \frac{2\lambda +(d-3)}{r} F(r)F^\prime(r) \bigg) \bigg], \label{Eq:L_epsilon_lambda-2}
\eeqa
assuming that $\Lambda\ne 0$ for $d=2$, which is reasonable as shown in Eq.(\ref{Eq:c-Lambda-d=2}); Eq.(\ref{Eq:L_epsilon_lambda-2}) becomes
\beqa
\ii\ii
&& \ii U_{2}(r,\lambda) \!=\! \frac{1}{2\Lambda}\frac{1}{r}\bigg[\epsilon^2 \Big( 4\Lambda^2 r^2 + 2\Lambda r F^\prime(r) \Big) \!+\! (1 \!-\! \epsilon^2) \label{Eq:L_epsilon_lambda-d=2}\\
&& \ii \times  \bigg( \frac{\lambda \!-\! 1}{2}F^\prime(r)^2  \!-\! \frac{2\lambda \!-\! 1}{r} F(r)F^\prime(r) \!+\! \frac{2(2\lambda \!-\! 1)}{2r^2}F(r)^2\bigg) \bigg]. \quad \nn
\eeqa

By combing Eqs.(\ref{Eq:L_epsilon_lambda}) and (\ref{Eq:S-Maxwell-Einstein-RSS}) together and choosing constant $\alpha$ to be
\beqa
\ii
\alpha^{-1}= \frac{c^3}{16\pi G_N}, \label{Eq:kappa-G_N-alpha}
\eeqa
we obtain the $(d+1)$-dimensional HL-Maxwell action at $z=3$ with $\lambda=1$,
\beqa
S &=& S_\epsilon + S^{EM}  = \frac{\Omega_{d-1,k}c^3}{16\pi G_N} \int dt dr \bigg( \tilde{N}(r)  ( U_d(r,\lambda) \nn\\
&-& \frac{1}{2} r^{d-1} p^2 )  + \phi (r^{d-1}p)^\prime  \bigg)  + S_B^{EM}.   \label{Eq:I-Maxwell-HL-z=1}
\eeqa

Varying the action $S$ with respect to $F(r)$, one obtains
\beqa
&& \bigg[ (\lambda-1)\bigg( \frac{2(d-2)}{r^2}F(r) - \frac{d-3}{r}F^\prime(r) - F^{\prime\prime}(r) \bigg) \tilde{N}(r) \nn\\
&& + \bigg( \frac{2\lambda+(d-3)}{r}F(r)  - (\lambda - 1)F^\prime(r) \bigg) \tilde{N}^\prime(r)  \bigg](1-\epsilon^2) \nn\\
&& + \bigg( -\frac{2\Lambda_W}{d-2}r \tilde{N}^\prime(r) \bigg)\epsilon^2 =0.
\eeqa
For the $d=3$ with $\epsilon=0$ case, it just reproduces to Eq.(\ref{Eq:EOM_Nt}).

For the $\lambda=1$ case, one obtains
\beqa
\ii (1-\epsilon^2)  \frac{d-1}{r}F(r)   \tilde{N}^\prime(r)   - \epsilon^2\frac{2\Lambda_W}{d-2}r \tilde{N}^\prime(r)  =0.
\eeqa
The general solution of $\tilde{N}(r)$ for the $\lambda=1$ case becomes
\beqa
\tilde{N}(r) = \text{const.}.
\eeqa
Varying the action $S$ with respect to $\phi(r)$, $p(r)$, and $\tilde{N}(r)$, respectively, one has the equations of motion
\beqa
&&  (r^{d-1} p)^\prime = 0, \quad  -r^{d-1} (\tilde{N}p+\phi^\prime) = 0, \nn\\
&& U_d(r,1) - \frac{1}{2}r^{d-1} p^2 = 0. \label{Eq:Nt-p-phi-z=1}
\eeqa
For neutral black holes in $(3+1)$ dimensions with $p=0$ and $\epsilon=0$, above EOM, i.e., $U_3(r,\lambda)=0$, just reproduce Eq.(\ref{Eq:EOM_Fr}).

For the $\lambda=1$ case, the EOMs above give the solutions below
\beqa
d\ge 3:  \tilde{N}(r) &=& N_0, ~ p(r) = -\frac{q_0}{r^{d-1}}, ~ U_d(r,1) = \frac{q_0^2}{2 r^{d-1}}, \nn\\
 \phi(r) &=& -\frac{N_0 q_0}{(d-2)r^{d-2}} + \phi_0. \label{Eq:Gauge-Field-d3}\\
d=2: \tilde{N}(r) &=& N_0, ~ p(r) = -\frac{q_0}{r}, ~ U_2(r,1) = \frac{q_0^2}{2r}, \nn\\
 \phi(r) &=& -N_0 q_0 \log{r} + \phi_0 . \label{Eq:Gauge-Field-d2}
\eeqa
$N_0$, $q_0$, $\phi_0$, and $c_0$ are the integration constants, where $N_0\equiv N(\infty)$ and $\phi_0\equiv\phi(\infty)$ are the values of $N$ and $\phi$ at infinity; $q_0$ is electric charge. It will be convenient for the metric to set $N_0 = \mu$ and $\phi_0 \equiv \mu$ at infinity; in this case one has
\beqa
d\ge 3:  \tilde{N}(r)&=&\mu, ~ \phi(r)= \mu\bigg(1 -\frac{r_0^{d-2}}{r^{d-2}}\bigg), \nn\\
				 p(r)&=& -\frac{q_0}{r^{d-1}}, \quad q_0 = (d-2) r_0^{d-2}. \label{Eq:phi=A0}\\
 d= 2:  \tilde{N}(r) &=& \mu, ~ \phi(r) = \mu(1 - q_0 \ln{r}) = \mu\bigg(1 - \frac{\ln{r}}{\ln{r_0}} \bigg), \nn\\
				p(r) &=& - \frac{q_0}{r}, \quad q_0 = {(\ln{r_0})^{-1}}. \label{Eq:phi=A0-d=2}
\eeqa
The scaling dimension of the conjugate momentum and charge are respectively $[p]=[L^{-1}]$ and $[q_0]=[L^{d-2}]$.

For $d\ge 3$, the solution of $F(r)$ becomes
\begin{widetext}
\beqa
\ii
F(r) \!=\!  \frac{2\epsilon^2}{1-\epsilon^2}\frac{\Lambda_W}{(d-1)(d-2)} r^2 \!+\! r^{2-\frac{d}{2}}\sqrt{\frac{c_0}{1-\epsilon^2} \!+\! \frac{\epsilon^2}{(1-\epsilon^2)^2}\frac{4\Lambda_W^2r^d}{[(d-1)(d-2)]^2} \!+\! \frac{1}{1-\epsilon^2}\frac{2\Lambda_W}{[(d-1)(d-2)]^2}\frac{q_0^2}{r^{d-2}}  }. \label{Eq:Fr-d-lambda=1}
\eeqa
According to Eq.(\ref{Eq:Fr-fr}), we obtain
\beqa
f(r)&=& k - \frac{2}{1-\epsilon^2}\frac{\Lambda_W}{(d-1)(d-2)} r^2 - r^{2-\frac{d}{2}}\sqrt{\frac{c_0}{1-\epsilon^2}+ \frac{\epsilon^2}{(1-\epsilon^2)^2}\frac{4\Lambda_W^2r^d}{[(d-1)(d-2)]^2} + \frac{1}{1-\epsilon^2}\frac{2\Lambda_W}{[(d-1)(d-2)]^2}\frac{q_0^2}{r^{d-2}}  } \nn \\
&& \ii\ii\ii \left\{ \begin{aligned}
&\overset{\epsilon = 1}{=} k - \frac{\Lambda_W }{(d-1)(d-2)}r^2 + \frac{(d-1)(d-2)}{4\Lambda_W }\frac{c_0}{r^{d-2}} + \frac{1}{2(d-1)(d-2)}\frac{q_0^2}{r^{2d-4}}, \label{Eq:ds2-HL-topological-charged-black-brane-L-epsilon-1}  \\
&\overset{\epsilon = 0}{=} k - \frac{2\Lambda_W}{(d-1)(d-2)} r^2 - r^{2-\frac{d}{2}}\sqrt{c_0 + \frac{2\Lambda_W}{[(d-1)(d-2)]^2} \frac{q_0^2}{r^{d-2}} }.
           \end{aligned} \right.
\eeqa
with $[c_0]=[L^{d-4}]$ and $[q_0]=[L^{d-2}]$. 
By using Eq.(\ref{Eq:Lambda_W-ell}), we have
\beqa
f(r) &=&  k + \frac{2}{1-\epsilon^2}\frac{ r^2}{\ell^2}\bigg[ 1 - \epsilon \sqrt{1 +
\frac{1-\epsilon^2}{\epsilon^2}\bigg(\frac{ c_0\ell^4  }{ 4 r^d  } - \frac{\ell^2}{2(d-1)(d-2)}\frac{ q_0^2 }{r^{2d-2}}  \bigg) } \bigg] \label{Eq:ds2-HL-topological-charged-black-brane-L-epsilon-2} \\
&& \ii\ii\ii \left\{ \begin{aligned}
&\overset{\epsilon=1}{=} k + \frac{ r^2}{\ell^2}\bigg( 1 - \frac{c_1 \ell^2 }{r^{d}} + \frac{\ell^2}{2(d-1)(d-2)} \frac{q_0^2}{ r^{2d-2}}\bigg), \nn \\
&\overset{\epsilon=0}{=}  k + 2 \frac{r^2}{\ell^2} - r^{2-\frac{d}{2}}\sqrt{c_0 - \frac{2}{(d-1)(d-2)\ell^2}\frac{q_0^2 }{r^{d-2}}},  \nn
           \end{aligned} \right.
\eeqa
\end{widetext}
where
\beqa
&& \ii c_1 \2i \equiv \2i - \frac{(d-1)(d-2)}{4\Lambda_W}c_0 \!=\! \frac{\ell^2 }{4}c_0 \! \equiv \! \ell^{d-2} \nn\\
&& \Rightarrow c_0 \!=\! 4\ell^{d-4}. \label{Eq:c0-c1}
\eeqa
In the limit $\epsilon\to 1$, we obtain the solution of topological charged black holes in $(d+1)$-dimensional Einstein gravity. While in the limit of $\epsilon\to 0$, we obtain the solution of charged topological black holes in $(d+1)$-dimensional HL gravity at $z=3$ with the $\lambda=1$ case, which is the exact solution one is most interested in. Thus the solution of the redshift factor in the charged black hole solution becomes
\beqa
f(r) &=&  k + \frac{2}{1-\epsilon^2}\frac{ r^2}{\ell^2}\bigg[ 1 - \epsilon  \label{Eq:ds2-HL-topological-charged-black-brane-L-epsilon-3} \\
&& \times \sqrt{1 +
\frac{1-\epsilon^2}{\epsilon^2}\bigg(\frac{\ell^{d}  }{  r^d  } - \frac{\ell^2}{2(d-1)(d-2)}\frac{ q_0^2 }{r^{2d-2}}  \bigg) } \bigg] \nn \\
&& \ii\ii\ii\, \left\{ \begin{aligned}
&\overset{\epsilon=1}{=} k + \frac{ r^2}{\ell^2}\bigg( 1 - \frac{\ell^d }{r^{d}} + \frac{\ell^2}{2(d-1)(d-2)}\frac{q_0^2 }{r^{2d-2}}\bigg) . \nn \\
&\overset{\epsilon=0}{=} k + 2 \frac{r^2}{\ell^2} - r^{2-\frac{d}{2}}\sqrt{4\ell^{d-4} - \frac{2}{(d-1)(d-2)\ell^2}\frac{q_0^2 }{r^{d-2}}}.  \nn
           \end{aligned} \right.
\eeqa
As expected, it is just the AdS( since $\Lambda_W<0$ for $\lambda>1/3$ ) RN black hole solution in the IR limit.

In $(3+1)$ dimensions, one has
\beqa
f(r) &=&   k + \frac{2 }{1-\epsilon^2}\frac{r^2}{\ell^2}\bigg[ 1 - \epsilon \sqrt{1 + \frac{1-\epsilon^2}{\epsilon^2}\frac{\ell^3}{r^3}\bigg( 1 - \frac{1}{4\ell }\frac{q_0^2}{r} \bigg)} \bigg] \nn\\
&& \ii\ii\ii\, \left\{ \begin{aligned}
&\overset{\epsilon=1}{=} k + \frac{r^2}{\ell^2}- \frac{\ell}{r} + \frac{1}{4}\frac{q_0^2}{r^2}  \label{Eq:fr-eps-1} \\
&\overset{\epsilon=0}{=} k + 2\frac{r^2}{\ell^2} -  2\sqrt{ \frac{r}{\ell} - \frac{q_0^2 }{4\ell^2}}. \label{Eq:fr-eps-0}
           \end{aligned} \right. \label{Eq:ds2-fr-L-epsilon-q0}
\eeqa
In the IR limit at large distance, we have
\beqa
\ii\ii
f(r) \overset{r\to\infty}{=} k + \frac{2}{1+\epsilon}\frac{r^2}{\ell^2}-\frac{1}{\epsilon}\frac{\ell}{r} + \frac{1}{\epsilon} \frac{q_0^2}{4r^2} + o(\frac{1}{\epsilon^3 r^4}).
\eeqa
The solution has a finite mass $\sim \ell$ for nonvanishing $\epsilon$, while it diverges at $\epsilon=0$, namely, in the full detailed balance condition, the solution restores to be the corresponding ones in Eq.(\ref{Eq:fr-eps-0}), where the divergence becomes a square root $-2\sqrt{ r/\ell - q_0^2/(4\ell^2)}$. When $\epsilon=1$, $L_1$ vanishes, namely, no detailed balance condition is imposed; therefore, the solution restores to be those in Eq.(\ref{Eq:fr-eps-1}).

In $(d+1)$ dimensions, the topological black hole solutions correspond to those in Eqs.(\ref{Eq:ds2-HL-topological-charged-black-brane-L-epsilon-1}) and (\ref{Eq:ds2-HL-topological-charged-black-brane-L-epsilon-2}) with $q_0=0$, which becomes, respectively,
\beqa
f(r)&=& k - \frac{2}{1-\epsilon^2}\frac{\Lambda_W}{(d-1)(d-2)} r^2 \label{Eq:ds2-HL-topological-neutral-black-brane-L-epsilon-1}
\label{Eq:ds2-HL-topological-neutral-black-brane-L-epsilon-2}\\
&-& r^{2-\frac{d}{2}}\sqrt{\frac{c_0}{1-\epsilon^2}+ \frac{\epsilon^2}{(1-\epsilon^2)^2}\frac{4\Lambda_W^2r^d}{[(d-1)(d-2)]^2}   }  \nn \\
 &=& k + \frac{2}{1-\epsilon^2}\frac{ r^2}{\ell^2} - r^{2-\frac{d}{2}}\sqrt{\frac{c_0}{1-\epsilon^2}+ \frac{ \epsilon^2 }{(1-\epsilon^2)^2 }\frac{4  r^d}{\ell^4} }, \nn \\
&& \ii\ii\ii\, \left\{ \begin{aligned}
&\overset{\epsilon = 1}{=}  - \frac{\Lambda_W }{(d-1)(d-2)}r^2 + \frac{(d-1)(d-2)}{4\Lambda_W }\frac{c_0}{r^{d-2}} , \nn \\
&\overset{\epsilon = 0}{=} k - \frac{2\Lambda_W}{(d-1)(d-2)} r^2 - r^{2-\frac{d}{2}}\sqrt{c_0 }. \nn
           \end{aligned} \right.
\eeqa
For HL$_4$, one has
\beqa
f(r)  &\overset{d=3}{=}& k - \frac{\Lambda_W r^2}{1-\epsilon^2} - \sqrt{r}\sqrt{\frac{c_0}{1-\epsilon^2}+ \frac{\epsilon^2}{(1-\epsilon^2)^2}\Lambda_W^2 r^3   } \nn\\
&& \ii\ii\ii\3i \left\{ \begin{aligned}
&\overset{\epsilon=1}{=}  k - \frac{\Lambda_W }{2}r^2 - \frac{c_1}{r} %= k + \frac{ r^2}{\ell^2} - \frac{\ell}{r}
= k + \frac{r^2}{\ell^2}\bigg(1 - \frac{2M}{r^3}  \bigg), \\
&\overset{\epsilon=0}{=}  k - \Lambda_W r^2 - \sqrt{c_0 r}  = k + 2 \frac{r^2}{\ell^2} -2 \sqrt\frac{r}{\ell},   \label{Eq:AdS4-fr-neutral-L01}\\
           \end{aligned} \right. \\
  2M &=& \ell^3, \quad c_1 = \ell,  \quad \Lambda_W = - \frac{2}{\ell^2}, \quad c_0 = 4\ell^{-1},  \nn
\eeqa
while for HL$_5$, one has
\beqa
f(r) &\overset{d=4}{=}& k - \frac{1}{1-\epsilon^2}\frac{\Lambda_W r^2}{3}  - \sqrt{\frac{c_0}{1-\epsilon^2}+ \frac{\epsilon^2}{(1-\epsilon^2)^2}\frac{\Lambda_W^2r^4}{9}   }  \nn \\
&& \ii\ii\ii\3i \left\{ \begin{aligned}
&\overset{\epsilon=1}{=}  k - \frac{\Lambda_W }{6}r^2 - \frac{c_1}{r^2} %= k + \frac{ r^2}{\ell^2} - \frac{\ell^2}{r^2}
= k + \frac{r^2}{\ell^2}\bigg(1 - \frac{2M}{r^4}  \bigg),  \\
&\overset{\epsilon=0}{=}  k - \frac{\Lambda_W }{3}r^2 - \sqrt{c_0} = k + 2\frac{ r^2}{\ell^2} - 2,  \label{Eq:AdS5-fr-neutral-L01} \\
           \end{aligned} \right. \\
 2M &=& \ell^4 \quad c_1 = \ell^2,  \quad \Lambda_W = - \frac{6}{\ell^2}, \quad c_0 =4. \nn
\eeqa
In the limit $\epsilon\to 1$, the solutions are reduced to be the IR relevant solutions to the topological neutral black holes in HL gravity, which are exactly pure AdS$_4$ and AdS$_5$ space-time with negative cosmological constant, respectively; $\tilde{N}(r)=1$,
\beqa
\text{AdS}_4 : f(r) &=& k - \frac{\Lambda_W }{2}r^2 - \frac{c_1}{r} = k + \frac{r^2}{\ell^2} - \frac{\ell}{r}, \label{Eq:AdS4} \\
 c_1 &=& - \frac{c_0}{2\Lambda_W} = \frac{c_0\ell^2}{4} = \ell, \quad c_0 = 4\ell^{-1}; \quad\quad\quad \nn\\
\text{AdS}_5 : f(r) &=& k - \frac{\Lambda_W }{6}r^2 - \frac{c_1}{r^2} = k + \frac{ r^2}{\ell^2} - \frac{\ell^2}{r^2}, \label{Eq:AdS5}\\
 c_1 &=& - \frac{3c_0}{2\Lambda_W} = \frac{c_0\ell^2}{4} = \ell^2, \quad c_0=4. \nn
\eeqa
When the detailed balance condition is satisfied, namely, $\epsilon=0$, the solution reflects the detailed balance condition of the theory,
\beqa
\text{HL}_4 :  f(r) \3i &=& \3i k - \Lambda_W r^2 - \sqrt{c_0 r} = k + 2 \frac{r^2}{\ell^2} - 2\sqrt{\frac{r}{\ell}},  \label{Eq:HL4} \quad\quad\quad\\
\text{HL}_5 :  f(r) \3i &=& \3i k - \frac{\Lambda_W}{3} r^2 - \sqrt{c_0} = k + 2 \frac{r^2}{\ell^2} - 2. \label{Eq:HL5}
\eeqa
For the $d=2$ case, one needs to solve Eq.(\ref{Eq:Gauge-Field-d2}), with $U_{2}(r,1)$ given by Eq.(\ref{Eq:L_epsilon_lambda-d=2}) with $\lambda=1$, one needs to solve
\beqa
U_{2}(r,2) &=& \frac{1}{2\Lambda}\frac{1}{r}\bigg[\epsilon^2 \bigg( 4\Lambda^2 r^2 + 2\Lambda r F^\prime(r) \bigg) \nn\\
&+& (1-\epsilon^2)\bigg( \frac{1}{r^2}F(r)^2 - \frac{1}{r} F(r)F^\prime(r) \bigg) \bigg] = \frac{q_0^2}{2r}, \nn
\eeqa
which gives the solution
\beqa
f(r) &=& k - \frac{2\Lambda}{1-\epsilon^2}r^2 \label{Eq:ds2-HL-topological-charged-black-brane-L-epsilon-d=2}\\
&-& r \sqrt{\frac{c_0}{1-\epsilon^2}+ \frac{4\epsilon^2}{(1-\epsilon^2)^2}\Lambda^2 r^2 - \frac{2\Lambda}{1-\epsilon^2} q_0^2 \ln{r}} \nn \\
&& \ii\ii\ii \left\{ \begin{aligned}
&\overset{\epsilon=1}{=} k - \Lambda r^2 + \frac{c_0}{4\Lambda}- \frac{q_0^2}{2}\ln{r}, \\
&\overset{\epsilon=0}{=} k - 2\Lambda r^2 - r \sqrt{c_0- 2 \Lambda q_0^2 \ln{r}}.
           \end{aligned} \right.
\eeqa

\subsection{Topological charged black holes in generalized HL$_4$ gravity at $z=4$ with $\lambda=1$}
\label{sec:HL(3+1)-z=4}

For the case we considered in Appendix \ref{app:(3+1)HL-z=3}, when we only have $W_1$ and $W_2$ in Eqs.(\ref{Eq:W_1-3D}) and (\ref{Eq:W_2}), the full $W$ is the action for topological massive gravity~\cite{Deser:1981wh,Deser:1982vy,Bergshoeff:2009hq}, and resulting HL gravity just has a $z=3$ fixed point in the UV. To obtain a $z=4$ gravity, one not only needs to keep the $W_1$ and $W_2$, but also has to keep the $W_3$ term in Eq.(\ref{Eq:W_3-3D}). In the case we choose $\beta=-3/8$, $W$ represents the action of the Euclidean version of the new massive gravity, which is a renormalizable gravity theory in the Minkowski space-time~\cite{Bergshoeff:2009hq}. The action of $(3+1)$-dimensional HL gravity at $z=4$ in Eq.(\ref{Eq:S-HL(3+1)-z=4}) can be re-expressed as
\beqa
S &=&  \int dt \int d^3 x  ( {\mathcal L}_0 + {\mathcal L}_1 ) \nn\\
{\mathcal L}_0 &=&  \sqrt{g}  N  \bigg(\frac{2}{\kappa^2}(K_{ij}K^{ij}-\lambda K^2) +\frac{\kappa^2(\Lambda_W R-3\Lambda_W^2)}{8\kappa_W^4(1-3\lambda)} \bigg), \nn\\
{\mathcal L}_1 &=&  \sqrt{g}  N  \bigg[ - \frac{\kappa^2}{2\omega^4}C_{ij}C^{ij} + \frac{\kappa^2}{2\kappa_W^2 \omega^2}\epsilon^{ijk}R_{il}\nabla_j R^l_k
\nn\\
&-&\frac{\kappa^2}{8\kappa_W^4}R_{ij}R^{ij}+\frac{\kappa^2(1-4\lambda)}{32\kappa_W^4(1-3\lambda)}R^2 + \frac{\kappa^2}{2M\omega^2}C^{ij}L_{ij} \nn\\
&-&\frac{\kappa^2}{4M\kappa_W^2}\bigg(L_{kl}R^{kl} - \frac{1-2\lambda}{2(1-3\lambda)}LR + \frac{1}{1-3\lambda}\Lambda_W L \bigg) \nn\\
&-&\frac{\kappa^2}{8M^2}(L^{ij}L_{ij}-\tilde\lambda L^2) \bigg].  \label{Eq:L0-L1-4D-z4}
\eeqa

Assuming the metric ansatz for topological black holes in Eq.(\ref{Eq:ds2-topological-HL}), for simplicity, consider the $\lambda=1$ case, the action in Eq.(\ref{Eq:L0-L1-4D-z4}) becomes
\beqa
S &=& \frac{\kappa^2 \Omega_{2,k}}{8\kappa_W^4(3\lambda-1)} \int dt dr \tilde{N}(r) \bigg\{ \frac{1}{r}\bigg[ \Lambda_W r^2 + (f(r)-k) \nn\\
&+& \frac{1}{M} \bigg(\frac{(5+14\beta)\kappa_W^2 (f(r)-k)^2}{r^2} + \frac{(3+8\beta)\kappa_W^2}{4}f^\prime(r)^2 \nn\\
&-& (3+8\beta)\kappa_W^2 (f(r)-k) f^{\prime\prime}(r) \bigg) \bigg]^2  \bigg\}^\prime.
\eeqa
In the limit $M\to \infty$, the terms due to operators $\sim R^2$ in the potential $W_3$ in Eq.(\ref{Eq:W_3-3D}) vanish; thus, the action reduces to that in $z=3$ HL gravity for the $\lambda=1$ case in Eq.(\ref{Eq:L01_lambda}),
\beqa
S &=&  \frac{\kappa^2 \Omega_{2,k}}{16\kappa_W^4}\int dt dr \tilde{N}(r) \bigg(\frac{F(r)^2}{r} \bigg)^\prime \\
&=& \frac{\Omega_{2,k}c^3}{16\pi G_N}\frac{1}{-\Lambda_W} \int dt dr \tilde{N}(r)  \bigg(  \frac{(r^2 \Lambda_W + f(r)-k)^2}{r} \bigg)^\prime, \nn
\eeqa
which is nothing but Eq.(\ref{Eq:L0-L1-lambda=1}) with $d=3$.

When the term proportional to $M^{-1}$ is present and dominates at UV in $z=4$ HL gravity with $\lambda=1$, let us consider a special case called new massive gravity\cite{Bergshoeff:2009hq} with $\beta=-3/8$. In this case, the total action with $L_\epsilon+L_3$ becomes
\beqa
S &=& \frac{\Omega_{2,k}c^3}{16\pi G_N}\frac{1}{-\Lambda_W}  \int dt dr \tilde{N}(r) \bigg[ \frac{1}{r}\bigg( \Lambda_W r^2 + (f(r)-k) \nn\\
&-& \frac{\kappa_W^2 (f(r)-k)^2}{4Mr^2}   \bigg)^2 - \epsilon^2\frac{(f(r)-k)^2 }{r} \bigg]^\prime. \label{Eq:S-Lepsilon-L3}
\eeqa

By comparing Eq.(\ref{Eq:S-Lepsilon-L3}) with Eq.(\ref{Eq:S-Maxwell-Einstein-RSS}), and choosing the conversion to be that in Eq.(\ref{Eq:kappa-G_N-alpha}), then one obtains a EOM for charged black holes, in analogy to Eq.(\ref{Eq:Gauge-Field-d3}),
\beqa
&& \frac{1}{-\Lambda_W}\bigg[ \frac{1}{r}\bigg( \Lambda_W r^2 + (f(r)-k) - \frac{\kappa_W^2 (f(r)-k)^2}{4Mr^2}   \bigg)^2 \nn\\
&& - \epsilon^2\frac{(f(r)-k)^2 }{r} \bigg]^\prime = \frac{q_0^2}{2r^2}.
\eeqa
For HL gravity, one only needs to consider the $\epsilon=0$ case,
\beqa
c_0 - \frac{q_0^2}{2r}(-\Lambda_W) &=& \frac{1}{r}\bigg( \Lambda_W r^2 + (f(r)-k) \nn\\
&-& \frac{\kappa_W^2 (f(r)-k)^2}{4Mr^2}   \bigg)^2,
\eeqa
from which we obtain the topological charged black holes in HL$_4$ with $z=4$,
\beqa
\ii
f(r) \3i &=& \3i k \!+\! \frac{2 M}{\kappa_W^2}r^2 \Bigg[ 1 \!-\! \sqrt{1 \!+\! \frac{\kappa_W^2 \Lambda_W}{M}\bigg(1 \!+\! \frac{\sqrt{c_0 \!+\! \frac{\Lambda_W }{2} \frac{q_0^2}{r}}}{r^{3/2}\Lambda_W}\bigg)} \Bigg] \nn \\
\3i & = &  \3i k -\Lambda_W r^2 - \sqrt{c_0 r + \Lambda_W \frac{q_0^2}{2}}, \quad M\to\infty \label{Eq:HL-charged-z=4}\\
\3i & = &  \3i k + 2\frac{r^2}{\ell^2} - \sqrt{ \frac{4r}{\ell} - \frac{q_0^2}{\ell^2} }, \nn
\eeqa
When $M\to \infty$, the solution reduces to be neutral black holes in HL$_4$ gravity at critical point $z=3$ with $\epsilon=0$ and $\lambda=1$. $c_0$ is an integral constant, which is given by Eq.(\ref{Eq:c0-c1}), namely, $c_0=4\ell^{d-4}=4c_1/\ell^2$, and $\Lambda_W=-2/\ell^2$ for $d=3$. $c_0$ can also be determined by the horizon radius via $f(r_+)=0$, which gives
\beqa
c_0 &=& \frac{k^2}{r_+} - 2k \Lambda_W r_+ + \Lambda_W^2 r_+^3 - \Lambda_W\frac{q_0^2}{2r_+} - \frac{k^2\kappa_W^2\Lambda_W}{2M r_+}  \nn\\
& \times & \bigg( 1 - \frac{k}{\Lambda_W r_+^2} - \frac{k^2\kappa_W^2}{8M\Lambda_W r_+^4} \bigg).\label{Eq:c0-rp}
\eeqa
In the $M\to\infty$ limit, the solution just reproduces the solution in $z=3$ HL gravity in Eq.(\ref{Eq:fr-eps-0}). %While in the limit $q_0\to 0$ and $M\to\infty$, it just reduce to the topological neutral black holes in Eq.(\ref{Eq:HL-neutral=z=4}).
In the infinite boundary, the redshift factor becomes
\beqa
f(r) & = & k - \frac{2M}{\kappa_W^2}\bigg(\sqrt{1+\frac{\kappa_W^2\Lambda_W}{M}} -1 \bigg)r^2 \nn\\
&-& \sqrt{c_0 r \bigg( 1 + \frac{\kappa_W^2\Lambda_W}{M} \bigg)^{-1}} \nn\\
&+& \frac{q_0^2}{4\sqrt{c_0r}}(-\Lambda_W)\bigg( 1 + \frac{\kappa_W^2\Lambda_W}{M} \bigg)^{-1} + {\mathcal O} (\frac{1}{r}) \nn\\
&\overset{M\to\infty}{=}& k - \Lambda_W r^2 - \sqrt{c_0 r} - \Lambda_W \frac{q_0^2}{4\sqrt{c_0 r}} + {\mathcal O} (\frac{1}{r}) . \nn
\eeqa
For the Ricci flat case $k=0$ with $M$ fixed, the redshift factor in the infinite boundary becomes
\beqa
f(r) & \overset{r\to \infty}{\to} & - \frac{2M}{\kappa_W^2}\bigg(\sqrt{1+\frac{\kappa_W^2\Lambda_W}{M}} -1 \bigg)r^2 \nn\\
&\equiv & - \Lambda_W r^2 h(\frac{\kappa_W^2\Lambda_W}{M}) %= 2\frac{r^2}{\ell^2} h(\frac{-2\kappa_W^2}{M\ell^2}),
\label{Eq:z=4-fr-infity}
\eeqa
where $h(x) \equiv  2\big(\sqrt{1+x} - 1\big)/x = 1 - {x}/{4} + {x^2}/{8} - {\mathcal O}(x^3)$.
The temperature becomes
\beqa
T \ii &=& \ii \frac{1}{8\pi r_+} \frac{1}{{1+ k\frac{\kappa_W^2}{2M r_+^2}}}\bigg[ -3\Lambda_W r_+^2  - k-k^2\frac{5\kappa_W^2}{4Mr_+^2} \qquad \nn\\
\ii &-& \ii\bigg(-\Lambda_W r_+^2 + k + k^2\frac{\kappa_W^2}{2M r_+^2} \bigg)^{-1}(- \Lambda_W)\frac{q_0^2}{2}  \bigg] \\
\ii &\overset{M\to\infty}{=} & \! \frac{-3\Lambda_W r_+^2 - k - (-\Lambda_W r_+^2 + k)^{-1} (- \Lambda_W)\frac{q_0^2}{2} }{8\pi r_+} \nn\\
\ii &\overset{k=0}{=}& \ii %\frac{1}{8\pi r_+} \bigg( -3\Lambda_W r_+^2 - \frac{q_0^2}{2r_+^2} \bigg)
\frac{1}{8\pi r_+} \bigg( 6 \frac{r_+^2}{\ell^2} - \frac{q_0^2}{2r_+^2} \bigg). \nn
\eeqa
In Ricci flat case with $k=0$, the Hawking temperature is not affected by the $W_3$ term in Eq.(\ref{Eq:W_3-3D}). As to be expected, the charge will not affect the entropy of the black holes, which gives the result as that in neutral black holes.
From the temperature obtained above, one can define charge through extremal radius,
\beqa
q_0^2 \equiv   - 6\Lambda_W r_\star^4 =\frac{12 r_\star^4}{\ell^2}.
\eeqa
By substitute the $c_0$ from solving $f(r_+)=0$ in Eq.(\ref{Eq:c0-rp}) and the $q_0$ above into the redshift factor in Eq.(\ref{Eq:HL-charged-z=4}), one obtains
\beqa
f(r) \3i &=& \3i\frac{2 M}{\kappa_W^2}r^2 \bigg[ 1 \!-\! \sqrt{1 \!+\! \frac{\Lambda_W\kappa_W^2 }{M}\bigg(1 \!-\! \sqrt{\frac{r_+^3}{r^3} \!+\! \frac{\big(1 \!-\! \frac{r}{r_+} \big)q_0^2}{2\Lambda_W r^4}  } \bigg) } \bigg] \nn\\
\3i &=& \3i \frac{2 M}{\kappa_W^2}r^2 \bigg[ 1 \!-\! \sqrt{1 \!-\! \frac{2}{\ell^2}\frac{\kappa_W^2 }{M}\bigg(1 \!-\! \sqrt{\frac{r_+^3}{r^3} \!-\! 3 \frac{r_\star^4}{r^4}\Big(1 \!-\! \frac{r}{r_+} \Big)  } \bigg) } \bigg]. \nn
\eeqa
It is obvious that $f(r_+)=0$ is always satisfied. For finite temperature case $r_\star<r_+<r$, the near horizon behavior is
\beqa
f(r) \approx \frac{6(r-r_+)(r+r_+-2r_\star)}{\ell^2}\overset{r_+\to r_\star}{=} \frac{6(r-r_s)^2}{\ell^2}, \nn
\eeqa
where the last equality gives near horizon behavior of extremal black holes at the zero temperature case $r_+=r_\star$. It is independent of the UV parameter $M$ as expected.

\section{Conformal Symmetry on the Boundary of Topological Charged Black Holes in Generalized Ho\v{r}ava-Lifshitz Gravity}
\label{sec:CFT-HL}

\subsection{Matching generalized HL gravity with $\lambda=1$ to Einstein gravity with Maxwell action}

\subsubsection{Generalized HL$_{d+1}$ gravity with gauge field}

Consider the modified action of HL$_{d+1}$ gravity with $\epsilon=0$ due to the dynamics of $L_0+L_1$, as shown in Eq.(\ref{Eq:S-HL(d+1)-z=3}), which reduces to be the Einstein gravity at large distance in the IR limit, with only $L_0$ relevant, namely,
\beqa
\ii
S_0 = \frac{1}{2\kappa_G}\int dt d^d x \sqrt{g}N[(K_{ij}K^{ij}-K^2)+R-2\Lambda], \label{Eq:S_minimal}
\eeqa
where $\kappa_G = 8\pi G_N $, $G_N$ is the effective Newton's gravitational constant, $c$ is the effective speed of light in vacuum and $\Lambda$ is the effective cosmological constant. They can be expressed as UV parameters as below,
\beqa
\ii\3i
c \! = \! \frac{\kappa^2}{4\kappa_W^2}\sqrt\frac{(d-2)\Lambda_W}{1-d\lambda}, \label{Eq:c}
 G_N \!=\! \frac{\kappa^2 c}{32\pi},  \Lambda \!=\! \frac{d}{2(d-2)}\Lambda_W. ~\label{Eq:G_N-Lambda}
\eeqa
By using a scaling dimension of the couplings in Eqs.(\ref{Eq:scaling-kappa-lambda}) and (\ref{Eq:scaling-kappa_W}), one obtains the dimension for the effective speed of light, Newton's constant, and cosmological constant,
\beqa
\ii
 [c] = z-1, ~ [\kappa_G]=[G_N] %= (z-d) +(z-1)
= 2z -(d+1), ~  [\Lambda] = 2.
\eeqa
As a result, at large distance, when $d=3$ and $z=1$, we have $[\kappa_G] = -2$ as expected in general relativity, which leads to a nonrenormalizable theory in UV when quantum loops are taken into account. It means that the effective large distance speed of light originates microscopically from a relevant coupling in the UV theory describing the anisotropic dynamics of the space-time at short distance. Conversely, the UV parameters $\kappa$, $\Lambda_W$ can also be expressed by the physical measurable quantities,
\beqa
 \kappa = \sqrt{ \frac{32\pi G_N}{c}}, \quad \Lambda_W = \frac{2(d-2)}{d}\Lambda,  \label{Eq:LambdaW-Lambda} %\\ && \nn
\eeqa
with $ [\Lambda_W] = [\Lambda] = [R] = [L^{-2}] =  2$. At large distance all of them deduce from the UV complete nonrelativistic gravity theory with dynamical critical exponent $z\ne 1$, which dominates at short distance.

By using effective speed of light and Newton's constant in Eq.(\ref{Eq:c}), we have
\beqa
 \frac{c^3}{G_N} &=& \frac{2\pi(d-2)}{d\lambda-1}(-\Lambda_W)\frac{\kappa^2}{\kappa_W^4}.  \label{Eq:c/G_N}
\eeqa

From the expression in Eq.(\ref{Eq:c}), we know that in the space-time at large distance with $d\ge 3$ and $\lambda=1$, the cosmological constant can only be negative,
\beqa
c>0 \quad \Rightarrow  \quad \Lambda<0, \quad (d\ge 3,\lambda=1),
\eeqa
i.e., for the $d=3$ case, one has
\beqa
\ii\3i
c  \! = \! \frac{\kappa^2}{4\kappa_W^2}\sqrt\frac{\Lambda_W}{1-3\lambda}>0,
~ G_N \! = \! \frac{\kappa^2 c}{32\pi}, ~ \Lambda \! = \! \frac{3}{2}\Lambda_W<0.
\eeqa
For modified HL gravity with Lagrangian $L_\epsilon=L_0 +(1-\epsilon^2)L_1$ with $\lambda=1$, with a minimal extension by including the Maxwell action with Hamiltonian expression $S^{EM}$ defined in Eq.(\ref{Eq:S-Maxwell-Einstein}). The generic metric of topological charged black holes to the equations of motion of the actions $S_\epsilon+S^{EM}$ is given by
\beqa
ds^2 = -  f(r) dt^2 + \frac{dr^2}{f(r)} + r^2 d\Omega_{d-1,k}^2,  \label{Eq:ds2-fr}
\eeqa
with the redshift factor,
\beqa
f(r) = k + \a^{-2} \frac{r^2}{\ell^2} g(r), \quad ~ \a \equiv \sqrt\frac{1+\epsilon}{2}, \label{Eq:fr-gr}
\eeqa
with the $d\ge 3$ case as given in Eqs.(\ref{Eq:phi=A0}) and (\ref{Eq:ds2-HL-topological-charged-black-brane-L-epsilon-2})
\beqa
&& g(r) \overset{d\ge 3}{\equiv} \frac{ 1 - \epsilon \sqrt{1 + \frac{1-\epsilon^2}{\epsilon^2}\bigg(  \frac{c_1\ell^2}{r^d} - \frac{\ell^2}{2(d-1)(d-2)}\frac{ q_0^2 }{r^{2d-2}}  \bigg) } }{1-\epsilon}, \nn \\
&& \phi(r) \overset{d\ge 3}{=} \mu\bigg(1 -\frac{r_0^{d-2}}{r^{d-2}}\bigg),  \quad q_0 \overset{d\ge 3}{=}  (d-2) r_0^{d-2},  \label{Eq:ds2-RN-HL(d+1)}
\eeqa
or with $d=2$ as special case as given in Eqs.(\ref{Eq:phi=A0-d=2}) and (\ref{Eq:ds2-HL-topological-charged-black-brane-L-epsilon-d=2}),
\beqa
&& g(r)  \overset{d = 2}{\equiv}  \frac{1}{1-\epsilon}\bigg[ 1 - \epsilon \sqrt{1 + \frac{1-\epsilon^2}{\epsilon^2}\bigg(  \frac{c_1\ell^2}{r^2} +  \frac{\ell^2}{2} \frac{q_0^2}{r^2}\ln{r}\bigg) } \bigg], \nn \\
&& \phi(r) \overset{d=2}{=} \mu\bigg(1 - \frac{\ln{r}}{\ln{r_0}} \bigg), \quad q_0 \overset{d=2}{=} (\ln{r_0})^{-1}, \label{Eq:ds2-RN-HL(2+1)}
\eeqa
where we have used that $\Lambda=-d(d-1)/(2\ell^2) \overset{d=2}{=} - \ell^{-2} $ and $\Lambda_W = -(d-1)(d-2)/\ell^2$ to match with the AdS in the IR limit at large distance.

For the case of maximal detail balance violation with $\epsilon=1$ in generalized HL gravity with $\lambda=1$, its dynamics are completely determined by $L_0$, i.e., Einstein's gravity. As shown in Eq.(\ref{Eq:ds2-HL-topological-charged-black-brane-L-epsilon-2}), it just reproduces the topological charged black holes in AdS$_{d+1}$ geometric, namely, the RN metric,
\beqa
\ii\ii
\epsilon \!=\! 1, ~  g(r) \! = \! 1 \!-\! \frac{c_1 \ell^2}{r^d} \!+\! \frac{\ell^2}{2(d-1)(d-2) }\frac{q_0^2}{r^{2d-2}}. \label{Eq:ds2-gr-RN-AdS(d+1)}
\eeqa

For the case of no detail balance violation with $\epsilon=0$ in HL gravity with $\lambda=1$, its dynamics are determined by both $L_0$ and $L_1$, i.e., $L_0+L_1$ with $\lambda=1$. As shown in Eq.(\ref{Eq:ds2-HL-topological-charged-black-brane-L-epsilon-2}), it just reproduces the topological charged black holes in HL$_{d+1}$ gravity,
\beqa
\ii\ii
\epsilon \!=\! 0, ~ g(r) \! = \! 1 \!-\! \sqrt{ \frac{c_1 \ell^2}{r^{d}} \!-\! \frac{ \ell^{2}}{2(d-1)(d-2)}\frac{q_0^2}{r^{2d-2}}  }.  \label{Eq:ds2-gr-RN-HL(d+1)}
\eeqa

For Ricci flat case $k=0$, one obtains the charged black branes in generalized HL$_{d+1}$ gravity
\beqa
\ii\ii
ds^2 = - \a^{-2}\frac{r^2}{\ell^2} g(r) dt^2 + \a^2 \frac{\ell^2}{r^2}\frac{dr^2}{g(r)} + \frac{r^2}{\ell^2} dx_{d-1}^2 , \label{Eq:ds2-fr-k=0}
\eeqa
where we have used the notation that % $2M\equiv \ell^d$, and
$d\Omega_{d-1,0}^2 \equiv dx_{d-1}^2/\ell^2 = \sum_{i=1}^{d-1}dx_i^2/\ell^2$.

In the infinite boundary $r\to \infty$, $g(r)\to 1$. As $\a$ is a coordinate independent parameter, the metric in Eq.(\ref{Eq:ds2-fr-k=0}) just reduces to be the pure AdS$_{d+1}$ space-time, namely,
\beqa
\ii\ii
ds^2 = - \frac{r^2}{\ell_H^2} dt^2 +  \frac{\ell_H^2}{r^2}dr^2 + \frac{r^2}{\ell_H^2} dx_{d-1}^2,
%\label{Eq:ds2-fr-k=0}
\eeqa
by making the rescaling law behavior
\beqa
\ell \to \a^{-1}\ell \equiv \ell_H, \quad \vec{x} \to  \a^{-1} \vec{x}. \label{Eq:r-ell-H}
%r \to \a r , \quad \vec{x} \to  \a^{-1} \vec{x}.  \label{Eq:r-ell-H}
\eeqa
which implies that the effective cosmological constant $\Lambda\to \a^2 \Lambda$. This implies that in the momentum $k$ space-time, the high curvature due to Lagrangian $(1-\epsilon^2)L_1$ will contribute a rescaling law in momentum space
\beqa
|\vec{k}| \to \sqrt\frac{1+\epsilon}{2} |\vec{k}|.
\eeqa

\subsubsection{Conformal coordinate in infinite boundary}

The metric in Eq.(\ref{Eq:ds2-fr}) can also be expressed in inverse radial coordinate $u\equiv \ell^2/r$ as\footnote{The coordinate $u$ can be identified as the RG cutoff scale $\sim \Lambda_c$.}
\beqa
\ii\ii
ds^2 =  \frac{\ell^2}{u^2}\bigg( - \frac{u^2}{\ell^2} f(u)dt^2 + \frac{\ell^2}{u^2}   \frac{du^2}{f(u)} + \ell^2 d\Omega_{d-1,k}^2 \bigg),  \label{Eq:ds2-fu}
\eeqa
with Eq.(\ref{Eq:fr-gr}) becoming
\beqa
 f(u) = k + \a^{-2} \frac{\ell^2}{u^2}g(u).
\eeqa

For the Ricci flat case, Eq.(\ref{Eq:ds2-fr-k=0}) becomes
\beqa
ds^2 =  \frac{\ell^2}{u^2} \bigg( \a^2 \frac{du^2}{g(u)} - \a^{-2}  g(u) dt^2 +  dx_{d-1}^2 \bigg). \nn
\eeqa
The nontrivial function $g(u)$ indicates that the physics is changing with scale. The charged black branes provide a universal geometric description of many different systems at a finite length scale $r$, independent of specific microscopic details.

In the infinite boundary condition $r\to \infty$, one has $g(r)\to 1$. In the conformal coordinate, the infinite boundary is lying at $u\to 0$ with $g(u) \to 1$ too. One just obtains a generalized HL$_{d+1}$ space-time metric in conformal coordinate,
\beqa
ds^2 = \bigg( \frac{\ell}{u}\bigg)^2 ( \a^2 du^2 - \a^{-2} dt^2 + d\vec{x}_{d-1}^2), \label{Eq:HL-AdS-infinty}
\eeqa
where the signature of the ordinary space-time is chosen as $(-\a^{-1},\vec{1}_{d-1})$, with $(d-1)$ spatial dimensions. It is obvious that there is a scale invariance for the metric with a scale transformation $u\to e^{\alpha(x^\mu)} u$ along the $u$ coordinate. By making the rescaling law behavior in Eq.(\ref{Eq:r-ell-H}),
\beqa
u \equiv \frac{\ell^2}{r} \to \frac{\ell_H^2}{r} = \a^{-2} u, \label{Eq:u-ell-H}
\eeqa
then Eq.(\ref{Eq:HL-AdS-infinty}) becomes
\beqa
ds^2 \to \bigg( \frac{\ell_H}{u}\bigg)^2 (  du^2 -  dt^2 +  d\vec{x}_{d-1}^2).
\eeqa
As discussed before, this just reduces to the pure AdS$_{d+1}$ space-time, by just using the rescaling law on momentum
\beqa
|\vec{k}|\to \a |\vec{k}|. \label{Eq:k-ell-H}
\eeqa

For HL$_4$ gravity at $z=4$ fixed point, by using Eq.(\ref{Eq:z=4-fr-infity}), we obtain effective pure AdS$_4$ black branes,
\beqa
ds^2 = \bigg( \frac{\ell}{u}\bigg)^2 \bigg( \frac{du^2}{2h(\frac{\kappa_W^2\Lambda_W}{M})}  - 2h(\frac{\kappa_W^2\Lambda_W}{M}) dt^2 + d\vec{x}^2\bigg), \nn
\eeqa
where $\quad h(x)\equiv {2}\big(\sqrt{1+x} - 1 \big)/x$. This is equivalent to replacing $\epsilon$ by $h^{-1}(x)-1$, $x=\kappa_W^2\Lambda_W/M$. In this case, the scaling law in Eqs.(\ref{Eq:r-ell-H}), (\ref{Eq:u-ell-H}), and (\ref{Eq:k-ell-H}) can be calibrated by
\beqa
\a \to \frac{1}{\sqrt{2h(x)}}, \quad \text{or} ~\quad \epsilon \to \frac{1}{\sqrt{h(x)}} - 1 \approx \frac{1}{8}x.
\eeqa

\subsubsection{Matching HL gravity to Einstein gravity with AdS symmetry}

We consider Einstein gravity
\beqa
S_0 = \frac{1}{2\kappa_G} \int d^{d+1} x \sqrt{-g}(R-2\Lambda), \label{Eq:S0-Einstein}
\eeqa
which should match the leading order of generalized HL$_{d+1}$ gravity, in the IR limit at large distance, i.e., the leading order action $S_0$ in Eq.(\ref{Eq:S_minimal}). The Hamiltonian expression of the Maxwell action $S^{EM}$ as shown in Eq.(\ref{Eq:S-Maxwell-Einstein}), is equivalent to the action of $U(1)$ gauge field,
\beqa
 S^{EM} \3i &=& \3i \int d^{d+1}x  \sqrt{-g} {\mathcal L}_{em}, \nn\\
{\mathcal L}_{em} \3i &=& \3i - \frac{1}{4g_{em}^2}  g^{MN}g^{PQ}F_{MP}F_{NQ}+ A_M J^M , \qquad
\eeqa
where $[J^M]=d$, $[A_M]=1$ and $g_{em}^2$ is the electric charge coupling of the gauge field with $[g_{em}^2]=[L^{d-3}]$, thus $[g_{em}]=(3-d)/2$.

The full action of Einstein gravity in Eq.(\ref{Eq:S0-Einstein}) and Maxwell action at large distance becomes
\beqa
S \3i &=& \3i \frac{1}{2\kappa_G} \! \int \! d^{d+1}x \sqrt{-g} \bigg( R \!+\! \frac{d(d\!-\!1)}{\ell^2}  \!-\! \frac{\ell^2}{g_F^2}F_{MN}F^{MN} \!+\! \ldots \bigg),  \nn
%\label{Eq:S_minimal-S_EM-eff}
\eeqa
where $[\kappa_G]=[L^{d-1}]=1-d$(e.g. for $d=3$, $[\kappa_G]=-2$, since $\kappa_G^{-1}\sim \Lambda_{\text{planck}}$), $[R]=[\Lambda]=[L^{-2}]=2$ and $[F^2]=4$. The cosmological constant in Eq.(\ref{Eq:S0-Einstein}) can be defined by $\Lambda \equiv - {d(d-1)}/{2\ell^2}$ for AdS gravity at large distance, where $\ell$ is the AdS radius. In addition, we have introduced an effective dimensionless gauge coupling $g_F$ as a measure of the relative strength of the electromagnetic and gravitational forces,
\beqa
 \frac{1}{4g_{em}^2} &=& \frac{1}{2\kappa_G} \frac{\ell^2}{g_F^2}, \quad \Rightarrow  \quad   g_{em} = \frac{\sqrt{\kappa_G} g_F}{\sqrt{2}\ell}, \quad \text{or} \quad\nn\\
   g_F &=& \frac{\sqrt{2}\ell}{\sqrt{\kappa_G}}g_{em}, \quad [g_F] = 0,\label{Eq:gF-effective}
\eeqa
where $\ell$ is the curvature radius of AdS. It is obvious that the effective couplings of the bulk theory, $g_F$ are becoming stronger as gravitational force is becoming weaker, and it becomes divergent in the $\kappa_G \to 0$ limit . Thus the effective gauge coupling $g_F$ characterizes the relative strength of the $U(1)$ gauge and gravitational forces.

After doing variation with respect to the metric and the gauge field, respectively, on obtain the Einstein equations and Maxwell equation
\beqa
\ii\ii
 E_{MN} \3i &=& \3i R_{MN} - \frac{1}{2}g_{MN}R + \Lambda g_{MN} - \kappa_G \, T_{MN}^{em} = 0, \nn\\
\ii 0 \3i &=& \3i \nabla_N  F^{NM} + J^{M},   \label{Eq:EOM_Einstein-Maxwell} \\
\ii T_{MN}^{em} \3i &=& \3i \frac{1}{g_{em}^2} \bigg( F_{ML}F_{N}^{\,\,\,\,L} - \frac{1}{4}g_{MN}F^{PQ}F_{PQ} \bigg) \!+\! g_{MN} A_{P}J^{P} , \ii \nn\\
\ii T^{em} \3i &=& \3i \frac{1}{g_{em}^2}  \bigg(1 - \frac{d+1}{4}\bigg) F^{PQ}F_{PQ}  + (d+1)A_P J^P, \nn
\eeqa
or, equivalently, after subtracting the Ricci scalar, one obtains
\beqa
 E_{MN} &=& R_{MN} - \frac{2}{d-1}\Lambda g_{MN} - \kappa_G \, t_{MN}=0, \\
 t_{MN}  &\equiv & \frac{1}{g_{em}^2} \bigg( F_{ML}F_{N}^{\,\,\,L} - \frac{1}{2(d-1)}g_{MN}F^{PQ}F_{PQ} \bigg) \nn\\
&-& \frac{2}{d-1} g_{MN} A_{P}J^{P}. \nn
\eeqa

First, assume that the topological background metric in Eq.(\ref{Eq:ds2-fr}) has a Ricci flat hypersurface with index $k=0$, namely, Eq.(\ref{Eq:ds2-gr-RN-AdS(d+1)}).

Second, assume that there is neither external source $J^M=0$, nor magnetic field, but only a static electric field, then $B_k = \epsilon^{ijk} F_{ij}/2 =0$, $E_{i}=F_{ti}=-F_{it}\ne 0$. In terms of vector potential, it is equivalent to assume that the vector potential has only time component, $A(r) = (A_t(r),0,\ldots, 0)$. For simplicity, it is assumed that
\beqa
&& E_r = F_{tr} = - F_{rt} = A_t^\prime(r) , \nn\\
&& E_x = F_{tx} = - F_{xt} = 0, \quad B_M =0.
\eeqa
We have used the Faraday's law that
\beqa
E_r = - \nabla \phi = - \phi^\prime = p(r), \quad A_t(r) = - \phi,
\eeqa
where $p(r)$ is given in Eqs.(\ref{Eq:Gauge-Field-d3}) or (\ref{Eq:Gauge-Field-d2}).

By substituting the vector potential into the ansatz metric in Eq.(\ref{Eq:ds2-gr-RN-AdS(d+1)}), the equation of motion of Maxwell and Einstein in Eq.(\ref{Eq:EOM_Einstein-Maxwell}) becomes
\beqa
0 \3i &=& \3i \frac{d-1}{r}A_t^\prime(r) + A_t^{\prime\prime}(r), \nn\\
0 \3i &=& \3i g^\prime(r) + \frac{d}{r}(g(r) - 1) + \frac{2\kappa_G}{g_{em}^2}\frac{1}{2(d-1)}\frac{1}{r} A_t^\prime(r)^2, \nn\\
0 \3i &=& \3i g^{\prime\prime}(r) \!+\! \frac{d\!+\!3}{r}g^\prime(r) \!+\! \frac{2d}{r^2}(g(r) \!-\! 1) \!-\! \frac{2\kappa_G}{g_{em}^2}\frac{d\!-\!2}{d\!-\!1}\frac{1}{r^2}A_t^\prime(r)^2. \nn
\eeqa

By imposing the infinite boundary condition for vector potential, i.e., the Maxwell field is nonvanishing but has finite charge density $A_t(r\to \infty)=\mu\ne 0$, the equations of motion above give the Maxwell vector potential and the metric
\beqa
A_t \2i &=& \2i \mu\bigg(1 - \frac{q_0}{(d-2)r^{d-2}}\bigg), \nn\\
g(r) \2i &=& \2i 1 + \frac{\kappa_G}{g_{em}^2}\frac{  q_0^2}{(d-1)(d-2)r^{2(d-1)}} - \frac{c_1 \ell^2 }{r^d}, \qquad \label{Eq:At-gr-HL(d+1)-epsilon=0}
\eeqa
where $q_0$ and $c_1$ are integral constants, $q_0$ is physically equivalent to the bulk charge density of the Maxwell field in $(d+1)$-dimensional space-time and $c_1$ is physically equivalent to the interaction source of the geometry$-$the Mass$\sim 2M$.

By comparing with Eq.(\ref{Eq:kappa-G_N-alpha}), one can deduce that $\alpha = g_{em}^2\ell^2c^{-3}$ so that the IR relevant solutions to HL gravity with $\epsilon=1$ and Maxwell gauge field are consistent with the RN AdS$_{d+1}$ metric with the $k=0$ case.

In this case of HL gravity with $\epsilon=1$, the vector potential of the gauge field in Eq.(\ref{Eq:ds2-fr-L-epsilon-q0}) can be re-expressed as
\beqa
\ii\ii
A_t  =  \mu\bigg( 1 -  \frac{r_0^{d-2}}{r^{d-2}} \bigg) , ~ g(r) =  1  - \frac{2M}{r^d} + \frac{Q^2}{r^{2(d-1)}}, \label{Eq:At-fr-AdS(d+1)}
\eeqa
where
\beqa
 2M \equiv \ell^2 c_1 = \ell^d, \quad Q^2 \equiv  \frac{\ell^2}{2(d-1)(d-2)}q_0^2. \label{Eq:M-ell-Q-q0}
\eeqa
with scaling dimensions $[M]=[L^d]$, $[Q] = [L^{d-1}] =  [L][q_0] $; thus, $ [q_0] = [L^{d-2}]$. In the second equality of the first equation, we have used Eq.(\ref{Eq:c0-c1}).

To match with the RN AdS$_{d+1}$ metric, by comparing with Eqs.(\ref{Eq:At-gr-HL(d+1)-epsilon=0}) and (\ref{Eq:At-fr-AdS(d+1)}), one obtains the charge
\beqa
q_0 \2i & \equiv & \2i  (d-2) r_0^{d-2},
\label{Eq:q0-mu} \\
Q^2 \2i & \equiv & \2i  \frac{\kappa_G}{g_{em}^2}\frac{q_0^2}{(d-1)(d-2)}.
\label{Eq:Q-q0}
\eeqa
By comparing with Eq.(\ref{Eq:gF-effective}) and using Eq.(\ref{Eq:q0-mu}),
\beqa
Q^2 = \frac{2\ell^2 q_0^2  }{g_F^2(d-1)(d-2)} = \frac{2}{g_F^2}\frac{d-2}{d-1}{\ell^2}{r_0^{2(d-2)}},
\eeqa
and comparing with Eq.(\ref{Eq:M-ell-Q-q0}), we have
\beqa
\frac{1}{g_F^2} = \frac{1}{4}, \quad \Rightarrow \quad  \frac{\kappa_G}{g_{em}^2} &=& \frac{\ell^2}{2},
\eeqa
with $[{\kappa_G}/{g_{em}^2}]=[L^2]=-2$ and $[g_F]=0$, where in the last equality we have used Eq.(\ref{Eq:gF-effective}).

Since $g_F$ is dimensionless, the charge can also be expressed with characterize length $\ell_F$,
\beqa
Q = \sqrt\frac{2(d-2)}{d-1} \ell_F r_0^{d-2}, \quad \ell_F \equiv \frac{\ell}{g_F},
\label{Eq:Q-mu-AdS(d+1)}
\eeqa
with $[Q] = [L^{d-1}]$, where $r_0$ is the horizon radius determined by the largest positive root
of the redshift factor $f(r_0)=0$ and also $A_t(r_0)=0$.

\subsection{Topological charged black holes in generalized HL gravity with $\lambda=1$ and scale invariance on the boundary}

By using the matching conditions in Eq.(\ref{Eq:M-ell-Q-q0}), Eq.(\ref{Eq:ds2-RN-HL(d+1)}) can be re-expressed for the $d\ge 3$ case as
\beqa
g(r) \overset{}{\equiv} \frac{ 1 - \epsilon \sqrt{1 + \frac{1-\epsilon^2}{\epsilon^2}\bigg(  \frac{2M}{r^d} - \frac{ Q^2 }{r^{2d-2}}  \bigg) } }{1-\epsilon}.
\label{Eq:ds2-RN-HL(d+1)-match}
\eeqa

The radius of the horizon in $r$ coordinates is defined as the root of the function $f(r)=0$ for $k=\pm 1$ or $g(r)=0$ for the $k=0$ case in Eqs.(\ref{Eq:ds2-gr-RN-AdS(d+1)}) and (\ref{Eq:fr-gr}), from which we obtain the mass parameter of the charged black hole
\beqa
\ii
2M \! = \!
r_0^d\bigg( 1 \!+\! \frac{Q^2}{r_0^{2d-2}} \bigg) \!+\! k \ell^2 r_0^{d-2} \!+\! k^2 \frac{1\!-\!\epsilon^2}{4}  \ell^4r_0^{d-4} \3i , \quad \label{Eq:2M-HL-HL(d+1)}
\eeqa
where the charge is defined by character length $\ell_F$ via Eq.(\ref{Eq:Q-mu-AdS(d+1)}), which essentially reflects the competition through relative strength of gravity and electromagnetic forces.

By using Eq.(\ref{Eq:2M-HL-HL(d+1)}), $f(r)$ in Eq.(\ref{Eq:fr-gr}) with $g(r)$ defined in Eq.(\ref{Eq:ds2-RN-HL(d+1)}) can be re-expressed as
%\begin{widetext}
\beqa
g(r) &\overset{d\ge 3}{=}& \frac{ 1 - \epsilon \sqrt{1 + \frac{1-\epsilon^2}{\epsilon^2}\frac{r_0^d}{r^d} h(r) }  }{1-\epsilon}.  \label{Eq:gr-Q-epsilon} \\
h(r) &=& 1 + k \frac{\ell^2}{r_0^2} +  k^2 \frac{1-\epsilon^2}{4} \frac{\ell^4}{r_0^4}  + \frac{Q^2}{r_0^{2d-2}}\bigg( 1 - \frac{r_0^{d-2}}{r^{d-2}} \bigg),\nn
\eeqa
%\end{widetext}
from which it is straightforward to check that $r_0$ is the horizon radius determined by $f(r_0)=k+\hat{H}^{-2}r_0^2g(r_0)/\ell^2 =0$ given by Eq.(\ref{Eq:fr-gr}).

The Hawking temperature of a topological black hole carrying charge, which is by definition proportional to the surface gravity, turns out to be
\beqa
T &=& \frac{dr_0}{4\pi \ell^2}\bigg( \frac{d-4}{2d} k \frac{\ell^2}{r_0^2} + \frac{ 1 + k \frac{\ell^2}{2r_0^2}  - \frac{d-2}{d}\frac{Q^2}{r_0^{2d-2}} }{1 +  k \frac{\ell^2}{2r_0^2} (1-\epsilon^2) } \bigg) \nn \\
&& \ii\ii\ii\3i \left\{ \begin{aligned}
&\overset{\epsilon=0}{=} \frac{dr_0}{4\pi \ell^2}\bigg( \frac{d-4}{2d} k \frac{\ell^2}{r_0^2} +  1 - \frac{1}{1 + k \frac{\ell^2}{2r_0^2} }\frac{d-2}{d}\frac{Q^2}{r_0^{2d-2}}  \bigg) \quad \\
&\overset{\epsilon=1}{=} \frac{dr_0}{4\pi \ell^2}\bigg( 1 + \frac{d-2}{d} k \frac{\ell^2}{r_0^2} -  \frac{d-2}{d}\frac{Q^2}{r_0^{2d-2}}  \bigg),   \label{Eq:TH-HL(d+1)} \\
           \end{aligned} \right.
\eeqa
where to match with the results in HL$_{d+1}(\epsilon=1)$ and AdS$_{{d+1}}$,
we have used the matching relation in Eq.(\ref{Eq:Q-q0}).
One can observe that for the Ricci flat case, $T_H$ is independent of $\epsilon$, i.e.,
Hawking temperature is irrelevant to the detailed balance condition for the $k=0$ case.

\subsubsection{Finite temperature case}

According to Eq.(\ref{Eq:TH-HL(d+1)}), the extremal radius of the charged black holes
is achieved at a length scale where the temperature is vanishing; thus, one can
define a specific $Q=Q(r_\star)$ by choosing a specific extremal horizon radius $r_\star$,
\beqa
Q^2 \3i &=& \3i \frac{d ~ r_\star^{2d-2}}{d-2}  \bigg[ \frac{d-4}{2d} k \frac{\ell^2}{r_\star^2}\bigg( 1 \! + \! k \frac{\ell^2}{ 2r_\star^2 }(1 \!-\! \epsilon^2) \bigg)
\!+ \! \bigg( 1 \!+\! k \frac{\ell^2}{2r_\star^2 } \bigg)  \bigg]
\nn\\
&& \ii\ii\ii \left\{ \begin{aligned}
&\overset{\epsilon=0}{=} \frac{d}{d-2} r_\star^{2d-2}  \bigg(\frac{d-4}{2d} k \frac{\ell^2}{r_\star^2} + 1\bigg) \bigg(1 + k \frac{\ell^2}{2r_\star^2}  \bigg),  \\
&\overset{\epsilon=1}{=} \frac{d}{d-2} r_\star^{2(d-1)}\bigg( 1 + \frac{d-2}{d} k \frac{\ell^2}{r_\star^2} \bigg), \label{Eq:Qstar-HL(d+1)}
           \end{aligned} \right.
\eeqa
which means that assuming the charge parameter of the topological charged black hole is a conversed physical quantity, it can be defined by the horizon radius of its extremal black hole.

By substituting the charge back into the temperature in Eq.(\ref{Eq:TH-HL(d+1)}), it can be expressed with $r_\star$, %(or $u_\star$),
\begin{widetext}
\beqa
T \3i &=& \3i \frac{dr_0}{4\pi \ell^2} \bigg[ \frac{1 + k \frac{\ell^2}{2r_0^2}}{1 + k  \frac{\ell^2}{2r_0^2} (1-\epsilon^2) } + \frac{d-4}{2d}k \frac{\ell^2}{r_0^2}  - \frac{r_\star^{2(d-1)}}{r_0^{2(d-1)}}
\bigg(   \frac{ 1 + k \frac{\ell^2}{2r_\star^2 } }{1 + k  \frac{\ell^2}{2r_0^2} (1-\epsilon^2) } +  \frac{d-4}{2d} k \frac{\ell^2}{r_\star^2}   \frac{ 1 + k  \frac{\ell^2}{ 2r_\star^2 } (1-\epsilon^2) }{1 + k  \frac{\ell^2}{2r_0^2} (1-\epsilon^2) }   \bigg) \bigg]   \nn\\
&& \ii\ii\ii \left\{ \begin{aligned}
&\overset{\epsilon=0}{=} \frac{dr_0}{4\pi \ell^2}\bigg[ 1 + \frac{d-4}{2d} k \frac{\ell^2}{r_0^2} - \frac{ r_\star^{2(d-1)}}{r_0^{2(d-1)}}
\frac{ 1 + k \frac{\ell^2}{2r_\star^2}   }{1 + k \frac{\ell^2}{2r_0^2} }  \bigg( 1 + \frac{d-4}{2d} k \frac{\ell^2}{r_\star^2} \bigg)  \bigg],  \label{Eq:TH-HL-k!=0} \\
&\overset{\epsilon=1}{=} \frac{dr_0}{4\pi \ell^2}\bigg[ 1 +  \frac{d-2}{d} k \frac{\ell^2}{r_0^2} -  \frac{r_\star^{2(d-1)}}{r_0^{2(d-1)}} \bigg( 1 + k \frac{d-2}{d}\frac{\ell^2}{r_\star^2} \bigg)  \bigg].
           \end{aligned} \right.
\eeqa
\end{widetext}
It is easy to check that the temperature is vanishing at the horizon radius of its extremal black hole, i.e., $T_H(r_\star)=0$. While for any horizon radius $r_0 \ge r_\star $, one has $T_H(r_0)\ge 0$. Especially, for $r_0 \gtrsim r_\star$, it corresponds to a low but small finite temperature limit.

By substituting the new definition of the charge density of the topological black holes defined in Eq.(\ref{Eq:Qstar-HL(d+1)}) back into the redshift factor in Eq.(\ref{Eq:gr-Q-epsilon}) assuming $r_0 \gtrapprox r_\star$, namely, near the horizon of the extremal black holes with finite temperature, $g(r)$ with $d\ge 3$ can be re-expressed as
\begin{widetext}
\beqa
\ii
g(r) \3i &\overset{d\ge 3}{=}& \3i \frac{ 1 - \epsilon \sqrt{1 + \frac{1-\epsilon^2}{\epsilon^2} \frac{r_0^d}{r^d} \bigg[  1 + \frac{d}{d-2}\frac{r_\star^{2(d-1)}}{r_0^{2(d-1)}} h(r) + k \frac{\ell^2}{r_0^2}\bigg( 1 + \frac{r_\star^{2(d-2)}}{r_0^{2(d-2)}}h(r) \bigg) + k^2 \frac{(1-\epsilon^2)}{4} \frac{\ell^4}{r_0^4}\bigg( 1+ \frac{d-4}{d-2}\frac{r_\star^{2(d-3)}}{r_0^{2(d-3)}} h(r) \bigg) \bigg] } }{1-\epsilon} \quad \label{Eq:gr-At-star-HL(d+1)-T!=0-k!=0} \\
&& \ii\ii\ii\!\! \left\{ \begin{aligned}
&\overset{\epsilon=1}{=} 1 - \frac{r_0^d}{r^d} \bigg[ 1 + \frac{d}{d-2}\frac{r_\star^{2(d-1)}}{r_0^{2(d-1)}} h(r)  + k \frac{\ell^2}{r_0^2}\bigg( 1 + \frac{r_\star^{2(d-2)}}{r_0^{2(d-2)}}h(r) \bigg) \bigg],  \label{Eq:gr-At-star-AdS(d+1)-T!=0-k!=0}\\
&\overset{\epsilon=0}{=} 1 - \sqrt{\frac{r_0^d}{r^d} \bigg[  1 + \frac{d}{d-2}\frac{r_\star^{2(d-1)}}{r_0^{2(d-1)}} h(r)  + k \frac{\ell^2}{r_0^2}\bigg( 1 + \frac{r_\star^{2(d-2)}}{r_0^{2(d-2)}}h(r) \bigg) + k^2 \frac{1}{4} \frac{\ell^4}{r_0^4}\bigg( 1+ \frac{d-4}{d-2}\frac{r_\star^{2(d-3)}}{r_0^{2(d-3)}} h(r) \bigg) \bigg]},
          \end{aligned} \right.
\eeqa
\end{widetext}
where $h(r)\equiv 1 - (r_0/r)^{d-2}$. One can check that $f(r_0)=0$ through Eq.(\ref{Eq:fr-gr}).

By expanding the temperature defined in Eq.(\ref{Eq:TH-HL-k!=0}) around extremal horizon region $r\approx r_\star$, one obtains the finite temperature limit,
\beqa
T \overset{r_0 > r_\star}{=} \frac{r_0-r_\star}{2\pi\ell_{2,k}^2} + {\mathcal O}(r_0 - r_\star)^2 ,
\label{Eq:T!=0}
\eeqa
where $r_0 > r_\star$, and we have defined a new length scale,
\beqa
\ii\ii
\ell_{2,k}^{2} \equiv  \ell_2^2 \frac{1+ (1-\epsilon^2)\frac{k\ell^2}{2r_\star^2}} { 1+ \frac{(d-2)^2}{d(d-1)} \frac{k\ell^2}{r_\star^2} + (1-\epsilon^2) \frac{(d-3)(d-4)}{d(d-1)} \frac{k^2\ell^4}{4r_\star^4} }, \label{Eq:l2-l-k}
\eeqa
or, equivalently,
\beqa
\ell_{2,k} &\equiv & \ell_2 \sqrt\frac{1+ (1-\epsilon^2)\frac{k\ell^2}{2r_\star^2}} { 1+ \frac{(d-2)^2}{d(d-1)} \frac{k\ell^2}{r_\star^2} + (1-\epsilon^2) \frac{(d-3)(d-4)}{d(d-1)} \frac{k^2\ell^4}{4r_\star^4} } \nn\\
&& \ii\ii\ii \left\{ \begin{aligned}
 & \overset{\epsilon=0}{=}  \ell_2 \sqrt\frac{1 + \frac{k\ell^2}{2r_\star^2}} { 1+ \frac{(d-2)^2}{d(d-1)} \frac{k\ell^2}{r_\star^2} + \frac{(d-3)(d-4)}{d(d-1)} \frac{k^2\ell^4}{4r_\star^4} }, \\
 & \overset{\epsilon=1}{=}   \frac{\ell}{\sqrt{ d(d-1) + (d-2)^2 \frac{k\ell^2}{r_\star^2}}}, \label{Eq:l2-l-k-epsilon=1}
          \end{aligned} \right.
\eeqa
where
\beqa
\ell_2 \equiv \ell_{2,0} = \frac{\ell}{\sqrt{d(d-1)}}.  \label{Eq:l2}
\eeqa

One can express $r$ and $u$ coordinates in terms of new parameters $\eta_{k}$ and $\zeta_k$,
\beqa
 r- r_\star \! &=& \! \frac{\ell_{2,k}^2}{\eta_{k}}, \, r_0 - r_\star = \frac{\ell_{2,k}^2}{\eta_{0,k}},
 \nn\\
 u_\star - u \! &=& \! \frac{\ell_{2,k}^2}{\ell^2}\frac{u_\star^2}{\zeta_k}, \, u_\star - u_0 = \frac{\ell_{2,k}^2}{\ell^2}\frac{u_\star^2}{\zeta_{0,k}}, \label{Eq:r-u-zeta-eta-k!=0}
\eeqa
where the new parameters are defined through
\beqa
 \eta_k & \equiv & \frac{\ell_{2,k}^{2}}{r-r_\star},  \quad \zeta_k \equiv  \frac{\ell_{2,k}^2}{\ell^2}\frac{u_\star^2}{u_\star-u};
\label{Eq:zeta-eta-k!=0} \\
 \eta_{0,k} & \equiv & \frac{\ell_{2,k}^{2}}{r_0-r_\star},  \quad \zeta_{0,k} \equiv  \frac{\ell_{2,k}^2}{\ell^2}\frac{u_\star^2}{u_\star-u_0}.
\label{Eq:zeta0-eta0-k!=0}
\eeqa

As bulk geometry is approaching the extremal horizon $r_0 \to r_\star$, the temperature in Eq.(\ref{Eq:T!=0}) is approaching zero,
\beqa
T \overset{r_0\to r_\star}{=} \frac{r_0-r_\star}{2\pi \ell_{2,k}^2}.
\label{Eq:TH-(r0-rs)-(us-u0)}
\eeqa

Through Eq.(\ref{Eq:r-u-zeta-eta-k!=0}), the temperature in Eq.(\ref{Eq:T!=0}) can be re-expressed as
\beqa
T \overset{}{=} \frac{1}{2\pi\eta_{0,k}} , \label{Eq:TH-zeta0-eta0-k!=0}
\eeqa
and the gauge field in Eq.(\ref{Eq:At-fr-AdS(d+1)}) becomes
\beqa
A_t(r) & \overset{r_0 > r_\star}{=} & (d-2)\mu\frac{r-r_0}{r_0} + {\mathcal O} (r-r_0)^2 \label{Eq:At-T!=0-k!=0}\\
&¡¡\overset{}{\approx} & (d-2)\mu \frac{\ell_{2,k}^2}{r_0} \bigg(  \frac{1}{\eta_k} - \frac{1}{\eta_{0,k}} \bigg).
\nn %\label{Eq:At-T!=0-k!=0}
\eeqa

At near extremal horizon region,\footnote{Note, one needs to expand both $r$ and $r_0$ around the $r_\star$ separately. First, we expand $r$ around $r_0$; second, isolate the pole like $(r-r_0)$, and third, expand $r_0$ around $r_\star$ for the residues. } the redshift factor becomes\footnote{Note, special comments on $k=-1$ case are needed.
For the $k=-1$ case, one has a divergent radius $\ell_{2,k}$ according to Eq.(\ref{Eq:l2-l-k-epsilon=1}) at the $\epsilon=1$ case,
\beqa
r_\star^2 = -k \frac{(d-2)^2\ell^2}{d(d-1)} \ge 0, \quad \Rightarrow \quad r_\star \overset{k=-1}{=} (d-2)\ell_2.
\eeqa
By expanding the redshift factor $f(r)$ to the third order of $r$, and by substituting the condition above back, one obtains a redshift factor $f(r)$ with three roots,
\beqa
f(r)
&& \ii \left\{ \begin{aligned}
& \3i \overset{r > r_0}{=} \frac{2d(d-1)}{3\ell^2}\frac{r-r_0}{r_\star}[r^2 + (r_0 - 3 r_\star)r + r_0^2-3r_0 r_\star + 3 r_\star^2], \nn \\
& \3i \overset{r_0=r_\star}{=} \frac{2d(d-1)}{3\ell^2 }\frac{1}{r_\star}(r-r_\star)^3.  \quad (T=0) \nn
          \end{aligned} \right.
\eeqa
}
\beqa
f(r)  & \overset{r_0 > r_\star}{=} &  \frac{1}{\ell_{2,k}^2}(r-r_0)(r+r_0-2r_\star) \nn\\
&=&  \frac{\ell_{2,k}^2}{\eta_k^2} \bigg(1 - \frac{\eta_k^2}{\eta_{0,k}^2} \bigg).  \label{Eq:fr-k!=0-T!=0}
\eeqa
It is worthy of notice that the $\epsilon$ will not be present, which means that detailed balance parameter in the UV will not affect the near horizon behavior of the physics. The absence of $\epsilon\sim 0$ in the UV is important for HL gravity, since this leads HL gravity to flow to $z=3$ fixed point, which will be a renormalizable quantum gravity theory in the UV limit at short distance.

The near extremal horizon metric of topological charged black holes in HL$_{d+1}$ gravity becomes,
\beqa
\3i  \ii ds^2  \! &=& \! \frac{\ell_{2,k}^{2}}{\eta_k^2} \bigg[ \! - \! \bigg( 1 \! - \! \frac{\eta_k^2}{\eta_{0,k}^2} \bigg)  dt^2 \! + \! \bigg( 1 \! - \! \frac{\eta_k^2}{\eta_{0,k}^2} \bigg)^{-1} \ii d\eta_k^2 \bigg] \! \nn\\
&+& \!  r_\star^2  d\Omega_{d-1,k}^2, \label{Eq:ds2-k!=0-T!=0-AdS2}
\eeqa
and the corresponding gauge field at the finite temperature limit ($T \ne 0$) is
\beqa
\3i  \ii A_t(\eta_k) \! = \! \frac{e_{d,k}}{\eta_k} \bigg( 1 \! - \! \frac{\eta_{k}}{\eta_{0,k}} \bigg), \label{Eq:At-k!=0-T!=0-AdS2}
\eeqa
where we have introduced a dimensionless effective IR gauge coupling $e_{d,k}$ as defined below,
\beqa
e_{d,k} \equiv (d-2)\mu \frac{\ell_{2,k}^2}{r_0}. \label{Eq:edk}
\eeqa

\subsubsection{Zero temperature case}

For the topological charged black hole with zero temperature at its extremal horizon, at $r_0=r_\star$, $T=0$ as has been shown in Eq.(\ref{Eq:TH-HL-k!=0}).

By using Eq.(\ref{Eq:Qstar-HL(d+1)}), the characterize length $\ell_F$ in Eq.(\ref{Eq:Q-mu-AdS(d+1)}) can be re-expressed as,
\beqa
\ell_F = \frac{\sqrt{d(d-1)}}{\sqrt{2}(d-2)} r_\star  \sqrt{ 1+ \frac{d-2}{d} k \frac{\ell^2}{r_\star^2} + \frac{1-\epsilon^2}{4}\frac{d-4}{d} k^2 \frac{\ell^4}{r_\star^4}  }, \nn
\eeqa
from which the extremal horizon radius can be expressed as
\beqa
\ii
r_\star = \ell_F \sqrt{\frac{2(d-2)^2}{d(d-1)} - \bigg(\frac{d-2}{d}k + \frac{d-4}{d}k^2 \bigg)g_F^2 }.
\label{Eq:rstar_gF}
\eeqa
The factor $g(r)$ in Eqs.(\ref{Eq:gr-At-star-HL(d+1)-T!=0-k!=0}) and (\ref{Eq:gr-At-star-AdS(d+1)-T!=0-k!=0}) becomes
\begin{widetext}
\beqa
g(r) &\overset{r=r_\star}{=}& \frac{ 1 - \epsilon \sqrt{1 + \frac{1-\epsilon^2}{\epsilon^2}\frac{r_\star^d}{r^d} \bigg[  1 + \frac{d}{d-2} h(r)  + k \frac{\ell^2}{r_\star^2}\bigg( 1 + h(r) \bigg) +  \frac{(1-\epsilon^2)}{4} k^2 \frac{\ell^4}{r_\star^4}\bigg( 1+ \frac{d-4}{d-2} h(r) \bigg) \bigg] } }{1-\epsilon}, \label{Eq:gr-At-star-HL(d+1)-T=0-k!=0} \\
&& \ii\ii\ii\3i \left\{ \begin{aligned}
&\overset{\epsilon=1}{=} 1 - \frac{r_\star^d}{r^d} \bigg[ 1 + \frac{d}{d-2}  h(r)  + k \frac{\ell^2}{r_\star^2}\bigg( 1 + h(r) \bigg) \bigg],  \label{Eq:gr-At-star-AdS(d+1)-T=0-k!=0} \\
&\overset{\epsilon=0}{=} 1 - \sqrt{\frac{r_\star^d}{r^d} \bigg[  1 + \frac{d}{d-2} h(r)  + k \frac{\ell^2}{r_\star^2} \bigg( 1 + h(r) \bigg) +  \frac{1}{4} k^2 \frac{\ell^4}{r_\star^4}\bigg( 1+ \frac{d-4}{d-2} h(r) \bigg) \bigg]},
           \end{aligned} \right.
\eeqa
\end{widetext}
with $ h(r)\equiv 1 - ({r_\star}/{r})^{d-2}$. The gauge field in Eq.(\ref{Eq:At-fr-AdS(d+1)}) becomes
\beqa
A_t(r) = \mu \bigg( 1 - \frac{r_\star^{d-2}}{r^{d-2}} \bigg). \label{Eq:At-fr-AdS(d+1)-T=0}
\eeqa

At near extremal horizon radius, the gauge field in Eq.(\ref{Eq:At-T!=0-k!=0}) becomes
\beqa
A_t(r) & \overset{r_0 = r_\star}{=} & (d-2)\mu\frac{r-r_\star}{r_\star} + {\mathcal O} (r-r_\star)^2 \label{Eq:At-T=0-k!=0}\\
&¡¡\overset{}{\approx} &
(d-2)\mu \frac{\ell_{2,k}^2}{r_\star}  \frac{1}{\eta_k},
\nn
\eeqa
where in the last equality, we have used the new parameters defined in Eq.(\ref{Eq:r-u-zeta-eta-k!=0}).

The redshift factor in Eq.(\ref{Eq:fr-gr}) with $g(r)$ in Eq.(\ref{Eq:gr-At-star-HL(d+1)-T=0-k!=0}) becomes
\beqa
f(r)   \overset{r_0 = r_\star}{=}   \frac{1}{\ell_{2,k}^2}(r-r_\star)^2 =  \frac{\ell_{2,k}^2}{\eta_k^2} .  \label{Eq:fr-k!=0-T=0}
\eeqa
Thus, the near extremal horizon metric of topological charged black holes in HL$_{d+1}$ gravity and corresponding gauge field at zero temperature limit ($T = 0$) are
\beqa
 &&  ds^2   =  \frac{\ell_{2,k}^{2}}{\eta_k^2} (  -    dt^2  +    d\eta_k^2 )  +   r_\star^2  d\Omega_{d-1,k}^2, \nn\\
 &&  A_t(\eta_k)  =  \frac{e_{d,k}}{\eta_k} ,
\eeqa
where $e_{d,k}$ is defined as in Eq.(\ref{Eq:edk}) except that $r_0=r_\star$ here,
\beqa
e_{d,k} \equiv (d-2)\mu \frac{\ell_{2,k}^2}{r_\star}. \label{Eq:edk-r_star}
\eeqa
Therefore, at very close to the extremal horizon, the geometry of topological charged black holes becomes AdS$_2\times \Omega_{d-1,k}$ with the curvature radius of AdS$_2$ given by $\ell_{2,k}$ defined in Eq.(\ref{Eq:l2-l-k-epsilon=1}), which applies to the region $(r-r_\star)\ll r $ [or $(u_\star-u)\ll u_\star$]. As $r\to r_\star$, $\eta_k\to \infty$, the time direction shrinks to zero, and the spatial direction approaches a constant; the Maxwell field approaches zero.

It implies that the boundary theory of the bulk gravity at low energy in the IR limit at large distance flows to a fixed point with AdS$_2\times \Omega_{d-1,k}$ symmetry. The AdS$_2$ symmetry appears in the near extremal horizon region. The AdS$_2$ is isomorphic to a full $SL(2,R)$ group, which owns the scaling isometry:
\beqa
(t, \eta_k) \to \lambda \,  (t, \eta_k) , \quad x_{d-1,k} \to x_{d-1,k}.
\eeqa
The long time limit just corresponds to the low frequency limit, since $\omega$ is conjugate to $t$, while the residue $(d-1)$-dimensional hypersurface $ \Omega_{d-1,k} $ is scaling irrelevant.

For example, in $(3+1)$-dimensional momentum space-time, we have
\beqa
(\omega, r-r_\star )  \to \frac{1}{\lambda}(\omega ,r-r_\star ), \quad \kappa \to  \kappa,
\eeqa
where the momentum in a Ricci flat brane($k=0$) is $\kappa = |\vec{k}|$, with the corresponding spatial coordinate boundary geometry given by ${\mathbb R}^{2}$, i.e., Eq.(\ref{Eq:ds2_d=3-k=0}), while for those in a sphere($k=1$) or hyperbolic($k=- 1$) hypersurface are $\kappa_l=l,-(l+1)$, where $l \in {\mathbb Z}$, with the corresponding spatial coordinate boundary geometry given by Eqs.(\ref{Eq:ds2_d=3-k=1}) and (\ref{Eq:ds2_d=3-k=-1}), respectively.

Therefore, in the low frequency limit, the $d$-dimensional boundary theory with finite charge density should be described by an IR CFT$_1$, which is a conformal symmetry only in the time direction(of course, one has take a notice that the spatial direction still has important physical consequence along the transverse sector), i.e., a $(0+1)$-dimensional conformal quantum mechanics, including the scale invariance along the time direction. This IR CFT $_1$ is a new conformal symmetry due to collective behavior of a large number of degrees of charged excitation; thus, it is distinguished from the UV CFT$_d$ in the asymptotic infinite boundary, which is broken by the finite charge density.

\subsection{Topological charged black branes and effective IR gauge couplings}

For the Ricci flat case with $k=0$, the mass parameter in Eq.(\ref{Eq:2M-HL-HL(d+1)}) becomes
\beqa
2M  \overset{k=0}{=}  r_0^d + \frac{Q^2}{r_0^{d-2}},
\label{Eq:2M-HL-AdS(d+1)}
\eeqa
and Eq.(\ref{Eq:gr-Q-epsilon}) becomes
\beqa
g(r) &\overset{d\ge 3}{\equiv} &\frac{ 1 - \epsilon \sqrt{1 + \frac{1-\epsilon^2}{\epsilon^2}\Big[  \frac{r_0^d}{r^d}\Big( 1 + \frac{Q^2}{r_0^{2d-2}} \Big) - \frac{ Q^2 }{r^{2d-2}}  \Big] } }{1-\epsilon}  \nn \\
&& \ii\ii\ii\3i \left\{ \begin{aligned}
&\overset{\epsilon=1}{=}
1 - \frac{r_0^d}{r^d} - \frac{Q^2}{r^d}\bigg( \frac{1}{r_0^{d-2}} - \frac{1}{r^{d-2}}\bigg),   \label{Eq:gr-Q}  \\
&\overset{\epsilon=0}{=}
1 - \sqrt{ \frac{r_0^d}{r^{d}} + \frac{Q^2}{r^{d}}\bigg( \frac{1}{r_0^{d-2}} - \frac{1}{r^{d-2}} \bigg)  },
           \end{aligned} \right.
\eeqa
from which it is obvious that $r_0$ is the horizon radius determined by $g(r_0)=0$, since $f(r_0) = \a^{-2} (r_0/\ell)^2 g(r_0) =0 $.

In this case, the temperature in Eq.(\ref{Eq:TH-HL(d+1)}) becomes
\beqa
T \overset{k=0}{=}  \frac{d r_0}{4\pi \ell^2}\bigg( 1 - \frac{d-2}{d}\frac{Q^2}{r_0^{2d-2}} \bigg), \label{Eq:TH-AdS(d+1)}
\eeqa
where the charge of the black hole, according to Eq.(\ref{Eq:Qstar-HL(d+1)}),
can be measured by the its extremal horizon radius $r_\star$, e.g., for the $k=0$ case,
\beqa
\ii\ii
Q  =  \sqrt\frac{d}{d-2} r_\star^{d-1}, \label{Eq:Qstar-AdS(d+1)}
\eeqa
where $[Q]= [L^{d-1}]$ is consistent with Eq.(\ref{Eq:Q-q0}). %which has dimension of $[L]^{d-1}$.
Alternatively, this is equivalently that the extremal radius of horizon is defined by the charge of black brane, e.g., for $k=0$ case,
\beqa
\ii\ii
r_\star^{2(d-1)} \equiv \frac{d-2}{d}Q^2, \, \Rightarrow \, u_\star^{2(d-1)} \overset{k=0}{\equiv} \frac{d\ell^{4(d-1)}}{d-2}Q^{-2},
\label{Eq:r_star-Q:u_star-Q}
\eeqa
where we have introduced the conformal coordinate $u_\star\equiv {\ell^2}/{r_\star}$. From Eq.(\ref{Eq:r_star-Q:u_star-Q}), we have that
\beqa
r_\star = r_0^{\text{min}} \le r_0, \quad \Leftrightarrow \quad
 u_\star = u_0^{\text{max}} \ge u_0,
\eeqa
where $u_0$ is the horizon radius and $u_\star$ is the extremal horizon radius. Thus the singularity at $r_\star$ is covered by the horizon radius $r_0$. In this case, the temperature in Eq.(\ref{Eq:TH-HL-k!=0}) becomes\footnote{For briefness, we will just neglect the expression for results on $f(u)$,$A_t(u)$, etc. in the conformal coordinate $u$, which can be obtained by making the changes such as $r_{0}/r \to u/u_{0}$ and $r_{\star}/r \to u/u_{\star}$.}
\beqa
T
&=& \left\{ \begin{aligned}
& \frac{d r_0}{4\pi \ell^2}\bigg( 1 - \frac{r_\star^{2(d-1)}}{r_0^{2(d-1)}} \bigg), \quad r_0 \ge r_\star; \\
& \frac{d}{4\pi u_0}\bigg(1 -   \frac{u_0^{2(d-1)}}{u_\star^{2(d-1)}} \bigg), \quad u_0 \le u_\star.  \label{Eq:TH-HL-k=0}
           \end{aligned} \right.
\eeqa
By using Eqs.(\ref{Eq:2M-HL-AdS(d+1)}) and (\ref{Eq:Qstar-AdS(d+1)}) the mass of the black holes becomes
\beqa
\frac{2(d-1)}{d-2} r_\star^d \le 2M = r_0^d\bigg( 1 + \frac{d}{d-2}  \frac{r_\star^{2(d-1)}}{r_0^{2(d-1)}} \bigg)\le \frac{2(d-1)}{d-2} r_0^d. \nn
\eeqa
The lower bound of the mass just corresponds to the zero temperature case when $r_0=r_\star$.

For the Ricci flat case with $k=0$, by using Eq.(\ref{Eq:Qstar-AdS(d+1)}), the charged black brane in Eq.(\ref{Eq:gr-Q}) can be re-expressed as
\beqa
g(r)  \3i &\overset{d\ge 3}{\equiv} & \3i \frac{\bigg[ 1 - \epsilon \sqrt{1 + \frac{1-\epsilon^2}{\epsilon^2}\bigg[  \frac{r_0^d}{r^{d}} + \frac{d}{d-2}\frac{r_\star^{d}}{r^{d}}\bigg( \frac{r_\star^{d-2}}{r_0^{d-2}} - \frac{r_\star^{d-2}}{r^{d-2}} \bigg)  \bigg] } \bigg]}{1-\epsilon}   \nn \\
&& \ii\ii\ii\!\! \left\{ \begin{aligned}
&\overset{\epsilon=1}{=}  1 - \frac{r_0^d}{r^d} - \frac{d}{d-2}\frac{r_\star^{d}}{r^d}\bigg( \frac{r_\star^{d-2}}{r_0^{d-2}} - \frac{r_\star^{d-2}}{r^{d-2}}\bigg),  \label{Eq:gr-At-star-HL(d+1)-T!=0} \\
&\overset{\epsilon=0}{=}  1 - \sqrt{ \frac{r_0^d}{r^{d}} + \frac{d}{d-2}\frac{r_\star^{d}}{r^{d}}\bigg( \frac{r_\star^{d-2}}{r_0^{d-2}} - \frac{r_\star^{d-2}}{r^{d-2}} \bigg)  }. \label{Eq:gr-At-star-AdS(d+1)-T!=0}
           \end{aligned} \right.
\eeqa
It also worthy of noticing that $g(r_0)=0$ is still true.

At the finite temperature case, the finite new coordinates defined in Eqs.(\ref{Eq:zeta-eta-k!=0}) and (\ref{Eq:zeta0-eta0-k!=0}) become
\beqa
 \eta & \equiv & \frac{\ell_{2}^{2}}{r-r_\star},  \quad \zeta \equiv  \frac{\ell_{2}^2}{\ell^2}\frac{u_\star^2}{u_\star-u};
\label{Eq:zeta-eta-k=0} \\
 \eta_{0} & \equiv & \frac{\ell_{2}^{2}}{r_0-r_\star},  \quad \zeta_{0} \equiv  \frac{\ell_{2}^2}{\ell^2}\frac{u_\star^2}{u_\star-u_0},
\label{Eq:zeta0-eta0-k=0}
\eeqa
where $\ell_2$ is defined in Eq.(\ref{Eq:l2}). The metric near the horizon in Eq.(\ref{Eq:ds2-k!=0-T!=0-AdS2}) becomes a black brane in AdS$_2\times{\mathbb R}^{d-1}$,
\beqa
\ii
ds^2 \! = \! \frac{\ell_2^2}{\eta^2} \!\bigg[ \! - \! \bigg(1 \!-\! \frac{\eta^2}{\eta_0^2} \bigg)dt^2 \! + \! \bigg(1 \!-\! \frac{\eta^2}{\eta_0^2} \bigg)^{-1} \ii d\eta^2 \! \bigg] \! + \! \frac{r_\star^2}{\ell^2} dx_{d-1}^2,    ~~
\label{Eq:ds2-T!=0-AdS2}
\eeqa
where $d\Omega_{d-1,0}^2 \equiv dx_{d-1}^2/\ell^2  $ is used. The gauge field near the horizon in Eq.(\ref{Eq:At-k!=0-T!=0-AdS2}) is expressed as
\beqa
A_t(\eta) = \frac{e_{d,0}}{\eta} \bigg( 1 - \frac{\eta}{\eta_0} \bigg), \label{Eq:At-T!=0-AdS2}
\eeqa
where $e_{d,0}$ is given by Eq.(\ref{Eq:edk}),
\beqa
e_{d,0} = (d-2)\mu \frac{\ell_{2}^2}{r_0}.  \label{Eq:ed0}
\eeqa
The temperature(with respect to $t$) near the horizon in Eq.(\ref{Eq:TH-zeta0-eta0-k!=0}) becomes
\beqa
T = \frac{1}{2\pi \eta_0}.  \label{Eq:TH-zeta0-eta0-k=0}
\eeqa
It implies that at finite charge density, the bulk geometry near horizon boundary becomes AdS$_2\times {\mathbb R}^{d-1}$. The scale invariance of the AdS$_2$ implies that at low energies the corresponding dynamics on the boundary will be controlled by a $(0+1)$-dimensional CFT.

Let us consider the zero temperature case with $T=0$ at the extremal horizon of charged black holes, where $r_0 = r_\star$(or in conformal coordinates $u_0 = u_\star$). In this case $g(r)$ in Eq.(\ref{Eq:gr-At-star-HL(d+1)-T!=0}) can be re-expressed as
\beqa
g(r) &\overset{d\ge 3}{\equiv} &\frac{ 1 - \epsilon  \sqrt{1 + \frac{1-\epsilon^2}{\epsilon^2}\bigg(  \frac{2(d-1)}{d-2}\frac{r_\star^d}{r^{d}} - \frac{d}{d-2} \frac{r_\star^{2d-2}}{r^{2d-2}}  \bigg) } }{1-\epsilon},  \nn \\
&& \ii\ii\ii \left\{ \begin{aligned}
&\overset{\epsilon=1}{=}   1 - \frac{2(d-1)}{d-2}\frac{r_\star^d}{r^d} + \frac{d}{d-2} \frac{r_\star^{2d-2}}{r^{2d-2}},  \label{Eq:gr-At-star-AdS(d+1)-T=0} \\
&\overset{\epsilon=0}{=}  1 - \sqrt{ \frac{2(d-1)}{d-2}\frac{r_\star^d}{r^{d}} - \frac{d}{d-2} \frac{r_\star^{2d-2}}{r^{2d-2}}   }. \label{Eq:gr-At-star-HL(d+1)-T=0} %\\
           \end{aligned} \right.
\eeqa
In the near extremal horizon limit $r_0 \to r_\star$(or $u_0 \to u_\star$, $\zeta_{0} \to \infty$), Eq.(\ref{Eq:zeta0-eta0-k=0}) becomes
\beqa
 \eta_{0} \to \infty,
\eeqa
which stands for the zero temperature limit, since by definition the temperature in Eq.(\ref{Eq:TH-zeta0-eta0-k=0}) becomes
\beqa
T  \overset{\eta_0\to \infty}{=}0.
\label{Eq:TH-zeta0-eta0}
\eeqa
In the zero temperature limit, the metric in Eq.(\ref{Eq:ds2-T!=0-AdS2}) reduces to the $T=0$ case,
\beqa
ds^2 = \frac{\ell_2^2}{\eta^2}\big( - dt^2 + d\eta^2 \big) + \frac{r_\star^2}{\ell^2}dx^2, \label{Eq:ds2-T=0-AdS2}
\eeqa
with the corresponding gauge field in Eq.(\ref{Eq:At-T!=0-AdS2}) becoming
\beqa
A_t(\eta) = \frac{e_d}{\eta},  \label{Eq:At-T=0-AdS2}
%A_t(\zeta) = \frac{e_d}{\zeta}.
\eeqa
where $e_{d}$ is given by Eq.(\ref{Eq:edk}) at extremal horizon radius, or the extremal limit case of Eq.(\ref{Eq:ed0})
\beqa
e_d = \lim_{r_0 \to r_\star} e_{d,0} = (d-2)\mu \frac{\ell_{2}^2}{r_\star}. \label{Eq:ed-r_star}
\eeqa

At extremal horizon of black holes with $r_0=r_\star$, Eq.(\ref{Eq:rstar_gF}) becomes
\beqa
\ii\ii
r_\star  = \frac{\sqrt{2}(d-2)}{\sqrt{d(d-1)}}\ell_F, \quad
u_\star \equiv \frac{\ell^2}{r_\star} =  \frac{\sqrt{d(d-1)}}{\sqrt{2}(d-2)} g_F \ell. \label{Eq:u-star-HL(d+1)} \label{Eq:u-star-AdS(d+1)}
\eeqa
The $u_\star$ corresponds to a single scale, which is proportional to the effective dimensionless gauge coupling $g_F$. The scale is increasing with the relative strength of the electromagnetic forces with respect to that of the gravitational force.

By using Eq.(\ref{Eq:u-star-HL(d+1)}), the effective dimensionless IR gauge coupling defined in Eq.(\ref{Eq:edk}) becomes,
\beqa
e_d = \mu \ell_2 \frac{g_F}{\sqrt{2}} .  \label{Eq:ed-gF}
\eeqa
which implies that the chemical potential $\mu$ can be defined in terms of effective dimensionless IR gauge couplings $e_d$ and effective dimensionless UV gauge coupling $g_F$ at Ricci flat case, i.e., $\mu \equiv \sqrt{2} e_d/(g_F\ell_2) $. Therefore, $\mu$ is defined inverse proportional to the UV gauge couplings $g_{F}$. Consequently, the chemical potential is measured by the relative strength of the gravitational forces and electromagnetic forces at UV, which is obvious by observing Eq.(\ref{Eq:gF-effective}).

\section{Thermodynamics of Topological Charged Black Holes in Generalized Ho\v{r}ava-Lifshitz Gravity}
\label{sec:thermodynamics}

In the section, we would like to explore the thermodynamics of the topological charged black hole solutions in generalized HL$_{d+1}$ gravity, by using the Hamiltonian approach as those used in dimensional continued gravity in ~\cite{Banados:1993ur},\cite{Cai:1998vy}. The thermodynamics of topological neutral black holes in HL gravity with generic $\lambda$ has been explored by using the canonical Hamiltonian formulation~\cite{Cai:2009qs}.

\subsection{Euclidean action and partition function}
The partition function for a thermodynamical ensemble is identified with the Euclidean path integral in the saddle point approximation around Euclidean continuation of the classical solution\footnote{After Wick-rotating to the imaginary time $t \to \tau = it$, one has the consequence relations as below
\beqa
\dot{g}_{ij}\to i \,\dot{g}_{ij}, \quad N_j \to i \,N_j, \quad K_{ij}\to i\, K_{ij}, \quad K\to -K.  \label{Eq:Wick-Rotation} \nn
\eeqa
Then the action in Eq.(\ref{Eq:S}) becomes $S(dt, dx) \to i S_E(d\tau, dx)$, from which one obtains the Euclidean action of HL gravity, then the partition function becomes $Z \sim \exp{(-S_E)}  \label{Eq:Z_E}$. Therefore, in the imaginary time, the Euclidean action $S_E$ has to be real.}. Consider the Euclidean continuation of the action of topological black holes for general $\lambda$ in Hamiltonian from $S \to -S_E$,
\beqa
S_E = \int dt d^3x \big(\pi^{ij}\dot{g}_{ij}-N {\mathcal H} - N^i { \mathcal H}_i \big) + S_{EB},  \label{Eq:S_HL_Euclidean}
\eeqa
where $S_{EB}$ is a boundary term. $N$ and $N_i$ are lapse function and shift variable respectively. For the metric ansatz in Eq.(\ref{Eq:ds2-topological-HL}), we have lapse function $N^2 = \tilde{N}(r)^2 f(r)$. On one hand, the shift variable $N_i=0$, on the other hand, we do not need to give the explicit form of the momentum ${\mathcal H}_i$ and conjugate $\pi^{ij}$ of $\dot{g}_{ij}$ since we are considering static black holes case with $\pi^{ij}=0$. Then the Euclidean action is reduced to be
\beqa
S_E &=& - \int dt d^3x N {\mathcal H} + S_{EB} \nn\\
&=& - \int dt \int_{r_+}^\infty dr \tilde{N}(r) {\mathcal H}(r) + S_{EB} \nn\\
&=& - \beta \int_{r_+}^\infty dr \tilde{N}(r) {\mathcal H}(r) + S_{EB},
\eeqa
where for $d=3$ case with $\epsilon=0$, according to Eq.(\ref{Eq:L_epsilon_lambda}), or Eq.(\ref{Eq:L01_lambda}), ${\mathcal H}(r)$ is given by
\beqa
{\mathcal H}(r) &=& \frac{\Omega_{2,k}c^3}{16\pi G_N}\frac{1}{\Lambda_W}\bigg( \frac{2\lambda-1}{r^2}F(r)^{2} - \frac{2\lambda}{r}F(r)F^\prime(r) \nn\\
&+& \frac{\lambda-1}{2}F^{\prime }(r)^2 \bigg).
\eeqa
$\beta=\int_{0}^{T} dt$ is the period of Euclidean time and $r_+$ is the radius of the black hole horizon defined by the largest root of $f(r_+)=0$. The Euclidean black holes are static and satisfy the constraint ${\mathcal H}(r)=0$; thus, the Euclidean action is just the boundary term $S_{EB}$. In the canonical ensemble, the temperature should be kept fixed under the variation of the action. By doing variation of the Euclidean action, one finds that the variation of the boundary term is given by the total derivative term after the variation\footnote{Note:
$\delta F^\prime = - \overleftarrow\partial_r \delta F$, we have also used the identity that ${\mathcal H}(r)=0$, in addition,
The coordinate $r$ is invariant under the variation.},
%{ \cb
\beqa
\delta S_{EB} \3i &=& \3i \delta S_{EB}|^\infty_{r_+} \nn\\
\3i &=& \3i \beta \delta[\int_{r_+}^{\infty}dr \tilde{N}(r){\mathcal H}(r) ]= \beta \int_{r_+}^{\infty}dr \tilde{N}(r)\delta{\mathcal H}(r),  \nn\\
\3i &=& \3i \beta \frac{\Omega_{2,k}c^3}{16\pi G_N}\frac{1}{\Lambda_W} \int_{r_+}^{\infty} dr \tilde{N}(r)\bigg( 2\frac{2\lambda-1}{r^2}F\delta F \nn\\
\3i &-& \3i \frac{2\lambda}{r}F^\prime \delta F - \frac{2\lambda}{r}F \delta F^\prime+(\lambda-1)F^\prime \delta F^\prime  \bigg), \nn\\
\3i &=& \3i \beta \frac{\Omega_{2,k}c^3}{16\pi G_N}\frac{1}{\Lambda_W} \int_{r_+}^{\infty} dr \bigg[ \bigg( \frac{2\lambda}{r}F - (\lambda-1)F^\prime \bigg)\tilde{N}^\prime(r) \nn\\
\3i &+& \3i (\lambda - 1)\bigg( \frac{2}{r^2}F - F^{\prime\prime} \bigg)\tilde{N}(r) \bigg]\delta F + \beta \frac{\Omega_{2,k}c^3}{16\pi G_N}\frac{1}{\Lambda_W}  \nn\\
\3i &\times & \3i \bigg( -\frac{2\lambda}{r}\tilde{N}(r)F\delta F + (\lambda-1)\tilde{N}(r)F^\prime \delta F \bigg)_{r_+}^{\infty}.
\eeqa
%}
The equations in the first integral of the last identity is just one of the equations of motion related $F(R)$ and $\tilde{N}(R)$, thus is zero before the integral. Therefore, we obtain the variation of the boundary term
%{ \cb
\beqa
\delta S_{EB} &=&
\beta \frac{\Omega_{2,k}c^3}{16\pi G_N}\frac{1}{\Lambda_W} \bigg( -\frac{2\lambda}{r}\tilde{N}(r)F\delta F \nn\\
&+& (\lambda-1)\tilde{N}(r)F^\prime \delta F \bigg)_{r_+}^{\infty}.
\eeqa

To obtain equations of motion, we need not to know the explicit form of $\delta F$ or $\delta \tilde{N}$.
But to calculate the boundary term, we will need the explicit forms of the solutions,
which have been obtained in
% Eq.(\ref{Eq:F-Nt-HL-z})
Eq.(\ref{Eq:Fr}) and Eq.(\ref{Eq:Nr}). The thermodynamics of topological neutral black hole solutions in HL gravity with generic $\lambda$ have been explored in Ref.~\cite{Cai:2009qs}.

\subsection{Thermodynamics of the topological charged black holes}

In the following, we will mainly focus on the thermodynamics of $(3+1)$-dimensional topological charged black holes in generalized HL gravity with $\lambda=1$. More generically, let us consider the case with $\epsilon^2\ne 0$ case. According to Eq.(\ref{Eq:L_epsilon_lambda}) and Eq.(\ref{Eq:kappa-G_N-alpha}), the action of the full Lagrangian of the HL gravity at $d=3$ with $\lambda=1$ can be written as
\beqa
&& S_\epsilon = \int dt L_\epsilon = \frac{\Omega_{2,k}c^3}{16\pi G_N}\int dt \int dr \tilde{N}(r)  U_3(r,1) + S_B, \nn \\
&& U_3(r,1) = \frac{1}{\Lambda_W}\bigg(\frac{\epsilon^2[F(r)+\Lambda_W r^2]^2 - F(r)^2}{r}\bigg)^\prime. \label{Eq:L_epsilon_lambda=1}
\eeqa
where $\alpha^{-1} = {c^3}/{(16\pi G_N)}$ and $S_B$ is a surface term, which must be chosen so that the action has an extremum under variations of the fields with appropriate boundary conditions. One demands that the fields approach the classical solutions at infinity. Varying the action, one finds,
\beqa
\ii
\delta S_B \! = \! - (t_2 \!-\! t_1)N_0\delta M, ~ S_B \!=\! -(t_2 \!-\! t_1)N_0 M + S_0,
\eeqa
where the boundary term $S_B$ is the conserved charged associated action to the improper gauge transformations produced by time evolution. Here $M$ and $N_0$ are a conjugate pair, thus, when one varies $M$, $N_0$ must be fixed. Therefore, the boundary term should be in the form of an integral as above, where $S_0$ is an arbitrary constant fixed by some physical degree of freedom~\cite{Banados:1993ur}, i.e., the black holes mass vanishes when black holes horizon goes to zero. For the above action in HL gravity, one identify
\beqa
\tilde{N}(r) = N_0 = 1, \quad U_3(r,1) = \partial_r c_0, \label{Eq:U3r1-dr-c0}
\eeqa
where $N_0$ could be normalized to be one due to the arbitrariness of $\tilde{N}(r)$, i.e., it can be absorbed by rescaling time coordinate. $c_0$ is an integration constant, which can be expressed in terms of black holes horizon radius $r_+$ via $f(r_+)=0$.

For topological charged black holes, the corresponding redshift factors are
\beqa
f(r) \3i &=& \3i k \!-\! \frac{\Lambda_W}{1\!-\!\epsilon^2}r^2 \!-\! \frac{\sqrt{\!-\!\Lambda_W(1-\epsilon^2)\big(c_0 r \!-\! \frac{q_0^2}{2}\big) \!+\! \epsilon^2 \Lambda_W^2 r^4}}{1\!-\!\epsilon^2}  \nn\\
&& \ii\ii\ii \left\{ \begin{aligned}
&\overset{\epsilon=0}{=} k - \Lambda_W r^2 - \sqrt{-\Lambda_W\big(c_0 r - \frac{q_0^2}{2}\big)}, \\
&\overset{\epsilon=1}{=} k - \frac{\Lambda_W}{2}r^2 - \frac{c_0}{2 r} + \frac{q_0^2}{4 r^2},    %\quad c_0
          \end{aligned} \right.
\eeqa
from which the constant $c_0$ can be expressed in terms of black holes horizon radius $r_+$ via $f(r_+)=0$,
\beqa
c_0  &=& \frac{\epsilon^2 k^2 - (k - r_+^2 \Lambda_W)^2 }{r_+ \Lambda_W} + \frac{q_0^2}{2r_+} \nn \\
&& \ii\ii\ii \left\{ \begin{aligned}
&\overset{\epsilon=0}{=} - \frac{(k - r_+^2 \Lambda_W)^2 }{r_+ \Lambda_W} + \frac{q_0^2}{2r_+},  \label{Eq:C0} \\
&\overset{\epsilon=1}{=} 2k r_+ - \Lambda_W^2 r_+^3  + \frac{q_0^2}{2r_+}.
          \end{aligned} \right.
\eeqa
When $\epsilon=1$, the situation reduces to be the case of topological AdS$_4$ Schwazschild black holes.

\subsubsection{Mass}
For topological charged black holes in generalized HL$_{d+1}$ gravity, according to the Hamiltonian approach similar to Eq.(\ref{Eq:U3r1-dr-c0}), one obtain $U_{d}(r,1)=\partial_r c_0$, where $U_d(r,\lambda)$ is given in Eq.(\ref{Eq:L_epsilon_lambda}) and $F(r)$ with $\lambda=1$ given by Eq.(\ref{Eq:Fr-d-lambda=1}). We get the mass of the black holes in $(d+1)$ dimensions
\begin{widetext}
\beqa
M &=& \frac{\Omega_{d-1,k}c^3 }{16\pi G_N} \frac{1}{(d-2)\Lambda_W} \int dr \partial_r c_0 = \frac{\Omega_{d-1,k}c^3 }{16\pi G_N} \frac{1}{(d-2)\Lambda_W} c_0 \nn\\
&=& \frac{\Omega_{d-1,k}c^3 }{16\pi G_N} \frac{1}{(d-2)\Lambda_W} \frac{r_+^{d-4}(d-2)\big[ [(d-1)(d-2)k-2\Lambda_W r_+^2]^2 -(d-1)^2(d-2)^2k^2\epsilon^2 \big] +2 r_+^{2-d} q_0^2}{(d-1)^2(d-2)^3 }. \label{Eq:mass-HL(d+1)-lambda=1}
\eeqa
\end{widetext}
In $(3+1)$ dimensions, one has
\beqa
M &\overset{d=3}{=}& \frac{\Omega_{2,k}c^3 }{16\pi G_N} \frac{1}{\Lambda_W} \frac{(k-\Lambda_W r_+^2)^2 - \epsilon^2 k^2 + q_0^2/2}{r_+}  \nn \\
&& \ii\ii\ii\!\! \left\{ \begin{aligned}
&\overset{\epsilon=1}{=} \frac{\Omega_{2,k}c^3 }{16\pi G_N} \frac{1}{\Lambda_W} \frac{ -\Lambda_W r_+^2(2k-\Lambda_W r_+^2)  + q_0^2/2}{r_+},  \\
&\overset{\epsilon=0}{=} \frac{\Omega_{2,k}c^3 }{16\pi G_N} \frac{1}{\Lambda_W} \frac{(k-\Lambda_W r_+^2)^2 + q_0^2/2}{r_+}, \label{Eq:mass-HL-z=1}
          \end{aligned} \right.
\eeqa
where the black hole mass is always positive. The topological black hole obtain its minimal masses at
\beqa
r_{\text{m}} \3i &=& \3i \left\{ \begin{aligned}
&\sqrt{\frac{\sqrt{2} \sqrt{3 q_0^2+k^2 \left(8-6 \epsilon ^2\right)}-2k}{-6\Lambda_W}}, \quad k\ge 0, \nn\\
&\sqrt{\frac{-2 k-\sqrt{2} \sqrt{3 q_0^2+k^2 \left(8-6 \epsilon ^2\right)}}{-6\Lambda_W}}. \quad k < 0. \nn
          \end{aligned} \right.
\eeqa
For $\epsilon=0$, neutral black holes ($q_0=0$), one can obtain its minimal masses \footnote{Note: $M^{\prime\prime}=\kappa^2(k^2+3r^4\Lambda_W^2)\Omega_k/(8r^3\kappa_W^4)>0$.}
\beqa
M_{\text{min}} \3i  &=& \3i \left\{ \begin{aligned}
&\frac{\Omega_{2,k}c^3 }{16\pi G_N} \frac{1}{\Lambda_W} \frac{16k^2}{9r_+}, \quad r_+ = \sqrt{-\frac{k}{3\Lambda_W}}; \quad k >0, \\
& 0, \quad r_+ = \sqrt{\frac{k}{\Lambda_W}}. \quad k \le 0.
          \end{aligned} \right.
\eeqa
Considering that $\Lambda_W<0$, the first kind of minimum is achieved for the case of $k=1$ or $k=0$, while the second kind of minimum is achieved for the case of $k=-1$,
\beqa
 M_{\text{min}1}^{k=1} &=& \frac{\Omega_{2,1}c^3}{9\pi G_N \Lambda_W r_+}, \quad c_0 = \frac{16}{9}, \quad r_+ = \frac{\ell}{\sqrt{6}}; \nn\\
 M_{\text{min}1}^{k=0} &=& 0, \quad c_0 = 0, \quad r_+ = 0;  \\
 M_{\text{min}2}^{k=-1} &=& 0, \quad c_0 = 0,\quad r_+ = \frac{\ell}{\sqrt{2}}. \nn
\eeqa
The minimum means that for $k=1$, the AdS Schwartzschild black hole has an minimum mass; for $k=0$, the masses of the black branes and their horizons are both vanishing, thus, there are no black branes; for $k=-1$, the mass of the black holes are zero but the horizon are nonvanishing, Therefore, there exist massless black holes in AdS hyperbolic space, with minimal horizons at $r_+ = 1/\sqrt{-\Lambda_W}$ for the topological neutral black holes HL gravity with $\lambda=1$.

For $\epsilon=1$, the topological black hole solutions to HL gravity reduce to be the IR relevant ones with Lagrangian $L_0$ in Eq.(\ref{Eq:L0-HL}) with redshift factor as shown above and reduce to be the Einstein's GR solutions in Eq.(\ref{Eq:ds2-topological-Einstein}) by setting $q_0=0$. The positive of the black holes masses require that
\beqa
r_+ \ge \sqrt{\frac{k}{\Lambda_W}\bigg( 1 \pm \sqrt{1 - \frac{q_0^2}{2k^2}} \bigg)}.
\eeqa
For $k=1$ or $k=0$, this is always true, since $\Lambda_W<0$; for $k=-1$ case, this implies that
\beqa
 k=0,1: && r_+ \ge 0;  \\
 k= -1: && r_+ \ge \sqrt{\frac{1 + \sqrt{1-q_0^2/2}}{-\Lambda_W}} \overset{q_0 = 0}{=} \sqrt{-\frac{2}{\Lambda_W}}. \nn
\eeqa

According to the equations above, the mass of the topological black hole in Einstein general relativity becomes
\beqa
M \overset{\epsilon=1}{=} \frac{\Omega_{2,k}c^3 }{16\pi G_N} \bigg( r_+ (r_+^2 \Lambda_W - 2 k) + \frac{q_0^2}{2\Lambda_W r_+} \bigg)  \label{Eq:mass-EH}
\eeqa
from which one can obtain the minimal mass for neutral black hole\footnote{Note: $M^{\prime\prime}={3r_+ \kappa^2\Lambda_W^2 \Omega_k}/{(8\kappa_W^4)}>0$, note also that $r_+=0$ is not a local minimal point since $M^{\prime\prime}=0$.}
\beqa
M =  \frac{\Omega_{2,k}c^3 }{16\pi G_N}\bigg( - \frac{4}{3}k \bigg)  r_+ , \quad  r_+ = \sqrt{\frac{2k}{3\Lambda_W}}.
\eeqa
For $k=1$ case, the minimal mass becomes\footnote{Note: For $k=1$ case, $M$ is an monotonically increasing functions, since $M^{\prime}={\kappa^2 \Omega_k}(-\Lambda_W )(2 - 3\Lambda_W r_+^2)/({16\kappa_W^4})>0$.}
\beqa
 M_{\text{min}}^{k=1} \3i &\overset{q_0\ne 0}{=}& \3i\frac{\Omega_{2,k}c^3 }{16\pi G_N} \frac{2}{3\Lambda_W}\frac{1}{r_+} \bigg[ \sqrt\frac{2}{3}\bigg( \sqrt{\frac{2}{3}+q_0^2} - \sqrt\frac{2}{3} \bigg) + q_0^2 \bigg] , \nn\\
 r_+ \3i &=& \3i \sqrt{-\frac{\sqrt{1+3q_0^2/2}-1}{3\Lambda_W}}; \nn\\
 M_{\text{min}}^{k=1} \3i &\overset{q_0=0}{=}& \3i 0, \quad r_+ = 0;
\eeqa
which violates the positive mass theorem, but the square of the mass are still positive $(M_{\text{min}}^{k=1})^2 = -\frac{4}{9\Lambda_W}>0$. For $k=0$ case, the minimal mass becomes
\beqa
 M_{\text{min}}^{k=0} &\overset{q_0\ne 0}{=}& \frac{\Omega_{2,k}c^3 }{16\pi G_N} \frac{2}{3 \Lambda_W} \frac{q_0^2}{r_+}, \quad r_+ =\frac{1}{6^{1/4}}\sqrt{-\frac{q_0}{\Lambda_W}}; \nn\\
 M_{\text{min}}^{k=0} &\overset{q_0 = 0}{=}& 0, \quad r_+ =0;
\eeqa
which means this is no black hole. For $k=-1$ case, the minimal mass becomes
\beqa
&& M_{\text{min}}^{k=-1} \ge \text{max}\bigg[0,\frac{\Omega_{2,k}c^3 }{16\pi G_N}\frac{2}{3\Lambda_W}\frac{1}{r_+}\bigg( q_0^2 + \frac{2}{3} \times \nn\\
&& \big( \sqrt{1+\frac{3}{2}q_0^2}-1 \big) \bigg) \bigg] =0 , \quad r_+ \ge \sqrt{-\frac{\sqrt{1+3q_0^2/2}+1}{3\Lambda_W}}; \nn\\
&& M_{\text{min}}^{k=-1} \overset{q_0=0}{=} 0, \quad r_+ \ge \sqrt{-\frac{2}{3\Lambda_W}};
\eeqa
which corresponds to the AdS$_4$ black hole with minimum mass and nonvanishing horizon.

\subsubsection{Temperature}

For generic solution of topological neutral black holes in HL gravity with arbitrary $\lambda$ in Eq.(\ref{Eq:ds2-topological-HL}) with Eq.(\ref{Eq:ds2-HL-generic-lambda}), one can obtain the temperature by imposing periodic boundary conditions to diminish the conical singularity at horizon of the Euclidean black hole solutions, e.g, one can set the time period $\beta$ as below \footnote{Alternatively, one can directly obtain the Hawking temperature by calculating surface gravity of the metric, \beqa
T = \frac{\kappa_0}{2\pi} = \lim_{r\to r_+}\frac{1}{4\pi}\frac{g_{tt}^\prime(r)}{\sqrt{g_{rr}(r)g_{tt}(r)}}.
\eeqa
}
\beqa
\beta(\tilde{N}(r)f^\prime(r))|_{r=r_+}=4\pi, \label{Eq:T-periodicBCs-conical-singularity}
\eeqa
which gives the temperature of the black holes
\beqa
T &\equiv &\frac{1}{\beta} = \frac{\tilde{N}(r)f^\prime(r)}{4\pi}|_{r_+} \nn\\
&=& \frac{C_F}{4\pi r_+^{2\lambda_\pm}}[(-\Lambda_W)r_+^2 (2-\lambda_\pm)-k\lambda_\pm ].
\eeqa
In the limit $\lambda\to 1$, by using the L'Hospital rule, we have $\lambda_+\to \pm \infty$ and $\lambda_-\to 1/2$, the temperature of the negative branches are
\beqa
T= \frac{C_F}{8\pi r_+}[3(-\Lambda_W)r^2 -k],
\eeqa
which just reduce to the topological neutral black hole solutions in HL gravity with $\lambda=1$ when $C_F=1$.

Similarly, the Hawking temperature of the topological charged black holes in $(d+1)$-dimensional HL gravity with $\lambda=1$ can be obtained by directly calculating the surface gravity $\kappa_0$ at the horizon, according to the metric in Eq.(\ref{Eq:ds2-topological-HL}) with $\tilde{N}(r)=1$ and redshift factor $f(r)$ given in Eq.(\ref{Eq:ds2-HL-topological-charged-black-brane-L-epsilon-1}),
\begin{widetext}
\beqa
\ii
T = \frac{[(d-1)(d-2)k-2\Lambda_W r_+^2][(d-1)(d-2)(d-4)k-2d \Lambda_W r_+^2 ]-(d-1)^2(d-2)^2(d-4)k^2\epsilon^2-2r_+^{6-2d}q_0^2}{8(d-1)(d-2)\pi r_+[ (d-1)(d-2)k - 2\Lambda_W r_+^2 -(d-1)(d-2) k\epsilon^2 ]}.  \label{Eq:TH-HL(d+1)-lambda=1}
\eeqa
\end{widetext}
For $(3+1)$ dimensions, one has
\beqa
T &=& \frac{3 \Lambda_W^2 r_+^4  + 2k(-\Lambda_W) r_+^2 -(1-\epsilon^2)k^2-q_0^2/2}{8\pi r_+[  ( -\Lambda_W) r_+^2  + (1-\epsilon^2)k ]} \nn \\
&& \ii\ii\ii \left\{ \begin{aligned}
&\overset{\epsilon=1}{=} \frac{2k-3 \Lambda_W r_+^2 +q_0^2/(2\Lambda_W r_+^2)}{8\pi r_+}, \label{Eq:temperature-HL-lambda=1} \\
&\overset{\epsilon=0}{=} \frac{-k - 3 \Lambda_W r_+^2 + q_0^2/(2\Lambda_W r_+^2 -2k)}{8\pi r_+}, \nn \ 			 						 \end{aligned} \right.
\eeqa
where $r_+$ is the largest root of the equation $f(r)=0$. Note that the cosmological constant $\Lambda_W$ is negative to make the speed of light real. Since the appearance of the electric charge, the extremal black holes with vanishing temperature always exist within reasonable parameter regime.
\beqa
x_+^2 \3i &=& \3i \frac{\sqrt{2(4-3\epsilon^2)k^2+ 3 q_0^2}}{3\sqrt{2}} \! - \! \frac{k}{3} \! \overset{q_0=0}{=} \! \frac{1}{3}(\sqrt{(4-3\epsilon^2)}|k|-k) \nn \\
&& \ii\ii\ii \left\{ \begin{aligned}
&\overset{\epsilon=1}{=}  \frac{\sqrt{2k^2+3q_0^2}}{3\sqrt{2}}-\frac{k}{3} \overset{q_0=0}{=}\frac{1}{3}(|k|-k),  \\
&\overset{\epsilon=0}{=} \frac{\sqrt{8 k^2+ 3 q_0^2}}{3\sqrt{2}} - \frac{k}{3}\overset{q_0=0}{=}\frac{1}{3}(2|k|-k),
						\end{aligned} \right.
\eeqa
where $r_+^2 = - x_+^2/ \Lambda_W = x_+^2 \ell^2/2 $. For neutral black holes, the extremal solution $T=0$ exists if and only if $k=0$ or $k=+1$, with $r_+=0$ or $r_+=1/\sqrt{-3\Lambda_W}$ respectively. In the extremal shown above, the Hawking temperature vanishes, and it corresponds to an extremal black hole.

\subsubsection{Entropy}

In the following, we will obtain the black entropy by using the first law of black hole thermodynamics with assumption that the first law always keeps valid:
\beqa
dM=TdS.
\eeqa
By integrating the relation and by using the mass in Eq.(\ref{Eq:mass-HL(d+1)-lambda=1}) and the temperature in Eq.(\ref{Eq:TH-HL(d+1)-lambda=1}) we have for $(d+1)$ dimensions case,
\beqa
S &=& \int T^{-1}\frac{dM}{dr_+} dr_+ + S_0 \nn\\
&& \ii\ii\ii \left\{ \begin{aligned}
&\overset{d\ge 4}{=}
\frac{\Omega_{d-1,k}c^3}{G_N(d-1)^2(d-2)^2}r_+^{d-1}\bigg( 1 - k (1-\epsilon^2) \times\\
& \quad\quad \frac{(d-2)(d-1)^2}{2(d-3)}\frac{1}{\Lambda_W r_+^2} \bigg) + S_0, \label{Eq:entropy-HL(d+1)}\\
&\overset{d=4}{=} \frac{\Omega_{3,k}c^3}{36 G_N} r_+^{3}\bigg( 1 - k(1-\epsilon^2)\frac{9}{\Lambda_W r_+^2} \bigg) + S_0,
					\end{aligned} \right.
\eeqa
where we have used Eq.(\ref{Eq:c/G_N}).
it is worthy of notice that for $d\ge 4$ case, the entropy has no logarithmic terms. The curvature term is always proportional to $r_+^{-2}$. While in $(3+1)$ dimensions, we have
\beqa
S \3i &=& \3i \int T^{-1}\frac{dM}{dr_+} dr_+ + S_0 \nn\\
\3i & \overset{d=3}{=} & \3i
\frac{\Omega_{2,k}c^3}{4G_N}r_+^2\bigg( 1 - k  \frac{1-\epsilon^2}{\Lambda_W r_+^2} \ln{r_+^2} \bigg) + S_0. \quad \label{Eq:entropy-HL(3+1)}
\eeqa
where $S_0$ is an integration constant, which should be fixed by physical degree of freedom. In conclusion, the canonical Hamiltonian formulation allows us to define the entropy that satisfies the first law of thermodynamics. In general the integration constant $S_0$ is calculated by invoking the quantum theory of the gravity, thus, it is not fixed at the moment, as a result, one cannot determine whether the black holes are thermodynamic stable or not.

For the Ricci flat black branes with $k=0$, the logarithmic term is absent thus the entropy is proportional to the horizon area. In this case, one can set $S_0=0$, and assuming that the black hole entropy vanishes when horizon goes to zero. While for $k\ne 0$ case, $S_0$ in nonvanishing. For example, for $d=3$ case, by using Eq.(\ref{Eq:c/G_N}), the entropy can be expressed as
\beqa
S = \frac{\Omega_{2,k} c^3}{4G_N}r_+^2\bigg( 1 - k \frac{(1-\epsilon^2)}{\Lambda_W r_+^2}\log(r_+^2)  \bigg) + S_0, \label{Eq:entropy-HL(3+1)}
\eeqa
where we have used that $\Lambda_W = -2/\ell^2$, and $S_0$ is chosen to be
\beqa
S_0 \equiv -\big( c^3 \frac{A_{2,k}^0}{4G_N} \big) \big( k\frac{1-\epsilon^2}{2} \big) \log\ell^2.
\label{Eq:S0}
\eeqa
It is worth of noticing that the absolute value of entropy is nonvanishing even at $T=0$, i.e., independent of temperature.

According to Eq.(\ref{Eq:entropy-HL(d+1)}) and Eq.(\ref{Eq:entropy-HL(3+1)}), the entropy becomes
\beqa
S \3i &\overset{d\ge 4}{=}& \3i \frac{ A_{d-1,k} c^3}{G_N(d-1)^2(d-2)^2} \bigg( 1 + k (1-\epsilon^2) \frac{(d-1)}{2(d-3)}\frac{A_{2,k}^{0}}{A_{2,k}} \bigg), \nn  \\
\3i &\overset{d=3}{=}& \3i \frac{A_{2,k} c^3}{4G_N} \bigg( 1 + k(1-\epsilon^2)\frac{A_{2,k}^{0}}{ 2 A_{2,k}} \ln\frac{A_{2,k}}{A_{2,k}^{0}} \bigg),
\eeqa
where $A_{d-1,k}\equiv \Omega_{d-1,k} r_+^{d-1}$ are the boundary surface area of the black holes, and $A_{2,k}^{0}\equiv \Omega_{2,k}\ell^{2}$. In the nature unit $c=1,G_N=1$, the leading term is just one quarter of horizon area, i.e., $A_{2,k}/4$, which originates from the contribution of $L_0$ by setting $\epsilon=1$. The second term is a logarithmic term, where $A_{2,k}^0=\Omega_{2,k}\ell^2$ is a constant of dimension of length squared which are introduced to fix the integration constant $S_0$.

Basically speaking, $S_0$ should be fixed by counting micro degrees of freedom in some QFT of gravity in the UV. The logarithmic term of entropy often appears in the quantum correction of black hole entropy. For the topological black holes in the HL gravity, the logarithmic term disappears for the black branes with Ricci flat horizon, which can be verified by setting $k=0$,
\beqa
S_{EH} = \frac{c^3A}{4G_N}.  \label{Eq:entropy-EH}
\eeqa
In this case, the area formula of black hole entropy in Einstein gravity is recovered in topological black holes with flat $2$-dimensional spatial hypersurface. While for Ricci curved black holes, the logarithmic term will be present, which can be verified by setting $\epsilon=0$.
\beqa
S = \frac{c^3 A}{4G_N}\bigg(1 + \frac{k}{2}\frac{A_0}{A}\log\frac{A}{A_0} \bigg),
\eeqa
where the logarithmic term comes from the contribution of $L_1$, which originates from the detailed balance condition of HL gravity. While if the condition is completely violated, i.e, $\epsilon=1$, then the entropy reduces to the well-know area formula again, since the effect of higher derivative terms disappears.

Moreover, it is worthy of notice that the charge $q_0(r_+)$ does not appear explicitly in the expression of charged topological black holes entropy in terms of horizon radius $r_+$, in other words, the entropy is not of a function of $q_0$ explicitly. This is due to the fact that black holes entropy is a function of horizon geometry.

\subsubsection{Heat capacity}
To discuss the local stability of the black holes, one need to calculate the heat capacity of the black holes. The black holes are locally thermodynamically stable if the heat capacity is {always positive}. While they are locally thermodynamically unstable if the heat capacity is always negative.

The capacity of a black hole is defined as
\beqa
C= \frac{dM}{dT} = \frac{dM}{dr_+}/\frac{dT}{dr_+}.
\eeqa
From which, by using Eq.(\ref{Eq:mass-HL(d+1)-lambda=1}) and Eq.(\ref{Eq:TH-HL(d+1)-lambda=1}), we obtain the heat capacity of the topological charged/neutral black in HL gravity with $\lambda=1$\footnote{Note: In the following, we have kept the expression of $\kappa^2/\kappa_W^4$, in order to observe the sign of the heat capacity and Free energy. In $(3+1)$ dimensions, by using Eq.(\ref{Eq:kappa2-kappaW4}), one has
\beqa
\frac{\kappa^2}{\kappa_W^4} = \frac{c^3}{\pi G_N }\frac{1}{-\Lambda_W}>0.
\eeqa
},
\beqa
C &\overset{\epsilon=0}{=}& -\frac{\pi \kappa ^2 {\Omega_k}}{ 2\kappa_W^4}\frac{ \left(k+x_+^2\right)^2 \left(2 k^2+q_0^2-4 k x_+^2-6 x_+^4\right) }{\left(k+3 x_+^2\right) \left(2 k^2+q_0^2+4 k x_+^2+2 x_+^4\right) } \nn \\
&\overset{q_0=0}{=}& \frac{\pi \kappa^2 \Omega_k}{2\kappa_W^4}\frac{3x_+^2 - k}{3x_+^2 + k}(x_+^2 + k),\nn\\
C &\overset{\epsilon=1}{=}&  \frac{\pi \kappa^2 \Omega_k}{2\kappa_W^4}\frac{  -q_0^2+4 k x_+^2+6 x_+^4 }{3 q_0^2-4 k x_+^2+6 x_+^4} x_+^2 \nn\\
&\overset{q_0=0}{=}& \frac{\pi \kappa^2 \Omega_k}{2\kappa_W^4}\frac{  3 x_+^2 + 2k }{ 3 x_+^2 -2k } x_+^2,  \label{Eq:HeatCapacity-HL-z=1}
\eeqa
where we have used the notation that $x_+ \equiv \sqrt{-\Lambda_W}r_+$.

\subsubsection{Free energy}

By studying the Euclidean action or the free energy, $S_E=\beta F = \beta M- S$, one can obtain the information on the global stability of the black hole thermodynamics. The free energy of the black holes $F$ is given by
\beqa
F = M - TS,
\eeqa
which can be used to determine whether there exists the Hawking-Page transition~\cite{Hawking:1982dh} associated with the black holes in HL gravity with $\lambda=1$ in $(3+1)$ dimensions. The global stability of black hole is determined by the signature of free energy $F$, if $F<0$, then the black hole is globally thermodynamically stable.
By using Eq.(\ref{Eq:mass-HL-z=1}), Eq.(\ref{Eq:temperature-HL-lambda=1}) and Eq.(\ref{Eq:entropy-HL(3+1)}), we have
\beqa
F \3i &=& \3i \frac{\kappa^2 \Omega_k}{32\kappa_W^4 r_+} \bigg( 2[(k-\Lambda_W r_+^2)^2 -\epsilon^2 k^2] \nn\\
\3i &+& \3i \frac{(k-\Lambda_W r_+^2)(k+3\Lambda_W r_+^2)[ 2(1-\epsilon^2)k\log{r_+}  - \Lambda_W r_+^2 ]}{(1-\epsilon^2)k-\Lambda_W r_+^2} \nn\\
\3i &+& \3i \frac{2(1-\epsilon^2)k(2+\log{r_+})-3\Lambda_W r_+^2}{2[(1-\epsilon^2)k-\Lambda_W r_+^2]}q_0^2 \bigg) - T S_0  \label{Eq:F-HL-z=1} \\
&& \ii\ii\ii \left\{ \begin{aligned}
&\overset{\epsilon=1}{=} \frac{\kappa^2 \Omega_k}{32\kappa_W^4 r_+}\bigg( 2k^2 \!-\! 5k \Lambda_W r_+^2 \!-\! \Lambda_W^2 r_+^4 \!+\! 2k(k + 3\Lambda_W r^2)\\
 & \qquad \times \log{r_+} + \frac{2k-3\Lambda_W r_+^2+k\log{r_+}}{2(k-\Lambda_W r_+^2)}q_0^2 \bigg) - T S_0  \nn \\
&\overset{\epsilon=0}{=}  \frac{\kappa^2 \Omega_k}{32\kappa_W^4 r_+}\bigg( -\Lambda_W r_+^2(2k + \Lambda_W r_+^2) + \frac{3}{2}q_0^2 \bigg) - T S_0,  \\
					\end{aligned} \right. \nn
\eeqa
Because $S_0$ is undetermined, we cannot determine the signature of the free energy. However if the term $S_0$ is vanishing, then the free energy is negative for large horizon distance($r_+\gg 1$), implying that large black branes in HL gravity are globally thermodynamically stable. For the case $\epsilon=0, q_0=0$, the result in HL just corresponds to the free energy of black holes in the Einstein's general relativity. If one choose the $S_0$ to be that defined in Eq.(\ref{Eq:S0}), i.e., in this choice, the absolute value of entropy will be zero for $\epsilon=1$ case as classical general relativity at large distance. Then the corresponding free energy of the black holes becomes
\beqa
F && \ii \left\{ \begin{aligned}
&\overset{\epsilon=0}{=}  \frac{\kappa^2 \Omega_k}{32\kappa_W^4 r_+}\bigg[ 2k^2 + 10 k \frac{r_+^2}{\ell^2} + 4 \frac{r_+^4}{\ell^4} + \frac{k}{2}\bigg(k -6 \frac{r_+^2}{\ell^2}\bigg)\\
& \quad \times \log{\frac{r_+^2}{\ell^2}} + \frac{2k + 6 (r_+/\ell)^2+k\log{({r_+}/{\ell})}}{2[k +2 (r_+/\ell)^2]}q_0^2 \bigg],    \\
&\overset{\epsilon=1}{=} \frac{\kappa^2 \Omega_k}{32\kappa_W^4 r_+}\bigg[ 4 \frac{r_+^2}{\ell^2}\bigg(k - \frac{r_+^2}{\ell^2}\bigg) + \frac{3}{2}q_0^2 \bigg].  %\\
					\end{aligned} \right. \nn
\eeqa

The phase transition of the topological charged black holes in Hor\v{a}va-Lifshitz gravity have been studied in Refs.~\cite{Majhi:2012fz,Mo:2013sxa}.

\section{Conclusions}
\label{sec:concl}

The traditional quantization of classical gravity theory suffers the non-renormalizability problems in the ultraviolet (UV) high energy scale due to that the dimensionfull garavitational coupling carries a negative mass dimensions $[G_N]=-2$. As one of candidate UV complete gravity theories, Ho\v{r}ava-Lifshitz(HL) gravity solves the problem from the aspect of scaling dimension of gravity couplings for both kinetic and slef-interactions terms. The theory treats time as a special dimension among the whole space-time, which carries a critical scaling dimension index $z\ne 1$,  indicating the deviation from the scaling dimension of pure space dimension, namely, $1$. Consequently, the theory is nonrelativistic but renormalizable at ultra-high energy with fixed point $z \ge d$ in the $(d+1)$-dimensional space-time, which can flow through geometric Ricci flow equations with parameter $\lambda$, to Einstein's general relativity at large distance with critical exponent $z=1$.

In this paper, we have investigated the topological black holes in the most general HL$_{d+1}$ gravity with critical exponent $z=d$ in ultra-high energy, which indicates anisotropy between time and space at short distance, while still assuming that the pure spatial space are still homogeneous with the topology characterized by topological index, e.g., $k= +1, -1, 0$ indicating the sphere, hyperbolic and plane hypersurface. We concentrated on the topological charged black holes in $z=d$ HL$_{d+1}$ gravity with Ricci flow parameter $\lambda=1$, which just reduces to be the Dirac-De Witt's canonical gravity. Through Hamiltonian approach, we have obtained the topological charged black holes in $z=d$ HL$_{d+1}$, $z=5$ HL$_6$, $z=4$ HL$_5$, $z=3,4$ HL$_4$, $z=2$ HL$_3$ gravity with parameter $\lambda=1$. The solutions reduce to be the RN AdS black holes from classical nonrenormalizable Einstein-Maxwell action in the IR limit at large distance with the detailed balance violation parameter $\epsilon=1$(or $\a=1$ as defined in Eq.(\ref{Eq:fr-gr})). In particular, the metric of the corresponding topological charged black holes in HL$_4$ with critical exponent $z=4$, and those in HL$_3$ with critical exponent $z=2$ for $\lambda=1$ are also solved. With these topological charged black holes, we have studied their thermodynamic statistical quantities, including temperature, entropy, heat capacity and free energy.

In addition, we have studied the asymptotic behavior of the topological charged black holes in HL gravity on infinite boundary, which gives the pure AdS geometry as expected and thus the corresponding conformal symmetry on the boundary. The near horizon behaviors of the topological black holes in HL gravity are also studied, especially at and near extremal horizon regions, which just correspond to the zero temperature and finite temperature limit. It has been found that in both zero and finite temperature limit, the topological charged black holes in HL gravity with $d\ge 3$, just reduce to be, e.g., the AdS$_2\times{\mathbb R}^{d-1} $ or AdS$_2\times {\mathbb S}^{d-1}$ for Ricci flat case with $k=0$ or curved case with topological indexes $k=1$, respectively. As the physical consequences, scale invariance
is present on the boundary of HL gravity in the scaling limit for both zero temperature and finite temperature case. For
the topological charged black holes in HL$_3$ with $d=2$, the redshift factor of the metric tensor and $U(1)$ gauge field are given in Eq.(\ref{Eq:ds2-RN-HL(2+1)}), which are consistent with the basic results of $(2+1)$-dimensional electromagnetic field theory. This case is more special, due to the logarithmic divergence at singularity $r=0$ point. In fact, HL$_3$ gravity is special by itself, since the spectral dimension of space-time in ultra-high energy at short distances is effectively $d_s=2$~\cite{Horava:2009if}, which has been studied via causal dynamical triangulations approach~\cite{Anderson:2011bj}.

\vspace{0.2in}  \centerline{\bf{Acknowledgements}} \vspace{0.2in}

We thank valuable discussions with Rong-Gen Cai, Yi Cao, Chao-Jun Feng, Xin Gao, Bin Hu, Li Li, Xiao-Dong Li, Yan Liu, Yuan Liu, Shi Pi, Yi Wang, Jia-Jun Xu, Run-Qiu Yang, Ying-Li Zhang, Yang Zhang, and Zhen-Hui Zhang. Particularly, Y.~-H.~Qi thanks Xiao-Jun Bi, Peng-Fei Yin and Qiang Yuan for hospitality during his short visiting at Key Laboratory of Particle Astrophysics, Institute of High Energy Physics, Chinese Academy of Sciences. Y.~-L.~Zhang thanks University of Southampton for hospitality. This work is supported in part by the National Natural Science Foundation of China(No.10821504, 10905084, 10975170, 11035008, 11075194, 11135003, 11275246 and 11375247), and in part by the Ministry of Science and Technology of China under Grant No. 2010CB833000.

\appendix

\section{Theory Structure of Ho\v{r}ava-Lifshitz Gravity}
\label{app:HL}

\subsection{Lagrangian formulation}

In the ADM metric defined in Eq.(\ref{Eq:ds2-ADM}), the generating function (or partition function in quantum statistics) of the quantum gravity can be expressed in the path integral formalism,
\beqa
\ii
Z=\int {\mathcal D} g_{ij}\, {\mathcal D} N_i\, {\mathcal D} N \exp{\big(iS\big)},  %\label{Eq:Z}
\quad S= S_K - S_V.   \label{Eq:S}
\eeqa
where $S$ is the action of Ho\v{r}ava-Lifshitz gravity, which consists of the kinetic sector $S_K$ and the potential sector $S_V$.

\subsubsection{Kinetic sector}
The kinetic sector of the action is\footnote{For the briefness, we will use $\sqrt{g}$ to stands for $\sqrt{\det{g_{ij}}}$ in the following hence and forth.}
\beqa
S_K \3i &=& \3i \int dt d^dx \sqrt{-\hg} {\mathcal L}_K, \\
{\mathcal L}_K \3i &=& \3i \frac{1}{2\kappa^2} \dot{g}_{ij}G^{ijkl}\dot{g}_{kl} = \frac{1}{2\kappa^2}[\dot{g}_{ij}\dot{g}^{ij}-\lambda (\dot{g}_{ij}g^{ij})^2], \qquad
\eeqa
where $G^{ijkl}$ is a generalization of Wheeler-De Witt metric on spatial foliation with spatial diffeomorphism invariance,
\beqa
G^{ijkl}=\frac{1}{2}(g^{ik}g^{jl}+g^{il}g^{jk})-\lambda\, g^{ij}g^{kl}. \label{Eq:Wheeler-DeWitt}
\eeqa
$\lambda$ is a free parameter, representing a dynamical dimensionless coupling constant, and it is susceptible to quantum corrections. In the relativistic theory, the full space-time diffeomorphism invariance fixes the value of $\lambda=1$ to restore Einstein's general relativity at large distance\footnote{Note:There are still difference between HL gravity and Einstein gravity at short distance, especially at near horizon region.}. Consequently, $\lambda$ is a coupling constant, which characterizes the deviation from general relativity(GR) in the IR. The ordinary derivative can be generalized to be general covariant derivatives and the time derivative of the metric is replaced by $\dot{g}_{ij}\to (\dot{g}_{ij}-\nabla_i N_j -\nabla_j N_i)/N=2 K_{ij}$, which transforms covariantly under foliation-preserving diffeomorphisms. Together with the general covariant volume elements $\sqrt{-\hg}=\sqrt{\det{g_{ij}}}N =\sqrt{|g_{ij}|}N  \equiv \sqrt{g}N $, the kinetic sector of the action becomes
\beqa
&& S_K = \int dt d^dx \sqrt{g} N {\mathcal L}_K, \label{Eq:S_K}\\
&& {\mathcal L}_K =  \frac{2}{\kappa^2} K_{ij}G^{ijkl}K_{kl} = \frac{2}{\kappa^2}(K_{ij}K^{ij}-\lambda K^2), \nn
\eeqa
where $K\equiv g^{ij}K_{ij}$ is the trace of $K_{ij}$, $K^{ij}=g^{ik}g^{jl}K_{kl}$, $\kappa$ and $\lambda$ are two couplings constants. Considering the scaling relations in Eq.(\ref{Eq:scaling_gij_Ni_N}) that $[\sqrt{g}N]=0$, the dimension of the volume $[dt d^dx]=-z-d$, and $[K^{ij}]=[K]=z$, one obtains the scaling dimension of the couplings constants at the fixed point,
\beqa
[\kappa]=\frac{z-d}{2}, \quad [\lambda]=0, \label{Eq:scaling-kappa-lambda}
\eeqa
since $[\kappa^2]=[dtd^dx K^2]=z-d$ and $[\lambda K^2]=[K_{ij}K^{ij}]=2z$. Note that the couplings $\kappa$ will be dimensionless in $(d+1)$ space-time if $z=d$, while the coupling $\lambda$ is always dimensionless, reflecting the fact that $K_{ij}K^{ij}$ and $K^2$ are separately spatial invariant under foliation differmorphism invariant.

In the UV energy scale, the renormalizability of HL gravity is completed determined by the dimensions of the couplings $\kappa$ as shown in Eq.(\ref{Eq:scaling-kappa-lambda}), if
\beqa
[\kappa]=\frac{z-d}{2}\ge 0. \label{Eq:kappa}
\eeqa
The theory will be power-counting renormalizable when $z=d$ and super-renormalizable when $z>d$.

\subsubsection{Potential sector}

The potential sector of the action is
\beqa
S_V = \int dt d^dx \sqrt{g} N {\mathcal V}[g],
\eeqa
where ${\mathcal V}[g]$ is the potential of the Lagrangian density, which is a function of metric $g_{ij}$ and is independent of its time derivative(so that it respects time-independent spatial diffeomorphisms). The volume elements $\sqrt{-\hg}=\sqrt{\det{g_{ij}}} N$ is incorporated to make the action general covariant.

One will first focus on the highest dimension term, which will be dominant at the short distance and determine the high energy behavior of the theory. Then one should consider all possible relevant terms of lower dimension in potential, which are induced in the infrared via RG flow from the UV fixed point. Considering the scaling dimensions of the coordinates given in Eq.(\ref{Eq:scaling_x_t}) and since that $R_{ij}\sim \partial_{i}\Gamma_{jk}^k-\partial_i\partial_j g$, one obtains the scaling dimension of the Riemann tensor $R_{ij}$ and scalar $R$, which are always defined in terms of the metric $g_{ij}$ on the $d$-dimensional leaves of the space-time foliation,
\beqa
[R]=[R_{ij}]=[R_{i\, kl}^{\,\,j}]=2.
\eeqa

The potential sector of the theory can be great simplified by imposing the detailed balance condition, which requires that the potential action to be of a special form
\beqa
S_V &=& \int dt d^dx \sqrt{g} N {\mathcal V}[g],  \label{Eq:SV} \\
{\mathcal V}[g] &=&  \frac{\kappa^2}{8} E^{ij}{\mathcal G}_{ijkl} E^{kl} = \frac{\kappa^2}{8} (E^{ij}E_{ij}-\tilde\lambda E^2), \nn
\eeqa
where ${\mathcal G}_{ijkl}$ denotes the inverse of the generalized Wheeler-De Witt metric in Eq.(\ref{Eq:Wheeler-DeWitt}), satisfying
\beqa
&& {\mathcal G}_{ijkl}=\frac{1}{2}(g_{ik}g_{jl}+g_{il}g_{jk})-\tilde\lambda g_{ij}g_{kl}, \quad \tilde\lambda = \frac{\lambda}{d\lambda-1}, \nn \\
&& {\mathcal G}_{ijmn}G^{mnkl}=\frac{1}{2}(\delta^k_i\delta^l_j+\delta^l_i\delta^k_j).  \label{Eq:Wheeler-DeWitt-inverse}
\eeqa
It is obvious that the scaling dimension of the inverse metric is $[{\mathcal G}_{ijkl}]=0$. it is worthy of notice that when $\lambda=1 $ in $(3+1)$-dimensional space-time, the generalized metric just reduce to be the original Wheeler-De Witt metric, ${\mathcal G}_{ijkl}^{\text{WD}}=(g_{ik}g_{jl}+g_{il}g_{jk}-\lambda g_{ij}g_{kl})/2 $. The expression in Eq.(\ref{Eq:Eij}), is called the detailed balance condition, which connects a $d$-dimensional relativistic theory described by the action $W$ to a $(d+1)$-dimensional theory described by the action $S_K-S_V$.

$E^{ij}$ has to be obtained from a variation of some spatial isotropic potential $W[g_{ij}]$, i.e., the Euclidean action of a relativistic theory in $d$-dimensional space with respect to the metric $g_{ij}$,
\beqa
E^{ij} \equiv \frac{1}{\sqrt{g}}\frac{\delta W[g]}{\delta g_{ij}}, \label{Eq:Eij}
\eeqa
with its trace $E\equiv E^{ij}g_{ij}$. According to Eq.(\ref{Eq:Eij}), we obtain
\beqa
E^{ij} &\equiv& \frac{1}{\sqrt{g}}\frac{\delta (\sqrt{g} {\mathcal L})}{\delta g_{ij}} %= \frac{1}{\sqrt{g}} \bigg( \sqrt{g} \frac{\delta {\mathcal L }}{\delta g_{ij}} + \frac{\delta \sqrt{g}}{\delta g_{ij}} {\mathcal L } \bigg) \nn\\
=  \frac{\delta {\mathcal L }}{\delta g_{ij}} + \frac{1}{2}g^{ij} {\mathcal L }, \\
E_{ij} &\equiv& \frac{1}{\sqrt{g}}\frac{\delta (\sqrt{g} {\mathcal L })}{\delta g^{ij}} %= \frac{1}{\sqrt{g}} \bigg( \sqrt{g} \frac{\delta {\mathcal L }}{\delta g^{ij}} + \frac{\delta \sqrt{g}}{\delta g^{ij}} {\mathcal L } \bigg) \nn\\
=  \frac{\delta {\mathcal L }}{\delta g^{ij}} - \frac{1}{2}g_{ij} {\mathcal L }.
\eeqa
Since on one hand, $g_{ij}g^{jk}=\delta^k_j$, thus, $\delta(g_{ij}g^{jk})=0$, which gives $\delta g^{ij} = - g^{ik}g^{jl}\delta g_{kl}$ and $\delta g_{ij} = - g_{ik}g_{jl}\delta g^{kl}$. On the other hand, $\ln g = \ln \det g_{ij} = \text{Tr}\ln g_{ij}$, by doing variation with respect to $\delta g_{ij}$, which gives $ \delta \ln g = g^{-1} \delta g = \text{Tr}[g^{ij}\delta g_{ij}]$, thus, $\delta g = g g^{ij} \delta g_{ij} = -g g_{ij}\delta g^{ij}$.

\subsubsection{The full action and Ricci flow equation}

The full action of the theory becomes
\beqa
S \3i &=& \3i \frac{1}{2}\int dt d^dx \sqrt{g} N\bigg(\frac{2}{\kappa}K_{ij}-\frac{\kappa}{2}{\mathcal G}_{ijmn}E^{mn}\bigg) \nn\\
& & \times G^{ijkl} \bigg(\frac{2}{\kappa}K_{kl}-\frac{\kappa}{2}{\mathcal G}_{klpq}E^{pq}\bigg) \label{Eq:S_K+V}
\! = \! \int dt  d^dx \sqrt{g} N {\mathcal L}, \quad\quad \nn\\
{\mathcal L} \3i &\equiv & \3i {\mathcal L}_K - {\mathcal V}[g] = \frac{2}{\kappa^2} K_{ij}G^{ijkl}K_{kl} - \frac{\kappa^2}{8} E^{ij}{\mathcal G}_{ijkl} E^{kl}.
\eeqa
Note that the second equality is true since the mixing term between $k_{ij}$ and $E_{ij}$ turns out to be a total derivatives in the bulk space-time. This can be seen as follows: by using the normalization condition $G^{ijkl}{\mathcal G}_{klpq}=(\delta^i_p \delta^j_q+\delta^i_q \delta^j_p)/2$, the definition of $K_{ij}$ in Eq.(\ref{Eq:Kij}) and $E_{ij}$ in Eq.(\ref{Eq:Eij}), we have the cross mixing term of kinetic and potential action, which are a sum of total derivatives,
\beqa
S_{\text{mix}} \3i  &=& \3i -\int dt d^d x \sqrt{g}N K_{ij} G^{ijkl}{\mathcal G}_{klpq}E^{pq} \nn\\
\3i &=& \3i-\int dt d^d x\sqrt{g}N K_{ij} E^{ij} \nn\\
\3i &=& \3i - \frac{1}{2}\int dt d^d x (\dot{g}_{ij}- 2 \nabla_{(i} N_{j)})\frac{\delta W}{\delta g_{ij}} \nn\\
\3i &=& \3i - \frac{1}{2}\int dt d^d x \bigg[\partial_t{W} \! - \! 2 \partial_i\bigg(N_j \frac{\delta W}{\delta g_{ij}}\bigg)  \! + \! 2 N_j \sqrt{g}\nabla_i E^{ij}  \bigg], \nn
\eeqa
where the last term will be absent by using the Bianchi identity $\nabla_i E^{ij}=0$ for Eq.(\ref{Eq:W}) or Eq.(\ref{Eq:scaling-Eij}), with $E^{ij}$ formally defined in (\ref{Eq:Eij}). Note that we have used
\beqa
\nabla_i N_j \frac{\delta W}{\delta g_{ij}}&=& \nabla_i \(N_j \frac{\delta W}{\delta g_{ij}}\)- N_j\nabla_i\( \frac{\delta W}{\delta g_{ij}}\) \nn\\
&=&\partial_i \bigg(N_j \frac{\delta W}{\delta g_{ij}}\bigg) -N_j{\sqrt{g}}\nabla_i\bigg(\frac{1}{\sqrt{g}}\frac{\delta W}{\delta g_{ij}}\bigg), \nn
\eeqa
since $\Gamma_{ij}^{j}=(\partial_{i}\sqrt{g})/\sqrt{g}$, and $\nabla_i V^i = \partial_i(\sqrt{g}V^i)/\sqrt{g}$. The advantage to write the expression in Eq.(\ref{Eq:S_K+V}) is that, one can rewrite it as an Gaussian integral $\int {\mathcal D} B^{ij} {\mathcal D} g_{ij}\, {\mathcal D }N_i\, {\mathcal D} N \exp\{iS\}$ by introducing an auxiliary field, i.e., $B^{ij}$, so that we have a new action
\beqa
S &=& \frac{1}{2}\int dt \int d^d x \sqrt{g} N \bigg[ -\frac{1}{2}B^{ij}{\mathcal G}^{ijkl}B^{kl} \nn\\
&+& \bigg(\frac{2}{\kappa}K_{ij}-\frac{\kappa}{2}{\mathcal G}_{ijkl}E^{kl}\bigg)B^{ij} \bigg],
\eeqa
The action with all terms are at least linear in the auxiliary field $B^{ij}$ and with the linear term proportional to a gradient flow equation, one can deduce that the scaling dimension of the auxiliary field as
\beqa
[B^{ij}] = \frac{z+d}{2}.
\eeqa
By making a translation of $B^{ij}$,  we can obtain the classical evolution equation of the gravity for $g_{ij}$,
\beqa
\ii
K_{ij}-\frac{\kappa^2}{4}{\mathcal G}_{ijkl}E^{kl}=0,  \quad E^{kl}\equiv \frac{1}{\sqrt{g}}\frac{\delta W[g]}{\delta g_{kl}}, \label{Eq:Kij-Ricci-Flow}
\eeqa
or more explicitly
\beqa
K_{ij} = \frac{\kappa^2}{4}(E_{ij}- \tilde\lambda g_{ij} E),
%\nn\\
%&=& \frac{\kappa^2}{4\sqrt{g}}\big( g_{ik}g_{jl}  - \tilde\lambda g_{ij} g_{kl} \big)\frac{\delta W[g]}{\delta g_{kl}},
\eeqa
with $\left[K_{ij} \right] = z$, and $ [E_{ij}]=d$. It means that the extrinsic curvature of the hypersurface of a $(d+1)$-dimensional bulk theory is determined by the gradient of a spatial isotropic potential in a $d$-dimensional space.

In addition, by using Eq.(\ref{Eq:Kij}), one can express it as an more explicit dynamical equation on the metric $g_{ij}$,
\beqa
%\dot{g}_{ij} = \nabla_i N_j + \nabla_j N_i + N  \frac{\kappa^2}{2\sqrt{g}}{\mathcal G}_{ijmn} \frac{\delta W}{\delta g_{mn}},
\dot{g}_{ij} = \nabla_i N_j + \nabla_j N_i + N  \frac{\kappa^2}{2}(E_{ij}-\tilde\lambda g_{ij}E),
\label{Eq:gij-Ricci-Flow-1}
\eeqa
which is a EOM with first order in time derivative and $z$ in spatial derivative. It is clear that the evolution of $g_{ij}$ is governed by a gradient flow $\delta W/\delta g_{ij}$ on the spatial foliations, and some gauge transformations represented by $N_i$ and $N$. If the action of a theory is associated with a gradient flow generated by some time-independent potential $W[g_{ij}(x)]$, we say that the theory satisfies the detailed balance condition. the detailed balance condition guarantee that the theory with action in $(d+1)$-dimension can inherit the renormalization properties from the theory with potential $W$ in $d$-dimensions in terms of quantum inheritance principle~\cite{Horava:2009uw}. Alternatively, the partition function of the $d$-dimensional theory described by $W$ yield a natural solution from the theory in $(d+1)$ dimensions.

One important issue is to analyze under what conditions the principle is valid in the HL gravity.

There is a simple static solution of the above equation in $(d+1)$ dimensions by choosing a popular gauge $N = 1$, $N_i = 0$ and taking $g_{ij} = g_{ij}(x)$ to be equivalent to a $d$-dimensional theory with action W, namely,
\beqa
\dot{g}_{ij}=0, \quad (N=1, \quad N_i =0) \quad \Leftrightarrow \quad \frac{\delta W}{\delta g_{ij}}=0.
\eeqa

\subsection{Hamiltonian formulation}
From the full action in Eq.(\ref{Eq:S_K+V}), the canonical momentum conjugate to the metric $g_{ij}$ can be easily obtained
\beqa
\pi^{ij} &\equiv & \frac{\delta S}{\delta \dot{g}_{ij}} = \frac{2}{\kappa^2}\sqrt{g}G^{ijkl}K_{kl}, \label{Eq:Pij}  \\
\Rightarrow K_{ij} &=& \frac{\kappa^2}{2\sqrt{g}}{\mathcal G}_{ijkl}\pi^{kl}, \label{Eq:Kij_Pij}
\eeqa
and the momentum conjugate to $N_i$ and $N$ are identically zero, i.e, $\pi^i\equiv \delta S/\delta N_i =0$, $\pi^0\equiv \delta S/\delta N = 0$. In this case, the Lagrangian of the kinetic term can be re-written as
\beqa
{\mathcal L}_K &=& \frac{2}{\kappa^2} K_{ij}G^{ijkl}K_{kl} = \frac{1}{\sqrt{g}}K_{ij}\pi^{ij} = \frac{\kappa^2}{2g}\pi^{ij}{\mathcal G}_{ijkl}\pi^{kl} \nn\\
&=& \frac{\kappa^2}{2g}\bigg( \pi^{ij}\pi_{ij}-\frac{\lambda}{d\lambda-1} (\pi_i^i )^2 \bigg).
\eeqa
With this and (\ref{Eq:gij-Ricci-Flow-1}), the Hamiltonian density of the theory becomes
\beqa
\ii
{\mathcal H}
 &\equiv &  \pi^{ij}\dot{g}_{ij} \!-\! \sqrt{-\hg} {\mathcal L} \! = \! (\pi^{ij}\dot{g}_{ij} \!-\! \sqrt{g}N {\mathcal L}_K) \!+\! \sqrt{g}N {\mathcal V}[g]
\nn\\
 &=&   N\pi^{ij}K_{ij} \!+\! 2\pi^{ij}\nabla_{(i} N_{j)}  \!+\! \sqrt{g}N {\mathcal V}[g],
\eeqa
where we have used the definition of extrinsic curvature in Eq.(\ref{Eq:Kij}).
Therefore, the Hamiltonian becomes
\beqa
H  & \equiv & \int d^d x {\mathcal H} \nn\\
 &=&  \int d^d x  \big( N\pi^{ij}K_{ij} \!-\! 2 N^i\nabla_j \pi^{ij} \big) \!+\! \int d^dx \sqrt{g}N {\mathcal V}[g] \nn\\
 &=&  \int d^d x  \bigg(\frac{\kappa^2}{2\sqrt{g}}N\pi^{ij}{\mathcal G}_{ijkl}\pi^{kl} \!- \! 2N_i\nabla_j \pi^{ij}\bigg) \nn\\
 &+& \int d^dx \sqrt{g}N {\mathcal V}[g]  ,
\eeqa
where in the second equality, we have imposed partial integration and dropped all the total derivative with respect to spatial coordinates, while in the third equality, we have used Eq.(\ref{Eq:Kij_Pij}) to express the extrinsic curvature as the conjugate momentum. $S_V$ is defined in Eq.(\ref{Eq:SV_1}). In the end, we obtain the Hamiltonian of the system
\beqa
H &=& \int d^d x  \bigg[\frac{\kappa^2}{2\sqrt{g}}N \bigg( \pi^{ij}\pi_{ij}-\frac{\lambda}{d\lambda-1} (\pi_i^i )^2 \bigg) \nn\\
&-& 2N_i\nabla_j \pi^{ij}\bigg] +  \int d^dx \sqrt{g}N {\mathcal V}[g]. \label{Eq:H}
\eeqa
From above equation, one reads the Hamiltonian densities associated with lapse function $N$ and shift variable $N_i$ respectively,
\beqa
&& {\mathcal H}_{\bot} = \frac{\kappa^2}{2\sqrt{g}} \bigg( \pi^{ij}\pi_{ij}-\frac{\lambda}{d\lambda-1} (\pi_i^i )^2 \bigg) + \sqrt{g}{\mathcal V}[g], \nn\\
&& {\mathcal H}_i = -2 \nabla_j \pi_{i}^{~j}.
\eeqa
The Hamiltonian density of the Einstein gravity at the long distance is
\beqa
{\mathcal H} &=& \frac{2\kappa_G}{\sqrt{g}}\pi^{ij}{\mathcal G}_{ijkl}\pi^{kl}-\frac{\sqrt{g}}{2\kappa_G}(R-2\Lambda) \nn\\
&=& \frac{1}{\sqrt{g}}\bigg(\pi^{ij}\pi_{ij}-\frac{1}{2}(\pi^i_i)^2\bigg) - \sqrt{g}(R-2\Lambda), \nn
\eeqa
where ${\mathcal G}_{ijkl}=(g_{ik}g_{jl}+g_{il}g_{jk}-g_{ij}g_{kl})/2$, and in the last identity, we have imposed the normalization condition that $2\kappa_G \equiv 16\pi G_N=1$. From the above equation, and by comparing with the kinetic term with that defined in Eq.(\ref{Eq:H}), we can obtain
\beqa
 N = c,  \quad & \Rightarrow & \quad \kappa^2 = \frac{4\kappa_G}{c}=\frac{32\pi G_N}{c}, \nn\\
 \tilde\lambda = 1/2  \quad & \Rightarrow & \quad \lambda = \frac{1}{d-2}.
\eeqa
The dimensionless coupling constant $\lambda$ is a classically marginal coupling that runs in the quantum field theory. In the $(3+1)$-dimensional case, we have $\lambda=1$ in general relativity as expected. Then ${\mathcal H}=0$ just gives the Wheeler-De Witt equation with cosmological constant,
\beqa
\frac{1}{\sqrt{g}}\pi^{ij}{\mathcal G}_{ijkl}\pi^{kl} - \sqrt{g}(R-2\Lambda)=0.
\eeqa
In $(d+1)$ dimensions, ${\mathcal H}=0$ gives
\beqa
&& \frac{\kappa^2}{2\sqrt{g}}N \bigg( \pi^{ij}\pi_{ij}-\frac{\lambda}{d\lambda-1} (\pi_i^i )^2 \bigg) \nn\\
&& -2N_i\nabla_j \pi^{ij}  + \sqrt{g}N {\mathcal V}[g] = 0,
\eeqa
which implies a generalized $(d+1)$-dimensional Wheeler-De witt equation in HL gravity
\beqa
\ii
\frac{\kappa^2}{2\sqrt{g}}N \bigg( \pi^{ij}\pi_{ij}-\frac{\lambda}{d\lambda-1} (\pi_i^i )^2 \bigg) + \sqrt{g}N {\mathcal V}[g] = 0,
\eeqa
and the Dirac's secondary constraint, i.e., momentum constraint,
\beqa
\nabla_j \pi^{ij} = 0,
\eeqa
in canonical quantum gravity approach in a generic gauge with $N_i\ne 0$. According to Eq.(\ref{Eq:Pij}), more explicitly, the conjugate momentum can be expressed as
\beqa
\pi^{ij} = \frac{2}{\kappa^2}\sqrt{g}(K^{ij}-\lambda K g^{ij}).
\eeqa

For the detail discussion on the physics of HL$_{d+1}$ Gravity in specific dimensions within $d\le 5$, we summarize the results in Appendix \ref{app:HL(d+1)}.

\section{Ho\v{r}ava-Lifshitz Gravity in Specific Dimensions}
\label{app:HL(d+1)}
In this section, we briefly summarize the relevant physics of Ho\v{r}ava-Lifshitz Gravity in specific dimensions, namely, HL$_3$ at $z=2$, HL$_{4}$ at $z=3,4$, HL$_5$ at $z=4$ and HL$_6$ at $z=5$ UV fixed point.

\subsection{$(2+1)$-dimensional HL gravity}
\label{app:HL(2+1)}

In $(2+1)$ dimensional HL gravity, the coupling $\kappa$ will be dimensionless at fixed point with $z=2$.
Therefore, in the UV limit at short distance, the theory will exhibit as a power-counting renormalizable UV theory at $z=2$ Lifshitz fixed point or a super-renormalizable theory at $z>2$, i.e., $z=3$ Lifshitz fixed point. The spatial potential action is the $d=2$ case of Eq.(\ref{Eq:W}),
\beqa
W_1 = \frac{1}{\kappa^2_W}\int d^2 x \sqrt{g}(R-2\Lambda_W), \label{Eq:W_1-2D}
\eeqa
According to Eq.(\ref{Eq:SV_1}) and Eq.(\ref{Eq:gij-Ricci-Flow-2}), the potential density and the Ricci flow equation for the theory becomes
\beqa
{\mathcal V}[g] \3i &=& \3i \frac{\kappa^2}{8\kappa_W^4} \bigg( R^{ij}R_{ij}-\frac{1}{2}R^2 + \frac{2\Lambda_W^2 }{1-2\lambda}\bigg), \label{Eq:SV_1_3D} \\
\dot{g}_{ij} \3i &=& \3i 2\nabla_{(i} N_{j)} \!-\! N  \frac{\kappa^2}{2\kappa_W^2}\bigg(R_{ij} \!-\! \frac{1}{2}R\,g_{ij} \!-\! \frac{1}{2\lambda\!-\!1}\Lambda_W g_{ij} \bigg), \nn
\eeqa
where $\tilde\lambda\equiv \lambda/(2\lambda-1)$, the pure spatial index runs $i,j=1,2$ and the scaling dimension of the parameters for the theory become
\beqa
[\kappa]=\frac{z-2}{2}, \quad [\lambda]=0, \quad [\kappa_W] = 0, \quad [E^{ij}]=2.
\eeqa
Since $[\kappa]=[\kappa_W]=0$ at $z=2$, thus, both the kinetic and potential couplings are dimensionless and they are renormalizable.

The full action of $z=2$ HL$_3$ gravity is the special case of Eq.(\ref{Eq:S-HL(d+1)-z=3}) with $d=2$,
\beqa
S \3i &=&\int dt(L_0+L_1), \label{Eq:S-HL(2+1)-z=2}\\
L_0 \3i &\equiv & \3i \int d^2x \sqrt{g} N \bigg( \frac{\kappa^2}{2}(K_{ij}K^{ij} - \lambda K^2) - \frac{\kappa^2}{8\kappa_W^4} \frac{2\Lambda_W^2}{1-2\lambda}  \bigg), \nn\\
L_1 \3i &\equiv & \3i \int d^2x \sqrt{g} N \frac{\kappa^2}{8\kappa_W^4} \bigg( \frac{\big( \frac{1}{2} - \lambda \big)}{1-2\lambda}R^2  - R^{ij}R_{ij} \bigg), \nn
\eeqa
from Eq.(\ref{Eq:S-HL(2+1)-z=2}), It is obvious that
\beqa
[dtd^2x]=-z-2 , \quad [R^2] = 4, \quad  [\frac{\kappa^2}{\kappa_W^4}] = z-2,
\eeqa
which is consistent with Eq.(\ref{Eq:kappa2-kappaW4}). At $z=2$, both kinetic and potential term will be renormalizable; thus, $z=2$ HL$_3$ gravity is power-counting renormalizable. While at $z>2$, although the kinetic density terms are still renormalizable or superrenormalizable, but the potential will not; thus, new potential action are called for to cure the UV divergence due to the original potential action.

In the UV limit at short distance, the theory will exhibit as renormalizable theory at Lifshitz fixed point with dynamic critical exponent $z=2$, since the dimension of the gravity couplings are dimensionless.

In the IR limit at large distance, the terms proportional to Ricci scalar $R$ and cosmological constant $\Lambda_W$ will dominate and the theory flows to the $3$-dimensional(without Lorentz theory) GR with dynamical critical exponent $z=1$. The dynamics of the metric at the fixed point is controlled by the Ricci flow equation in Eq.(\ref{Eq:gij-Ricci-Flow-2}) with $\lambda = 1$,
\beqa
\dot{g}_{ij}= -  \frac{\kappa^2}{2\kappa_W^2}\bigg(R_{ij}-\frac{1}{2}R\,g_{ij}-\Lambda_W g_{ij}\bigg).
\eeqa

The corresponding effective speed of light, Newtonian constant and cosmological constant are
\beqa
\ii
c  \equiv  0,~ G_N \! =  \! \frac{\kappa^2}{32\pi c} \!\overset{\kappa\ne 0}{=}\! \infty, ~ \Lambda \!=\! \frac{d}{2(d-2)}\Lambda_W \3i \overset{\Lambda_W\ne 0}{=}\3i \infty.  \label{Eq:c-Lambda-d=2}
\eeqa
Therefore, at $(2+1)$-dimensions, the effective speed of light of the HL gravity at large distance is absolutely zero. Meanwhile the cosmological constant term will be divergent if $\Lambda_W\ne 0 $, which implies that the whole space-time is dominated by the cosmological constant term. The vanishing effective speed of light in $(2+1)$-dimensional HL gravity at $z=1$ doesn't violates one of the basic assumption of special relativity: the speed of light is constant. However, it implies that the light can not escape from the gravity at any space-time point at all if $\Lambda_W\ne 0$, due to the dominant control of cosmological constant $\Lambda$.

The cosmological constant $\Lambda$ can be a finite constant if and only if $\Lambda_W=0$ in $d=2$ case, this is possible in black hole solution of the AdS gravity, e.g, if $\Lambda_W=-(d-1)(d-2)/\ell^2$, then $\Lambda = -d(d-1)/(2\ell^2) = -\ell^{-2}$.

\subsection{$(3+1)$-dimensional HL gravity}
\label{app:HL(3+1)}

In $(3+1)$ dimensions, the spatial potential action is the $d=3$ case of Eq.(\ref{Eq:W})
\beqa
W_1 = \frac{1}{\kappa^2_W}\int d^3 x \sqrt{g}(R-2\Lambda_W), \label{Eq:W_1-3D}
\eeqa
According to Eq.(\ref{Eq:SV_1}) and Eq.(\ref{Eq:gij-Ricci-Flow-2}), the potential density of the action and the Ricci flow equation for the HL$_3$ gravity become
\beqa
{\mathcal V}_1[g] \3i &=& \3i \frac{\kappa^2}{8\kappa_W^4}\bigg(R^{ij}R_{ij} \! - \! \frac{1/4 -\lambda }{1-3\lambda} R^2 \!-\!  \frac{\Lambda_W R - 3\Lambda_W^2}{1-3\lambda}\bigg) \! ,  \label{Eq:SV_1_4D} \\
\dot{g}_{ij}\3i &=& \3i  2 \nabla_{(i} N_{j)} \!-\! N  \frac{\kappa^2}{2\kappa_W^2}\bigg(R_{ij} \!-\! \frac{2\lambda\!-\! 1}{2(3\lambda\!-\!1)}R\,g_{ij} \!-\! \frac{\Lambda_W g_{ij}}{3\lambda\!-\!1}\bigg), \nn
\eeqa
where $\tilde\lambda\equiv \lambda/(3\lambda-1) $, with pure spatial index $i,j=1,2,3$ and the scaling dimension of the parameters for the theory become
\beqa
[\kappa]=\frac{z-3}{2}, ~ [\lambda]=0, ~ [\kappa_W] = -\frac{1}{2}, ~ [E^{ij}]=3.
\eeqa
it is worthy of notice that the dimension of coupling $\kappa_W<0$, i.e., becomes negative, which implies that the potential in Eq.(\ref{Eq:SV_1_4D}) is merely a low-energy effective field theory. It will be broken down at an energy scale set by the dimensionful couplings $\kappa_W$. Therefore, the potential density in Eq.(\ref{Eq:SV_1_4D}), which originates from Eq.(\ref{Eq:W}) is not a complete potential action for the UV theory, unless including a new potential term, i.e., the $d=3$ Chern-Simons gravity action present in Eq.(\ref{Eq:W_2}).

\subsubsection{$z=3$ HL$_4$ gravity}
\label{app:(3+1)HL-z=3}

In the Section, we will mainly focus on the $z=3$ HL gravity in $(3+1)$ dimensions.

In $(3+1)$ dimensions of HL gravity, the coupling $\kappa$ will be dimensionless at fixed point with $z=3$. Therefore, in the UV limit at short distance, the theory will exhibit as a power-counting renormalizable UV theory at $z=3$ Lifshitz fixed point or a super-renormalizable theory at $z>3$, i.e., $z=4$ Lifshitz fixed point.

For $d=3$ case, except for the potential action $W_1$ in Eq.(\ref{Eq:W_1-3D}), it will include a new term
\beqa
\ii
W_2 = \frac{1}{\omega^2}\int d^3x \sqrt{g}\varepsilon^{ijk}\Gamma^m_{il}\bigg(\partial_j\Gamma^l_{km}+\frac{2}{3}\Gamma^l_{jn}\Gamma^n_{km}\bigg),  \label{Eq:W_2}
\eeqa
with $[\omega] = 0$, where we have taken $\varepsilon_{ijk} = \sqrt{g}\epsilon_{ijk}$ and $\varepsilon^{ijk}= \epsilon^{ijk}/\sqrt{g}$ with $\epsilon_{123}=1$ for index under symmetric cycle.
Assuming that the coupling $\omega^2$ is positive, whose sign can be changed by flipping the orientation of the 3-manifold. The new potential term $W_2[g]$ is the gravitational Chern-Simons term~\cite{Cai:2012mg} with Christoffel symbols $\Gamma^i_{jk}[g]$ treated as function of the metric $g_{ij}$, but not as independent variables. It is essentially a $3$-dimensional Euclidean action of topologically massive gravity~\cite{Deser:1981wh,Deser:1982vy,Bergshoeff:2009hq} with a real Chern-Simons couplings $\omega^2$, which is a dimensionless constant. The action turns out to be a renormalizable relativistic quantum gravity in $(2+1)$-dimensions~\cite{Witten:1988hf}. From the three-dimensional action in Eq.(\ref{Eq:W_2}), one can construct a four-dimensional renormalizable gravity theory by imposing the detailed balance condition in Eq.(\ref{Eq:Eij}),
\beqa
E^{ij}_2  =  \frac{2}{\omega^2}C^{ij}, \quad C^{ij} = \frac{\varepsilon^{ikl}}{\sqrt{g}}\nabla_k (R^j_l \!-\! \frac{1}{4}R\delta^j_l),   \label{Eq:Eij_Cij}
\eeqa
with $[C^{ij}] = 3$, where $C_{ij}$ is the Cotton tensor, which is symmetric traceless and transverse(satisfies $\nabla_i C^{ij}=0$). $\epsilon^{ijk}$ is the bulk Levi-Civita contra-variant tensor defined through Levi-Civita tensor density $\varepsilon^{ijk}$.

The variation of the $W_2$ with respect to the metric $g_{ij}$ as defined in Eq.(\ref{Eq:Eij}) will leads to a term for action in Eq.(\ref{Eq:SV}),
\beqa
\ii
S_{V_2} = \int dt d^3x\sqrt{g}N  {\mathcal V}_2[g], \quad {\mathcal V}_2[g] = \frac{\kappa^2}{2\omega^4} C_{ij}C^{ij}, \label{Eq:SV_2}
\eeqa
where we have used $C_{ij}{\mathcal G}^{ijkl}C_{kl}=C_{ij}C^{ij}$.
In summary, the total $W$ potential action for $z=3$ HL gravity should be summation of both Eq.(\ref{Eq:W_1-3D}) and Eq.(\ref{Eq:W_2}), $W[g] = W_1 + W_2$, from which one obtains $E^{ij}$ according to the detailed balance condition in Eq.(\ref{Eq:Eij}),
\beqa
E^{ij} = \frac{2}{\omega^2}C^{ij}-\frac{1}{\kappa_W^2}\bigg(R^{ij}-\frac{1}{2}Rg^{ij}+\Lambda_W g^{ij}\bigg). \label{Eq:Eij-4D-z=3}
\eeqa

According to the field equation of motion defined in Eq.(\ref{Eq:Eij-4D-z=3}), the Ricci flow equation in Eq.(\ref{Eq:gij-Ricci-Flow-1}) and the potential density in Eq.(\ref{Eq:SV_1_4D}) becomes
\beqa
{\mathcal V}[g] \3i &=& \3i {\mathcal V}_1[g] + {\mathcal V}_2[g] - \frac{\kappa^2}{2\kappa_W^2 \omega^2}C^{ij}R_{ij} ,  \label{Eq:SV_1_4D-2}  \\
\dot{g}_{ij} \3i &=& \3i  2 \nabla_{(i} N_{j)} \!-\! N  \frac{\kappa^2}{2\kappa_W^2} \bigg( R_{ij} \!-\! \frac{2\lambda\!-\!1}{2(3\lambda\!-\!1)}R\,g_{ij} \!-\!\frac{\Lambda_W}{3\lambda\!-\!1} g_{ij}\bigg)  \nn\\
\3i &&  \3i + N\frac{\kappa^2}{\omega^2}C_{ij}, \nn
\eeqa
where we have used that $C^{ij}{\mathcal G}_{ijkl}R^{kl} = C^{ij}R_{ij}$. The first term and second term are due to $W_1$ and $W_2$ respectively, while the last term is the mixing term due to $W_1+W_2$, the last two terms together will improve the UV behavior of the HL gravity at short distance, comparing with ${\mathcal V}_1[g]$ only.

By using Eq.(\ref{Eq:Eij-4D-z=3}), the full action of HL$_4$ gravity at the $z=3$ UV fixed point are
\beqa
S \3i &=& \3i\int dt(L_0 + L_1 + L_2), \label{Eq:L0-L1-4D-z3}\\
L_0 \3i &\equiv & \3i \int d^3x \sqrt{g} N \bigg[ {\mathcal L}_K + \frac{\kappa^2}{8\kappa_W^4}\bigg( \frac{\Lambda_W}{1-3\lambda}R - \frac{3\Lambda_W^2}{1-3\lambda} \bigg) \bigg], \nn\\
L_1 \3i &\equiv & \3i \int d^3x \sqrt{g} N \frac{\kappa^2}{8\kappa_W^4} \bigg( \frac{\big( {1}/{4} - \lambda \big)}{1-3\lambda}R^2  - R^{ij}R_{ij} \bigg), \nn \\
L_2 \3i &\equiv & \3i \int d^3x \sqrt{g} N \bigg( \frac{\kappa^2}{2\kappa_W^2 \omega^2} C^{ij}R_{ij} - \frac{\kappa^2}{2\omega^4} C_{ij}C^{ij} \bigg), \nn
\eeqa
where ${\mathcal L}_K \equiv {2}(K_{ij}K^{ij} - \lambda K^2)/\kappa^2$. Comparing with the $z=3$ HL$_{d+1}$ gravity in Eq.(\ref{Eq:S-HL(d+1)-z=3})with $d=3$, there is an additional Lagrangian
\beqa
L_2 &=& \int d^dx \sqrt{g} N \bigg[ \frac{\kappa^2}{2\kappa_W^2 \omega^2}\epsilon^{ijk}R_{il}\nabla_j R^l_k \nn\\
\3i &-& \3i \frac{\kappa^2}{2\omega^4}\bigg( 2\nabla_i R_{jk} \nabla^{[i} R^{j]k} \! - \! \frac{1}{8}\nabla_i R \nabla^i R  \bigg)\bigg].
\eeqa
We have used the identity that
\beqa
C^{ij}R_{ij} &=& \epsilon^{ikl}(\nabla_k R^j_l)R_{ij} = \epsilon^{ijk}R_{il}\nabla_j R^l_k.\nn \\
C^{ij}C_{ij} &=& 2\nabla_i R_{jk} \nabla^{[i} R^{j]k} \! - \! \frac{1}{8}\nabla_i R \nabla^i R  , \nn
\eeqa
where we have used Eq.(\ref{Eq:Eij_Cij}) where $C^{ij} = \epsilon^{ikl}\nabla_k R^j_l - \varepsilon^{ikj} \partial_k R/4 $ and the antisymmetric property of $\epsilon^{[ij]k}$. To simplify the product $C^{ij}C_{ij}$, we have used the identity that $\epsilon_{imn}\epsilon^{ijk}=\delta^j_m \delta^k_n -\delta^k_m \delta^j_n$.

In the IR limit at large distance, the theory is dominated by the last term, which proportional to the spatial curvature Ricci scalar $R$ and the constant term. The theory with dynamical critical exponent $z=3$ should flow to the GR with critical exponent $z=1$, so that $\kappa_W^2\to \infty$, $\omega\to \infty$ while keep the ratio $\gamma\equiv\kappa^2/\kappa_W^2$ fixed.

Note that from Eq.(\ref{Eq:L0-L1-4D-z3}), It is obvious that
\beqa
[dtd^3x]=-z-3, \quad [R^2] = 4, \quad [\frac{\kappa^2}{\kappa_W^4}] = z-1,
\eeqa
which is consistent with Eq.(\ref{Eq:kappa2-kappaW4}). With the dimension of ${\kappa^2}/{\kappa_W^4}$ increasing, the UV behavior of the theory becomes more serious. For $z\ge 3$, the kinetic term will be renormalizable or super-renormalizable, while the original potential density term in Eq.(\ref{Eq:SV_1_4D}) will not be UV renormalizalbe; thus, new potential density term in Eq.(\ref{Eq:SV_2}) is called for to cure the UV divergence at short distance.

In the IR limit with large distance, the dynamics of the metric at the fixed point is controlled by the Ricci flow equation at $\lambda=1$ with gauge $N=1,N_i=0$ becomes
\beqa
\dot{g}_{ij} &=&  \frac{\kappa^2}{\omega^2}\epsilon_{ikl}\nabla^k \bigg(R^l_j-\frac{1}{4}R\delta^l_j \bigg) \nn\\
&-& \frac{\kappa^2}{2\kappa_W^2}\bigg(R_{ij}-\frac{1}{4}R\,g_{ij}-\frac{1}{2}\Lambda_W g_{ij}\bigg).
\eeqa

\subsubsection{$z=4$ HL$_4$ gravity}

In $(3+1)$ dimensions, if the theory is located at $z=4$ fixed point, the coupling $\kappa$ will have a scaling dimension $[\kappa]=1/2$, which implies that the UV theory defined in Eq.(\ref{Eq:S_K+V}) will be super-renormalizable by power-counting, because the kinetic term will be suppressed by scaling dimension $[\kappa^2]= 1$, which improve the short distance properties of the propagator. In three dimensions, the action of Euclidean gravity has two independent\footnote{ In $d=3$ dimensions, $R^{ijkl}R_{ijkl}$ is not independent of $R^{ij}R_{ij}$ and $R^2$, since the Weyl tensor vanishes identically at $d=3$($C_{ijkl}=0$); thus, according to Eq.(\ref{Eq:Weyl-anomaly-d}), $R^{ijkl}R_{ijkl}=4R^{ij}R_{ij}-R^2$.
} quadratic curvature terms is
\beqa
W_3 = \frac{1}{M}\int d^3x \sqrt{g}\big( R_{ij}R^{ij} + \beta R^2 \big), \label{Eq:W_3-3D}
\eeqa
with $ [\beta] = 0 $ and $ [M] = 1$. The two independent terms in $W_3$ are due to that when $d=3$, the Riemann tensor is completely determined in terms of the Ricci tensor since the Weyl tensor vanishes identically at $d=3$. The scaling dimensions of the two parameters $\beta$ and $M$ imply that the new term in potential action from $3$-dimension is super-renormalizable by power-counting, since the two couplings $[M]=[M/\beta]=1$ are dimensionful $1$. According to the detailed balance condition in Eq.(\ref{Eq:Eij}),
\beqa
E^{ij} = -\frac{1}{M}L^{ij},  \label{Eq:Eij_Lij}
\eeqa
where $L^{ij}$ is given in Eq.(\ref{Eq:Lij-MG}) with scaling dimension $[L^{ij}] = 1 + [E^{ij}] =4$ in $d=3$, and its trace in Eq.(\ref{Eq:L-MG}) turns out to be
\beqa
\ii
L\equiv g^{ij}L_{ij} =(\frac{3}{2}+4\beta)\nabla^2 R + \frac{1}{2}(\beta R^2 + R_{ij}R^{ij}),
\eeqa
which can be simplified by setting $\beta=-3/8$(note we have set $\alpha=M^{-1}$), so that we have $L = (\beta R^2 + R_{ij}R^{ij})/2$.

With $W_{1,2,3}$ respectively defined in Eq.(\ref{Eq:W_1-3D}), Eq.(\ref{Eq:W_2}) and Eq.(\ref{Eq:W_3-3D}), the total potential action becomes $W[g_{ij}] = W_1 + W_2 + W_3$, from which one obtains $E^{ij}$ according to the detailed balance condition in Eq.(\ref{Eq:Eij}),
\beqa
\ii
E^{ij} \! = \!\frac{2}{\omega^2}C^{ij} \!-\! \frac{1}{M}L^{ij} \!-\! \frac{1}{\kappa_W^2}\bigg(R^{ij} \!-\! \frac{1}{2}Rg^{ij} \!+\! \Lambda_W g^{ij}\bigg). \label{Eq:Eij-4D-z=4}
\eeqa
Consequently, the potential density in Eq.(\ref{Eq:SV_1_4D}) should be extended through Eq.(\ref{Eq:SV})
\beqa
{\mathcal V}[g] \3i &=& \3i \frac{\kappa^2}{8\kappa_W^4}\bigg(R^{ij}R_{ij}-\frac{1/4 -\lambda }{1-3\lambda} R^2 -  \frac{\Lambda_W R - 3\Lambda_W^2}{1-3\lambda}\bigg) \nn\\
\3i &+& \3i\frac{\kappa^2}{2\omega^4} C_{ij}C^{ij} - \frac{\kappa^2}{2\kappa_W^2 \omega^2}\epsilon^{ijk}R_{il}\nabla_j R^l_k  - \frac{\kappa^2}{2M\omega^2}C^{ij}L_{ij} \nn\\
\3i &+& \3i\frac{\kappa^2}{4M\kappa_W^2}\bigg(L_{kl}R^{kl} - \frac{1-2\lambda}{2(1-3\lambda)}LR + \frac{1}{1-3\lambda}\Lambda_W L \bigg) \nn\\
\3i &+& \3i\frac{\kappa^2}{8M^2}(L^{ij}L_{ij}-\tilde\lambda L^2). \label{Eq:SV_1_4D-2}
\eeqa

The scaling dimension $[{\mathcal V}[g]]=z+3\le 7$, and for different relevant operators with scaling dimensions
\beqa
\ii && [R]=2, \quad [C]=3, \quad [L]=4, \quad [K]=4, \quad [R\nabla R] = 5, \nn\\
\ii && [LR] = 6, \quad [CL]= 7, \nn\\
\ii && [R^2]= 4, \quad [C^2]=6, \quad [L^2]=8, \quad [K^2]=8.
\eeqa
As a practical example following the general discussion on the power-counting of the renormalizable field operators after Eq.(\ref{Eq:RE-operators}). For $z=4$ HL$_4$ gravity, in the UV limit, the operators with higher scaling dimensions in the potential density from $W_3$, i.e., $K,L$ dominates at short distance. While in the IR limit, the operators with lower dimensions in the potential density from $W_1$, i.e., $\Lambda,R$ dominates at long distance.

Combining Eq.(\ref{Eq:S_K+V}), Eq.(\ref{Eq:Eij_2}), and Eq.(\ref{Eq:Eij_Lij}) together, the total action can be written as
\beqa
S \3i &=& \3i \int dt \int d^3 x \sqrt{g}  N  \bigg[ {\mathcal L}_K  + \frac{\kappa^2(\Lambda_W R-3\Lambda_W^2)}{8\kappa_W^4(1-3\lambda)} \nn\\
\3i &-& \3i \frac{\kappa^2}{2\omega^4}C_{ij}C^{ij} \!+\! \frac{\kappa^2}{2\kappa_W^2 \omega^2} C^{ij}R_{ij} \! + \! \frac{\kappa^2}{2M\omega^2}C^{ij}L_{ij} \nn\\
\3i &-& \3i \frac{\kappa^2}{8M^2}(L^{ij}L_{ij}-\tilde\lambda L^2) \!-\! \frac{\kappa^2}{8\kappa_W^4}R_{ij}R^{ij} \!+\! \frac{\kappa^2(1-4\lambda)}{32\kappa_W^4(1-3\lambda)}R^2  \nn\\
\3i &-& \3i \frac{\kappa^2}{4M\kappa_W^2}\bigg( \! L_{kl}R^{kl} \!-\! \frac{1-2\lambda}{2(1-3\lambda)}LR \!+\! \frac{\Lambda_W}{1-3\lambda} L \bigg) \bigg]. \label{Eq:S-HL(3+1)-z=4}
\eeqa
where ${\mathcal L}_K\equiv {2}(K_{ij}K^{ij}-\lambda K^2)/{\kappa^2}$.
When $\beta=-3/8$ and $\omega\to \infty$, the action in Eq.(\ref{Eq:S-HL(3+1)-z=4}) just describes the dynamics of a new massive gravity~\cite{Bergshoeff:2009hq}. In the framework of new massive gravity, the calculation of the action can be simplified.
The Ricci flow equation in Eq.(\ref{Eq:gij-Ricci-Flow-1}) becomes,
\beqa
\dot{g}_{ij} \3i &=& \3i 2 \nabla_{(i} N_{j)}  + N\frac{\kappa^2}{\omega^2}C_{ij} - \frac{N\kappa^2}{2M}(L_{ij}-\tilde\lambda L g_{ij}) \nn\\
\3i &-& \3i N  \frac{\kappa^2}{2\kappa_W^2}\bigg(R_{ij}-\frac{2\lambda-1}{2(3\lambda-1)}R\,g_{ij}-\frac{1}{3\lambda-1}\Lambda_W g_{ij}\bigg). \nn
\eeqa
The corresponding effective speed of light, Newtonian constant and cosmological constant are
\beqa
c=\frac{\kappa^2}{4\kappa_W^4}\sqrt\frac{\Lambda_W}{1-3\lambda},\quad G_N = \frac{\kappa^2 c}{32\pi }, \quad \Lambda=\frac{3}{2}\Lambda_W.
\eeqa
Note that at $\lambda=1/3$, the effective speed of light is divergent, the dominant terms in Eq.(\ref{Eq:SV_1_4D-2}) are proportional to those involving only Ricci scalar $R$ and cosmological constant $\Lambda_W$, the theory exhibit an anisotropic Weyl symmetry, which is a new conformal anisotropic local scale invariance. The fact that the speed of light is almost infinite when $\lambda_c = 1/3$, the divergence in the UV opens an interesting possibility: that no inflation is needed at early times in the evolution of the universe. $\lambda$ represents a dynamical dimensionless coupling constant, susceptible to any quantum corrections. For $\lambda>\lambda_c$, $\Lambda_W<0$; while for $\lambda<\lambda_c$, $\Lambda_W>0$ so that the effective speed of light is non-negative. In addition it is worthy of notice that, the effective cosmological constant $\Lambda$ owns the same sign as $\Lambda_W$. While the effect of a positive cosmological constant corresponds to a negative pressure effect, and vice verse. This can be effectively understood by assuming that the universe is an ideal fluid~\cite{Weinberg:1972,Kiritsis:2009sh}, i.e., $T_{\mu\nu} \sim \Lambda g_{\mu\nu} \sim -p g_{\mu\nu}$. A positive cosmological constant will result a positive vacuum energy density, consequently it implies a negative pressure, which will drive an accelerated expansion of the space-time.

In summary, one has
\begin{enumerate}
\item $\lambda>1/3$: $\Lambda_W$ has to be negative, thus the cosmological constant $\Lambda<0$, which implies a positive pressure, that tend to decelerate the expansion of the universe, just like the effect of ordinary matter;
\item $\lambda<1/3$: $\Lambda_W$ has to be positive, thus the cosmological constant $\Lambda>0$, which implies a negative pressure, that tend to accelerate the expansion of the universe.
\end{enumerate}

\subsection{$(4+1)$-dimensional HL gravity}
\label{app:HL(4+1)}

In $(4+1)$ dimensions, the spatial potential action is the $d=4$ case of Eq.(\ref{Eq:W})
\beqa
W_1 &=& \frac{1}{\kappa^2_W}\int d^4 x \sqrt{g}(R-2\Lambda_W). \label{Eq:W_1-5D}
\eeqa
Naively, according to Eq.(\ref{Eq:SV_1}), the potential density of the action and the Ricci flow equation for the HL gravity become
\beqa
{\mathcal V}_1[g] \3i &=& \3i \frac{\kappa^2}{8\kappa_W^4}\bigg(R^{ij}R_{ij} \! + \! \frac{\lambda }{1 - 4\lambda} R^2 \!-\! \frac{2\Lambda_W R - 4\Lambda_W^2}{1-4\lambda}\bigg),  \label{Eq:SV_1_5D} \\
\dot{g}_{ij} \3i &=& \3i 2 \nabla_{(i} N_{j)} \!-\! N  \frac{\kappa^2}{2\kappa_W^2}\bigg(R_{ij}\!-\!\frac{2\lambda\!-\!1}{2(4\lambda\!-\!1)}R\,g_{ij} \!-\! \frac{\Lambda_W}{4\lambda\!-\!1} g_{ij}\bigg), \nn
\eeqa
where $\tilde\lambda\equiv \lambda/(4\lambda-1)$, and the scaling dimension of the parameters for the theory become
\beqa
\ii
[\kappa]=\frac{z-4}{2}, \quad [\lambda]=0, \quad [\kappa_W] = -1, \quad [E^{ij}]=4.
\eeqa
it is worthy of notice that the dimension of coupling $\kappa_W$ is negative, which implies that the potential density in Eq.(\ref{Eq:SV_1_5D}) is merely a low-energy effective field theory, which will be breakdown at an energy scale set by the dimensionful couplings $\kappa_W$. Therefore, Eq.(\ref{Eq:W_1-5D}) is not a complete potential action for the UV theory, unless including a new density term from the potential action shown in Eq.(\ref{Eq:W_4-5D}).

\subsubsection{$z=4$ HL$_5$ gravity}

In $(4+1)$ dimensions, the coupling $\kappa$ will be dimensionless if $z=4$; thus, the HL gravity will be a power-counting renormalizable UV theory at $z=4$ Lifshitz points at short distance. As discussed above, the potential density in Eq.(\ref{Eq:SV_1_5D}) is merely a low-energy effective field theory, which will be breakdown at an energy scale set by the dimensionful couplings $\kappa_W$. Consequently, the action in Eq.(\ref{Eq:W_1-5D}) will include a new term, a 4-dimensional action $W_4$, which consists of two independent quadratic curvature terms\footnote{ $R^{ijkl}R_{ijkl}$ is not independent of $R^{ij}R_{ij}$ and $R^2$ due to the topological invariant term, namely, Gauss-Bonnet term in $4$-dimensions~\cite{Horava:2009uw}. Meanwhile $R^{ij}R_{ij}$ is not independent of $C^{ijkl}C_{ijkl}$ and $R^2$ due to the ${\mathcal L}_{GB}$, as will shown in the following.  } with two dimensionless couplings $\alpha$ and $\beta$,
\beqa
W_4=\int d^4x \sqrt{g} (\alpha C_{ijkl}C^{ijkl} + \beta R^2),  \label{Eq:W_4-5D}
\eeqa
where $C_{ijkl}$ is $4$-dimensional Weyl tensor
\beqa
C_{ijkl}\equiv R_{ijkl} - \big(g_{i[k}R_{l]j}-g_{j[k}R_{l]i}\big)+\frac{1}{3}g_{i[k}g_{l]j}R. \nn
\eeqa
According to Eq.(\ref{Eq:Weyl-anomaly-d}) and Eq.(\ref{Eq:Weyl-square-d-GB}), one obtains Weyl anomaly for $d=4$ case as below,
\beqa
C^{ijkl}C_{ijkl} &=& R_{ijkl}R^{ijkl} - 2R_{ij}R^{ij} + \frac{1}{3}R^2 \label{Eq:Weyl-anomaly-4}\\
&=&{\mathcal L}_{GB} + 2R^{ij}R_{ij} - \frac{2}{3} R^2.   \label{Eq:Weyl-square-4-GB}
\eeqa
Consequently, there is no independent $R_{ij}R^{ij}$ term in the action in Eq.(\ref{Eq:W_4-5D}), since
\beqa
R^{ij}R_{ij} = \frac{1}{2}( C^{ijkl}C_{ijkl} - {\mathcal L}_{GB}) +\frac{1}{3}R^2,
\eeqa
is not independent of $C^{ijkl}C_{ijkl}$ and $R^2$, considering the Gauss-Bonnet term ${\mathcal L}_{GB}$ is a topological surface term in $d=4$.

Therefore, the action in Eq.(\ref{Eq:W_4-5D}) can be equivalently re-written as~\cite{Cai:2009ar}
\beqa
W_4= \frac{1}{M}\int d^4x \sqrt{g} (R_{ij}R^{ij} + \beta R^2),  \label{Eq:W_4-5D-2}
\eeqa
with $ [\beta] = 0 $ and  $[M] = 0$.

It is still renormalizable since the scaling dimension of the two parameters $M$ and $\beta$ are dimensionless. According to the detailed balance condition in Eq.(\ref{Eq:Eij}), one obtain
\beqa
E_4^{ij} = -\frac{1}{M}L^{ij}, \quad [L^{ij}]= [E^{ij}]=4,
\eeqa
where again $G^{ij} = R^{ij}- R g^{ij}/2$ and $L^{ij}$ is expressed in Eq.(\ref{Eq:Lij-MG}) by choosing $\alpha\equiv M^{-1}$. With $W_{1,4}$ respectively defined in Eq.(\ref{Eq:W_1-5D}), and Eq.(\ref{Eq:W_4-5D-2}), the total potential action are $W[g]= W_1 + W_4$, from which one obtain $E^{ij}$ according to the detailed balance condition in Eq.(\ref{Eq:Eij}),
\beqa
E^{ij} = -\frac{1}{M}L^{ij} - \frac{1}{\kappa_W^2}\bigg(R^{ij}-\frac{1}{2} R g^{ij}+\Lambda_W g^{ij}\bigg). \label{Eq:Eij-5D-z=4}
\eeqa
Consequently, the potential density in Eq.(\ref{Eq:SV_1_5D}) should be extended to be
\beqa
{\mathcal V}[g] \3i &=& \3i  {\mathcal V}_1[g]  \!+\! \frac{\kappa^2}{4M\kappa_W^2}\bigg(L_{kl}R^{kl} \!-\! \frac{1\!-\!2\lambda}{2(1\!-\!4\lambda)}LR \!+\! \frac{\Lambda_W}{1\!-\!4\lambda} L \bigg), \nn\\
\3i && \3i + {\mathcal V}_4[g], \quad {\mathcal V}_4[g] = \frac{\kappa^2}{8M^2}(L^{ij}L_{ij} \!-\! \tilde\lambda L^2), \label{Eq:SV_1_5D-2}
\eeqa
where the high curvature terms due to ${\mathcal V}_4[g]$ and its mixing term with ${\mathcal V}_1[g]$, will cure the UV divergence due to the low curvature terms in the potential density ${\mathcal V}_1$ of HL gravity at short distance.

Since the scaling dimension of $[{\mathcal V}_4]=z+4 \le 8$, the action can be modified by relevant operators with dimension less or equal than $8$. This is consistent with the discussion after Eq.(\ref{Eq:RE-operators}).

Combining Eq.(\ref{Eq:S_K+V}) and Eq.(\ref{Eq:Eij-5D-z=4}) together, the total action can be written as
\beqa
S \3i &=& \3i\! \int \! dt \! \int \! d^4 \! x \! \sqrt{g}  N \bigg[  \frac{2}{\kappa^2}(K_{ij}K^{ij} \!-\! \lambda K^2) \!+\! \frac{\kappa^2}{4\kappa_W^4}\frac{\Lambda_W R\!-\!2\Lambda_W^2}{(1\!-\!4\lambda)} \nn\\
\3i &-& \3i  \frac{\kappa^2}{8\kappa_W^4}R_{ij}R^{ij} \!-\! \frac{\kappa^2}{8\kappa_W^4}\frac{\lambda}{(1-4\lambda)}R^2 \!-\! \frac{\kappa^2}{8M^2}(L^{ij}L_{ij} \!-\! \tilde\lambda L^2) \nn\\
\3i &-& \3i  \frac{\kappa^2}{4M\kappa_W^2}\bigg(L_{kl}R^{kl} \!-\! \frac{1\!-\!2\lambda}{2(1\!-\!4\lambda)}LR \!+\! \frac{1}{1\!-\!4\lambda}\Lambda_W L \bigg) \bigg] ,
\label{Eq:S-HL(4+1)-z=4}
\eeqa
which is the special case of Eq.(\ref{Eq:S-HL(d+1)-z=4}) in $d=4$, by choosing the parameter in Eq.(\ref{Eq:alpha-1/M}).

In this case, the trace in Eq.(\ref{Eq:L-MG}) becomes
\beqa
L\equiv g^{ij}L_{ij}=2(1+3\beta)\nabla^2 R.
\eeqa
which can be simplified by setting $\beta=-1/3$, so that we have $L = 0$.

From Eq.(\ref{Eq:S-HL(4+1)-z=4}), It is obvious that
\beqa
[dtd^4x]=-z-4, \quad [R^2] = 4, \quad [\frac{\kappa^2}{\kappa_W^4}] = z.
\eeqa
This is consistent with Eq.(\ref{Eq:kappa2-kappaW4}), i.e., at $z=4$ fixed point, $[{\kappa^2}/{\kappa_W^4}] = 4$, $[R^2] = 4$, and $[dtd^4x]=-8$. Thus when $z\ge 4$, the kinetic term of the theory will be UV renormalizable, while the original potential density term in Eq.(\ref{Eq:SV_1_5D}) will not, unless the new potential density terms in Eq.(\ref{Eq:SV_1_5D-2}) are called for to cure the UV divergence of the potential term.

The Ricci flow equation in Eq.(\ref{Eq:gij-Ricci-Flow-1}) becomes,
\beqa
\dot{g}_{ij} \3i &=& \3i 2 \nabla_{(i} N_{j)} \!-\!  \frac{N\kappa^2}{2\kappa_W^2}\bigg(R_{ij} \!-\! \frac{2\lambda \!-\! 1}{2(4\lambda \!-\! 1)}R\,g_{ij} \!-\! \frac{\Lambda_W}{4\lambda \!-\! 1} g_{ij}\bigg)   \nn\\
\3i & & \3i - \frac{N\kappa^2}{2M}(L_{ij} - \tilde\lambda L g_{ij}) ,
\eeqa
where $\tilde\lambda = {\lambda}/{(4\lambda-1)}$.

The dynamics of the metric at the fixed point is controlled by the Ricci flow equation at $\lambda=1$($\tilde\lambda=1/(d-1)=1/3$) with gauge $N=1,N_i=0$ becomes
\beqa
\dot{g}_{ij} &=& - \frac{\kappa^2}{2\kappa_W^2}\bigg(R_{ij}-\frac{1}{6}R\,g_{ij}-\frac{1}{3}\Lambda_W g_{ij}\bigg) \nn\\
&& - \frac{\kappa^2}{2M}(L_{ij} - \frac{1}{3} L g_{ij}) .
\eeqa
The corresponding effective speed of light, Newtonian constant and cosmological constant are
\beqa
c=\frac{\kappa^2}{4\kappa_W^4}\sqrt\frac{2\Lambda_W}{1-4\lambda},\quad G_N = \frac{\kappa^2 c}{32\pi }, \quad \Lambda=\Lambda_W.
\eeqa

In summary, the physical consequence with $\lambda_c=1/4$ in $(4+1)$-dimensional space-time, is similar to that with $\lambda_c=1/3$ in $(3+1)$-dimensional space-time, as summarized at the end in Appendix \ref{app:HL(3+1)}, except that here $\lambda_c=1/4$. It is also worthy of noticing that, $\Lambda_W$ owns not only the same sign but also the same amplitude as $\Lambda$ for $(4+1)$-dimensional case.

\subsection{$(5+1)$-dimensional HL gravity}
\label{app:HL(5+1)}

In $(5+1)$ dimensions, the spatial potential action is the $d=5$ case of Eq.(\ref{Eq:W})
\beqa
W_1 &=& \frac{1}{\kappa^2_W}\int d^5 x \sqrt{g}(R-2\Lambda_W). \label{Eq:W_1-6D}
\eeqa
Naively, according to Eq.(\ref{Eq:W_1-6D}), the potential density of the action and the Ricci flow equation for the HL gravity becomes
\beqa
{\mathcal V}_1[g] \3i &=& \3i \frac{\kappa^2}{8\kappa_W^4}\bigg(R^{ij}R_{ij} \! + \! \frac{ {1}/{4} \!+\! \lambda }{1 \!-\! 5\lambda} R^2 \!-\! \frac{3\Lambda_W R \!-\! 5\Lambda_W^2}{1\!-\!5\lambda}\bigg),  \label{Eq:SV_1_6D} \\
\dot{g}_{ij} \3i &=& \3i 2\nabla_{(i} N_{j)} \!-\! N \frac{\kappa^2}{2\kappa_W^2}\bigg( R_{ij}\!-\!\frac{2\lambda\!-\!1}{2(5\lambda\!-\!1)}R\,g_{ij} \!-\! \frac{\Lambda_W}{5\lambda\!-\!1} g_{ij}\bigg), \3i \nn
\eeqa
where $\tilde\lambda \equiv \lambda/(5\lambda-1)$. The theory flows to the relativistic GR with critical exponent $z=1$.
The scaling dimension of the parameters for the theory become
\beqa
\ii
[\kappa]=\frac{z-5}{2}, \quad [\lambda]=0, \quad [\kappa_W] = -\frac{3}{2}, \quad [E^{ij}]=5.
\eeqa
Thus when $z\ge 5$, the kinetic term of the theory will be UV renormalizable, while the original potential term will not. As a result, new potential terms are called for to cure the UV divergence of the potential term.

In the UV limit at short distance, the theory will exhibit a $z=5$ Lifshitz fixed point. Since $[\kappa]=0$, the kinetic term of the theory is power-counting renormalizable at $z=5$, and is super-renormalizable at $z>5$, e.g., $z=6$ Lifshitz fixed point.

In the IR limit at large distance, the HL gravity in $(5+1)$ dimensions will be dominated by the lowest dimensional operators in the action, the last terms proportional to Ricci scalar curvature $R$ and constant term in Eq.(\ref{Eq:SV_1_6D}).

The corresponding effective speed of light, Newtonian constant and cosmological constant are
\beqa
c=\frac{\kappa^2}{4\kappa_W^4}\sqrt\frac{3\Lambda_W}{1-5\lambda},\quad G_N = \frac{\kappa^2 c}{32\pi }, \quad \Lambda= \frac{5}{6}\Lambda_W.
\eeqa

In summary, the physical consequence with $\lambda_c=1/5$ in $(5+1)$-dimensional space-time, is similar to that with $\lambda_c=1/3$ in $(3+1)$-dimensional space-time, as summarized at the end in Appendix \ref{app:HL(3+1)}, except that here $\lambda_c=1/5$. It is also worthy of noticing that, in $(5+1)$ dimensions, $\Lambda_W$ owns not only the same sign but also the same amplitude as $\Lambda$.

\subsubsection{$z=5$ HL$_6$ gravity}

In $(5+1)$ dimensions, the coupling $\kappa$ will be dimensionless if $z=5$; thus, the HL gravity will be a power-counting renormalizable UV theory at $z=5$ Lifshitz points at short distance. As discussed before, the potential density in Eq.(\ref{Eq:SV_1_6D}) is merely a low-energy effective field theory, which will be breakdown at an energy scale set by the dimensionful couplings $\kappa_W$, the potential action in Eq.(\ref{Eq:W_1-6D}) will include a new term, a 5-dimensional $W$ potential action consists of terms quadratic in curvature with three independent dimensionless couplings $\alpha$, $\beta$, $\gamma$,
\beqa
\ii
W_5 \!&=&\!  \int d^5x \sqrt{g} (   \alpha R^{ij}R_{ij} + \beta R^2 + \gamma {\mathcal L}_{GB} ) ,  \label{Eq:W_5-6D}
\eeqa
where $[\alpha]=[\beta]=[\gamma]=1$, the Gauss-Bonnet term ${\mathcal L}_{GB}=(R_{ijkl}R^{ijkl}-4R_{ij}R^{ij}+R^2)$ is not a topological invariant term, but a dynamical one in $d\ge 5$-dimensions.

With $W_{1,5}$ respectively defined in Eq.(\ref{Eq:W_1-6D}), and Eq.(\ref{Eq:W_5-6D}), the total potential action are $W[g]= W_1 + W_5$, from which one obtain field equations of motion $E^{ij}$ according to the detailed balance condition in Eq.(\ref{Eq:Eij}),
\beqa
E^{ij} = -\gamma J^{ij} - \frac{1}{\kappa_W^2}\bigg(R^{ij}-\frac{1}{2} R g^{ij}+\Lambda_W g^{ij}\bigg). \label{Eq:Eij-5D-z=4}
\eeqa
Consequently, the potential density in Eq.(\ref{Eq:SV_1_6D}) should be extended to be
\beqa
{\mathcal V}[g] \2i &=& \2i  {\mathcal V}_1[g] +  \frac{\kappa^2\gamma}{4\kappa_W^2}\bigg(J_{kl}R^{kl} \!-\! \frac{1\!-\!2\lambda}{2(1\!-\!5\lambda)}JR \!+\! \frac{\Lambda_W}{1\!-\!5\lambda} J \bigg)   \nn\\
\3i &+& \3i  {\mathcal V}_5[g], \quad
{\mathcal V}_5[g] \2i = \2i \frac{\kappa^2\gamma^2}{8}(J^{ij}J_{ij}-\tilde\lambda J^2).  \label{Eq:SV_1_6D-2}
\eeqa
The high curvature terms in the potential density ${\mathcal V}_5$ will cure the UV divergence due to the low curvature terms in the potential density ${\mathcal V}_1$, i.e., improve the UV behavior of the HL gravity at short distance in the UV limit. This is the special case in Eq.(\ref{Eq:SV_1_dD-2}) with $d=5$ and $p=2$.

While in the IR limit, the operators with lowest dimensions in the potential density, i.e., $R$ or $\Lambda_W$ will still dominate at long distance.

In this case, the scaling dimension of $[{\mathcal V}[g]]=z+5 \le 10$, therefore, the action can be modified by relevant operators with dimension less or equal than $10$. For different relevant operators with scaling dimensions
\beqa
\ii [R^2]=4,\quad [J^2]= 8, \quad [JR]=6,
\eeqa
the Ricci flow equation in Eq.(\ref{Eq:gij-Ricci-Flow-1}) becomes,
\beqa
\dot{g}_{ij} \3i &=& \3i 2 \nabla_{(i} N_{j)}  \!-\!  \frac{N\kappa^2}{2\kappa_W^2}\bigg(R_{ij} \!-\! \frac{2\lambda \!-\! 1}{2(5\lambda \!-\! 1)}R\,g_{ij} \!-\! \frac{\Lambda_W}{5\lambda \!-\! 1} g_{ij}\bigg)  \nn\\
\3i && \3i - \frac{N\kappa^2\gamma}{2}(J_{ij} - \tilde\lambda J g_{ij}),
\eeqa
where $\tilde\lambda = {\lambda}/{(5\lambda-1)}$. The theory flows to the relativistic GR with critical exponent $z=1$. The dynamics of the metric at $z=1$ fixed point is controlled by the Ricci flow equation at $\lambda=1$ by choosing a popular gauge $N=1$, $N_i=0$,
\beqa
\dot{g}_{ij} &=& -  \frac{\kappa^2}{2\kappa_W^2}\bigg(R_{ij}-\frac{1}{8}R\,g_{ij}-\frac{1}{4}\Lambda_W g_{ij}\bigg) \nn\\
&& - \frac{\kappa^2\gamma}{2}\bigg(J_{ij} - \frac{1}{4} J g_{ij}\bigg).
\eeqa

Combining Eq.(\ref{Eq:S_K+V}) and Eq.(\ref{Eq:Eij-5D-z=4}) together, the total action of $z=5$ HL$_6$ gravity can be written as
\beqa
\ii
S \3i &=& \3i \int \! dt \! \int \! d^5 \! x \sqrt{g}  N \! \bigg[  \frac{2}{\kappa^2}(K_{ij}K^{ij} \!-\! \lambda K^2) \!+\! \frac{\kappa^2}{8\kappa_W^4}\frac{3\Lambda_W R \!-\! 5\Lambda_W^2}{(1 \!-\! 5\lambda)}  \nn\\
\3i &-& \frac{\kappa^2}{8\kappa_W^4}R_{ij}R^{ij} - \frac{\kappa^2}{8\kappa_W^4}\frac{1/4+\lambda}{1-5\lambda}R^2 - \frac{\kappa^2\gamma^2}{8}(J^{ij}J_{ij}-\tilde\lambda J^2) \nn\\
\3i &-& \frac{\kappa^2\gamma}{4\kappa_W^2}\bigg(J_{kl}R^{kl} - \frac{1-2\lambda}{2(1-5\lambda)}JR + \frac{1}{1-5\lambda}\Lambda_W J \bigg) \bigg].
\label{Eq:S_6D-z=5}
\eeqa
where $J^{ij}$ is given in Eq.(\ref{Eq:Jij-GB}) and its trace in $d=5$ is given by Eq.(\ref{Eq:J-GB}),
\beqa
J \equiv  g^{ij}J_{ij} = -\frac{1}{2} {\mathcal L}_{GB}.
\eeqa
The lowest order terms of the action in the IR limit at large distance, is a special case of Eq.(\ref{Eq:S-HL(d+1)-z=3}) with $d=5$. Note in this case, it is obviously from Eq.(\ref{Eq:SV_1_6D}) that
\beqa
[dtd^4x]=-z-5, \quad [R^2] = 4, \quad [\frac{\kappa^2}{\kappa_W^4}] = z + 1 ,
\eeqa
which is consistent with Eq.(\ref{Eq:kappa2-kappaW4}), i.e., at $z=5$ fixed point, $[{\kappa^2}/{\kappa_W^4}] = 6$, $[R^2] = 4$, and $[dtd^4x]=-10$.

Alternatively, in $d=5$ case, according to Eq.(\ref{Eq:Weyl-anomaly-d}) and Eq.(\ref{Eq:Weyl-square-d-GB}), one has
\beqa
C^{ijkl}C_{ijkl} &=& R_{ijkl}R^{ijkl} \!-\! \frac{4}{3}R_{ij}R^{ij} \!+\! \frac{1}{6}R^2 \label{Eq:Weyl-anomaly-5} \\
&=& {\mathcal L}_{GB} + \frac{8}{3}R_{ij}R^{ij} - \frac{5}{6}R^2, \label{Eq:Weyl-square-5-GB}
\eeqa
where ${\mathcal L}_{GB}$ is the Gauss-Bonnet term defined in Eq.(\ref{Eq:GB}).
Thus by choosing appropriate parameters, i.e. a special case for Eq.(\ref{Eq:beta-over-alpha}) and Eq.(\ref{Eq:gamma-over-alpha}),
\beqa
\beta= - \frac{5}{16}\alpha, \quad \alpha = \frac{8}{3} \gamma,
\eeqa
the action in Eq.(\ref{Eq:W_5-6D}) can be re-expressed as Weyl anomaly action
\beqa
\ii
W_5 \!= \!  \int d^5x \sqrt{g}   \gamma C^{ijkl}C_{ijkl},  \label{Eq:W_5-6D_1}
\eeqa
where $C_{ijkl}$ are Weyl tensor defined as in Eq.(\ref{Eq:Weyl-d}), which is nonvanishing in $d\ge 4$.
In $5$-dimensions, the nonvanishing Weyl tensor in Eq.(\ref{Eq:Weyl-d}) becomes
\beqa
\ii\ii
C_{ijkl} \! \equiv  \! R_{ijkl} - \frac{2}{3}\big(g_{i[k}R_{l]j}-g_{i[l}R_{k]j}\big)+\frac{1}{6}g_{i[k}g_{j]l}R. \label{Eq:Weyl-5}
\eeqa

\section{  Differential Form of Lovelock Gravity }
\label{app:diff-form-Lovelock}

The action in Eq.(\ref{Eq:W_d_(d+1)D}) can also be expressed in the tetrad formulation(where one has chosen a set of coordinate independent basis $e_a \equiv  e^\mu_{~a}\partial_\mu$, whith $a=1,\ldots, d$, which span the tangent space at local point of space-time, i.e., $ds^2 \equiv g_{\mu\nu}dx^\mu dx^\nu = g_{ab}e^{a}e^{b}$, so that $g_{ab}=g_{\mu\nu}e^\mu_{~a}e^{\nu}_{~b}$.), i.e., as a functional of the vielbein\footnote{Note: Vielbein covers all dimensions while vierbein and tetrad are kept for $4$-dimensions.} $e^a=e^{~a}_\mu dx^\mu$(a local frame $1$-form) and a spin connection $\omega^{ab}$(which is defined the torsion free condition\footnote{Note: In the paper, we will assume the gravity theory is torsion free, i.e., the torsion tensor is not explicitly contained in the action. While the Bianchi identities of the torsion implies that $DR^{ab}=0$, but $DT^a= R^a_{~b}e^b$.}, $T^a=De^a=de^a+\omega^a_{~b}\wedge e^b=0$), with an associated gravity field strength $R^{ab}=D\omega^{ab}=d\omega^{ab}+\omega^{a}_{~c}\wedge\omega^{cb}$(or Cartan's second structural equation in terms of exterior differential), over a $d\le 2n$-dimensional differential manifold ${\mathcal M}$,
\beqa
\ii
W_d \2i = \2i \int_{{ \mathcal M} } \sum_{p=0}^{[d/2]} \alpha_p W_{(p)}, ~ W_{(p)} = ({\mathcal R}^p)^{a_1 \ldots a_{2p}} \star e_{a_1 \ldots a_{2p}}, \label{Eq:Wd-d-form}
\eeqa
where $W_d$ is a $d$-form and $W_{(p)}$ are the Euler density $2p$-forms. In dimension $d>2p$, $\{ W_{(i)} \}$ with $i=0, 1, \ldots, p$, are dimensional continued Euler densities.
${\mathcal R}^p$ means $p$ exterior products of the curvature $2$-form,
\beqa
({\mathcal R}^p)^{a_1\ldots a_{2p}} \2i & = &\2i  R^{a_1 a_2}\wedge \ldots \wedge R^{a_{2p-1}a_{2p}} \label{Eq:R2p-form}\\
&\equiv &  \frac{1}{2^p} R^{a_1 a_2}_{~~b_1 b_2} \ldots R^{a_{2p-1}a_{2p}}_{~~b_{2p-1}b_{2p}}e^{b_1}\wedge \ldots \wedge e^{b_{2p}}, \nn
\eeqa
the curvature $2$-form $R^{ab}\equiv \frac{1}{2}R^{ab}_{~~cd} e^{c} \wedge e^{d}$, ${e^{(a)}}$ is the dual basis of the non-coordinate basis ${e_{a}}$ with $e^a = e^a_{~\mu}dx^\mu$ defined by $\langle e^a, \hat{e}_b \rangle =\delta^a_{~b} $, and $e_a^{~\mu}$ are vielbeins(or vierbeins for $d=4$)(or $R^{a}_{~b}=D\omega^{a}_{~b}=d\omega^{a}_{~b}+\omega^{a}_{~c}\wedge \omega^{c}_{~b}$, where the curvature $2$-form $R^{a}_{~b}\equiv \frac{1}{2}R^{a}_{~bcd} e^c\wedge e^d$, and the Riemman tensor can be expressed as $R^a_{~bcd} = e^a((2 \nabla_{[c} \nabla_{d]} - f^e_{cd}\nabla_e)e_b)$, where $[e_a,e_b]=f^c_{ab}e_c$. ).  The $\star$ is the Hodge star defined through a $(d-2p)$ form dual to a given $2p$ form,
\beqa
\ii
\star e_{a_1 \ldots a_{2p}} \2i \equiv \2i \frac{\sqrt{g}}{(d-2p)!}\varepsilon_{a_1  \ldots a_{2p} a_{2p+1}\ldots a_{d}} e^{a_{2p+1}}\wedge \ldots \wedge e^{a_{d}}.
\eeqa
where $\varepsilon$ are the co-variant Levi-Civita tensor density.
By using Eq.(\ref{Eq:R2p-form}), the Euler density $2p$-forms in Eq.(\ref{Eq:Wd-d-form}) become
\beqa
W_{(p)} \3i &=& \3i \frac{1}{(d-2p)! } \epsilon_{a_1 \ldots a_d} R^{a_1 a_2}\wedge \ldots  \wedge R^{a_{2p-1} a_{2p}}  \nn\\
&\wedge &  e^{a_{2p+1}}\wedge \ldots \wedge e^{a_d},
\eeqa
where $ \epsilon_{a_1 \ldots a_d} \equiv \sqrt{g} \varepsilon_{a_1 \ldots a_d}$ is the co-variant Levi-Civita tensor, defined through the Levi-Civita tensor density $\varepsilon$.

Therefore, the differential form of the potential action in Eq.(\ref{Eq:Wd-d-form}) becomes
\beqa
W_d &=& \int_{{\mathcal M}}  \sum_{p=0}^{[d/2]} \frac{1}{(d-2p)!} \epsilon_{a_1 \ldots a_d} R^{a_1 a_2}\wedge \ldots  \wedge R^{a_{2p-1} a_{2p}}  \nn\\
&& \wedge   e^{a_{2p+1}}\wedge \ldots \wedge e^{a_d}. \label{Eq:Wd-d-form_1}
\eeqa
The above differential form of action in Eq.(\ref{Eq:Wd-d-form_1}) can be re-expressed in field theory as below,
\beqa
W_d &=& \int_{{\mathcal M}}  \sum_{p=0}^{[d/2]} \frac{1}{(d-2p)!} \epsilon_{a_1 \ldots a_d}  \frac{1}{2^p} R^{a_1 a_2}_{~~b_1 b_2} \ldots R^{a_{2p-1}a_{2p}}_{~~b_{2p-1}b_{2p}} \nn\\
&& e^{b_1}\wedge \ldots \wedge e^{b_{2p}} \wedge   e^{a_{2p+1}}\wedge \ldots \wedge e^{a_d}. \label{Eq:Wd-d-form_2}
\eeqa
one obtains
\beqa
&& W_d %&=& \int_{{\mathcal M}}  \sum_{p=0}^{[d/2]} \frac{\sqrt{g}}{(d-2p)!} \varepsilon_{a_1 \ldots a_d}  \frac{1}{2^p} R^{a_1 a_2}_{~~b_1 b_2} \ldots R^{a_{2p-1}a_{2p}}_{~~b_{2p-1}b_{2p}} \nn\\
= \int d^dx \sqrt{g} \sum_{p=0}^{[d/2]} \frac{1}{(d-2p)!}  \frac{1}{2^p} R^{a_1 a_2}_{~~b_1 b_2} \ldots R^{a_{2p-1}a_{2p}}_{~~b_{2p-1}b_{2p}} \nn\\
&& \qquad \times \varepsilon^{b_1 \ldots b_{2p} a_{2p+1} \ldots a_d} \varepsilon_{a_1 \ldots a_d} \nn\\
&& = \int d^dx \sqrt{g} \sum_{p=0}^{[d/2]}   \frac{1}{2^p} R^{a_1 a_2}_{~~b_1 b_2} \ldots R^{a_{2p-1}a_{2p}}_{~~b_{2p-1}b_{2p}} \delta^{b_1\ldots b_{2p}}_{a_1 \ldots a_{2p}},  \label{Eq:Wd-d-form_3}
\eeqa
where in the last equality, we have used that
\beqa
\varepsilon^{b_1 \ldots b_{2p} a_{2p+1} \ldots a_d} \varepsilon_{a_1 \ldots a_d} = (d-2p)!\delta^{b_1\ldots b_{2p}}_{a_1 \ldots a_{2p}}. \nn
\eeqa
Except this, in the above derivation, we have also used the definition of Vierbein $e^a=e^{~a}_\mu$ so that
\beqa
&& e^{b_1}\wedge \ldots \wedge e^{b_{2p}} \wedge   e^{a_{2p+1}}\wedge \ldots \wedge e^{a_d} \nn\\
&=& e^{~ b_1}_{\mu_1}  \ldots  e^{~ b_{2p}}_{\mu_{2p}}  e^{~a_{2p+1}}_{\mu_{2p+1}}  \ldots e^{~a_d}_{\mu_{d}}  dx^{\mu_1} \wedge dx^{\mu_2} \wedge \ldots \wedge dx^{\mu_d}. \nn
\eeqa
By definition, the wedge product in the integral becomes
\beqa
&& dx^{\mu_1} \wedge dx^{\mu_2} \wedge \ldots \wedge dx^{\mu_d} \nn\\
&=&  \varepsilon^{\mu_1 \mu_2 \ldots \mu_d} dx^{1} \wedge dx^{2} \wedge \ldots \wedge dx^d = \epsilon^{\mu_1 \mu_2 \ldots \mu_d} d^dx. \nn
\eeqa
The Levi-Civita tensor are related to the ordinary Levi-Civita tensor density as below:
\beqa
&& \epsilon^{a_1 \ldots a_d} \equiv \frac{1}{\sqrt{g}} \varepsilon^{a_1 \ldots a_d}, \quad \epsilon_{a_1 \ldots a_d} \equiv {\sqrt{g}} \varepsilon_{a_1 \ldots a_d},
\eeqa
where $ \epsilon^{a_1 \ldots a_d} $ and $ \epsilon_{a_1 \ldots a_d} $ are respectively the contra-variant and co-variant Levi-Civita tensor, which are both defined through the Levi-Civita tensor density $\varepsilon$.

To be concrete, let us consider the lower order terms of the $d$-form $W_d$ defined in Eq.(\ref{Eq:Wd-d-form}). The leading order term is a $0$-form
\beqa
W_{(0)} = 1,
\eeqa
which standards for the cosmological constant term, while the next leading order term is a $2$-form
\beqa
W_{(1)}=({\mathcal R}^{p})^{a_1 a_2} \star e_{a_1 a_2},
\eeqa
which is just the Hilbert action. The next leading order term is a $4$-form
\beqa
W_{(2)}=({\mathcal R}^{2})^{a_1 a_2 a_3 a_4} \star e_{a_1 a_2 a_3 a_4},
\eeqa
which is just the Gauss-Bonnet term.

Doing variation of the action in Eq.(\ref{Eq:Wd}) with respect to $e^a$ and $\omega^{ab}$ respectively gives the field equations,
\beqa
&& \sum_{p=0}^{[(d-1)/2]}\alpha_p(d-2p)\epsilon_{aa_1\ldots a_{d-1}}R^{a_1 a_2}\wedge \ldots \wedge R^{a_{2p-1}a_{2p}} \nn\\
&& \qquad \qquad \wedge e^{a_{2p+1}} \wedge \ldots \wedge e^{a_{d-1}} = 0, \nn\\
&& \sum_{p=0}^{[(d-1)/2]}\alpha_p p(d-2p)\epsilon_{ab a_3\ldots a_{d}}R^{a_3 a_4} \wedge \ldots \wedge R^{a_{2p-1}a_{2p}} \nn\\
&& \qquad \qquad \wedge T^{a_{2p+1}} \wedge e^{a_{2p+2}} \wedge \ldots \wedge e^{a_{d}} = 0.
\eeqa
where $T^a\equiv de^a + \omega^a_{~b}\wedge e^b$ is the torsion $2$-form.

In odd $d=2n-1$ dimensions, $[d/2]=n-1$, the potential density in Eq.(\ref{Eq:Wd}) becomes
\beqa
{\mathcal W}_{2n-1} &=&   \sum_{p=0}^{n-1} \alpha_p  \epsilon_{a_1 \ldots a_d} R^{a_1 a_2}\wedge \ldots  \wedge R^{a_{2p-1} a_{2p}}  \nn\\
&& \wedge   e^{a_{2p+1}}\wedge \ldots \wedge e^{a_{2n-1}}. \label{Eq:W_2n-1}
\eeqa
This is called the Euler-Chern-Simons $(2n-1)$-form, in the sense that its exterior derivative $d {\mathcal W}_{2n-1}$ is
\beqa
\ii
{\mathcal E}_{2n} =  \sum_{p=0}^{n} \alpha_p \epsilon_{A_1 \ldots A_{2n}} \tilde{R}^{A_1 A_2}\wedge \ldots \wedge \tilde{R}^{A_{2n-1} A_{2n}}.
\eeqa
It is nothing but the Euler characteristic class in $(d+1)$ dimensions, which is an exact $2n$-form, i.e., ${\mathcal E}_{2n}=d {\mathcal W}_{2n-1}$ is a total derivative in $2n$ dimensions thus doesn't contributes to the EOM(or cannot be used as Lagrangian), but ${\mathcal W}_{2n-1}$ is a Lagrangian in $d=2n-1$ dimensions. The $2$-form $\tilde{R}^{AB}=d\tilde\omega^{AB}+\tilde\omega^A_{~C}\wedge \tilde\omega^{CB}$ is constructed with the $SO(d,1)$ connection $\tilde\omega^{AB}$, which can also be decomposed into the $SO(d)$ connection $\omega^{ab}$ under fully spatial rotational symmetry and vielbein $e^a$ under translational symmetry,
\beqa
\tilde\omega^{AB}  = \left(
                  \begin{array}{cc}
                    \omega^{ab} & \ell^{-1} e^a \\
                    - \ell^{-1} e^b & 0 \\
                  \end{array}
                \right).
\eeqa
In the construction and decomposition, the $SO(d,1)$ curvature $2$-form are
\beqa
\tilde{R}^{AB}= \left(
                  \begin{array}{cc}
                    R^{ab}+ \ell^{-2} e^a\wedge e^b & \ell^{-1} T^a \\
                    - \ell^{-1} T^b & 0 \\
                  \end{array}
                \right).
\eeqa
The EOM of the action in Eq.(\ref{Eq:W_2n-1}) with respect to the vielbein is
\beqa
\epsilon_{a_1 a_2 \ldots a_{2n-1}} \tilde{R}^{a_1 a_2} \wedge \ldots \tilde{R}^{a_{2n-3} a_{2n-2}} =0.
\eeqa

In even $d=2n$ dimensions, $[d/2]=n$, then the last term in the summation, i.e.,($p=d/2=n$) is the Euler characteristic and does not contributes to the equations of motion, since it is a topological term in dimensions less than or equal to $d$. Thus the Lagrangian density in Eq.(\ref{Eq:Wd}) becomes\footnote{The summation index can be extended to $p=n$, but he Euler density ${\mathcal E}_{2n}$ doesn't contributes to the EOMs, since the integral becomes a topological invariant, the Euler characteristic. }
\beqa
{\mathcal W}_{2n} = \sum_{p=0}^{n-1} \alpha_p  \epsilon_{a_1 \ldots a_d} \tilde{R}^{a_1 a_2}\wedge \ldots  \wedge \tilde{R}^{a_{2n-1} a_{2n}} ,
\eeqa
which is called the Born-Infeld $2n$-form,
\beqa
{\mathcal W}_{2n}  =  \sum_{p=0}^{n-1} \alpha_p \sqrt{\det\tilde{R}^{ab}}  =  \sum_{p=0}^{n-1} \alpha_p \text{Pf}~\tilde{R}^{ab} ,  \label{Eq:W_2n}
\eeqa
where Pf denotes the Pfaffian(the square root of the determinant, i.e., $\det a = \text{Pf}(a)^2$) in the exterior product, and we have denoted
\beqa
\tilde{R}^{ab} = R^{ab} + \ell^{-2} e^a \wedge e^b.
\eeqa
The EOM of the action in Eq.(\ref{Eq:W_2n}) with respect to the vielbein is
\beqa
\ii
\epsilon_{a_1 a_2 \ldots a_{2n}}\tilde{R}^{a_1 a_2}\wedge \ldots \wedge \tilde{R}^{a_{2n-3} a_{2n-2}}\wedge e^{a_{2n-1}} = 0.
\eeqa

\section{ Causal Structure and Space-Time Singularity }
\label{app:causal-singulariy}

In the static topological invariant metric defined in Eq.(\ref{Eq:ds2-topological-HL}), the nonvanishing components of the Christoffel symbols of the metric tensor are
\beqa
&& \Gamma^{r}_{rr} = - \frac{g^\prime(r)}{2g(r)},  \quad \Gamma^m_{nr}=\Gamma^m_{rn} = \frac{1}{r} \delta^m_n,  \\
&& \Gamma^r_{mn} = \Gamma^r_{nm} =  rg(r) \delta_{nm} =   \frac{g(r)}{r}g_{mn}, \label{Eq:LeviCivita-rx}
\eeqa
and the only nonvanishing components of the spatial curvature tensor are
\beqa
R^{m_1 m_2}_{~~n_1 n_2} = \frac{g(r)}{r^2}\delta^{m_1 m_2}_{n_1 n_2}, \quad R^{r m}_{~~r n} = \frac{g^\prime(r)}{2r}\delta^m_n,  \label{Eq:Rijkl-Rrjrl}
\eeqa
where
\beqa
g(r)\equiv k - f(r) = - g^{rr}(r) .  \label{Eq:gr-fr}
\eeqa
While all residues are vanishing, e.g., $R^{rr}_{~~n_1 n_2}=R^{rr}_{~~n_1 r}=R^{rr}_{~~rr}=0$, $R^{mm}_{~~n_1 n_2}= R^{mm}_{~~rn_2}=R^{mm}_{~~rr}=0$, $R^{mm}_{~~mm}=0$, etc.
Then one obtains
\beqa
\ii
R^m_n \!=\! \bigg(  \! (d-2)\frac{g(r)}{r^2} \!+\! \frac{g^\prime(r)}{2r} \! \bigg) \delta^m_n, \, R^r_r \!=\! (d-1)\frac{g^\prime(r)}{2r}. \quad \label{Eq:Rmn-Rrr}
\eeqa
Thus some curvature invariants, e.g., the quadratic curvature invariants in pure space, are\footnote{Note: In order to consider the causal structure or singularity of the space-time, we need to use those in space-time in Appendix.\ref{app:causal-singulariy}.}
\beqa
R &=& \frac{(d-1)(d-2)}{r^2}g(r) + \frac{d-1}{r}g^{\prime}(r)\nn\\
R^{ij}R_{ij} &=& \frac{(d-1)(d-2)^2}{r^4}g(r)^2 ,\label{Eq:R-Rij2-Rijkl2}\\
&+& \frac{(d-1)(d-2)}{r^3}g^{\prime}(r)g(r) + \frac{d(d-1)}{4r^2}g^{\prime}(r)^2 ,\nn\\
R^{ijkl}R_{ijkl} &=& \frac{2(d-1)(d-2)}{r^4}g(r)^2+\frac{d-1}{r^2}g^\prime(r)^2 , \nn
\eeqa
where in the last equality, we have used that
\beqa
\delta^{m_1 m_2}_{n_1 n_2} \delta^{n_1 n_2}_{m_1 m_2} = 2!\delta^{m_1 m_2}_{m_1 m_2}= 2\frac{(d-1)!}{(d-3)!}. \nn
\eeqa

If one is interested in the causal structure or the singularity of the space-time but not the pure spatial space, then one needs to consider the time components. Except for those shown in Eq.(\ref{Eq:LeviCivita-rx}), the nonvanishing Christoffel symbols of the metric in Eq.(\ref{Eq:ds2-topological-HL}) are
\beqa
\Gamma^t_{tr} &=& \Gamma^t_{rt} = \frac{g^\prime(r)}{2g(r)} + \frac{\tilde{N}^\prime(r)}{\tilde{N}(r)}, \nn\\
\Gamma^r_{tt} &=& g(r)^2\tilde{N}(r)^2\bigg( \frac{g^\prime(r)}{2g(r)} + \frac{\tilde{N}^\prime(r)}{\tilde{N}(r)} \bigg), \label{Eq:LeviCivita-trx}
\eeqa
and except for those shown in Eq.(\ref{Eq:Rijkl-Rrjrl}), the nonvanishing components of the space-time curvature tensor are,
\beqa
R^{mt}_{~~nt} &=& \bigg( \frac{g^{\prime}(r)}{2r} + \frac{g(r)}{r}\frac{\tilde{N}^{\prime}(r)}{\tilde{N}(r)} \bigg)\delta^m_n = \frac{g^{\prime}(r)}{2r}\delta^m_n, \label{Eq:Rtjtl} \\
R^{rt}_{~~rt} &=& \frac{g^{\prime\prime}(r)}{2}  + g^{\prime}(r)\frac{3\tilde{N}^\prime(r)}{2\tilde{N}(r)} + g(r)\frac{\tilde{N}^{\prime\prime}(r)}{\tilde{N}(r)} = \frac{g^{\prime\prime}(r)}{2}, \nn
\eeqa
where in the last equalities of the two equation, we have made assumption that $\tilde{N}(r)=1$.
From both Eq.(\ref{Eq:Rijkl-Rrjrl}) and Eq.(\ref{Eq:Rtjtl}), one obtains($R^n_{~r}=R^r_{~n}=R^n_{~t}=R^t_{~n}=R^t_{~r}=R^r_{~t}=0$)
\beqa
\ii
R^m_{~n} \!&=&\! \bigg((d-2)\frac{g(r)}{r^2} \!+\! \frac{g^\prime(r)}{r} \bigg)\delta^m_n , \\
R^r_{~r} \!&=&\! (d-1)\frac{g^\prime(r)}{2r} +  \frac{g^{\prime\prime}(r)}{2}, ~ R^t_{~t} \!=\! (d-1)\frac{g^{\prime}(r)}{2r} + \frac{g^{\prime\prime}(r)}{2}. \nn
\eeqa
In this case, for example, the quadratic curvature invariants in Eq.(\ref{Eq:R-Rij2-Rijkl2}) are generalized to be those in the whole space-time,
\beqa
R \3i &=& \3i \frac{(d\!-\!1)(d\!-\!2)}{r^2}g(r) \!+\! \frac{2(d\!-\!1)}{r}g^{\prime}(r)+g^{\prime\prime}(r), \nn\\
R^{\mu\nu}R_{\mu\nu}
\3i &=& \3i \frac{(d\!-\!1)(d\!-\!2)^2}{r^4}g(r)^2 \!+\! \frac{2(d\!-\!1)(d\!-\!2)}{r^3}g^{\prime}(r)g(r) \nn\\
\3i && \3i  +  \frac{d^2\!-\!1}{2r^2}g^{\prime}(r)^2 \!+\! \frac{d\!-\!1}{r}g^{\prime\prime}(r)g^{\prime}(r) \!+\! \frac{1}{2}g^{\prime\prime}(r)^2 , \nn\\
R^{\mu\nu\rho\sigma}R_{\mu\nu\rho\sigma} \3i &=& \3i \frac{2(d\!-\!1)(d\!-\!2)}{r^4}g(r)^2 \!+\! \frac{2(d\!-\!1)}{r^2}g^\prime(r)^2 \!+\! g^{\prime\prime}(r)^2 , \nn
\eeqa
where $R = R^m_{~n}\delta^n_m + R^r_{~r} + R^t_{~t}$, $R^{\mu\nu}R_{\mu\nu} = R^m_{~n}R^n_{~m} + R^r_{~r}R^r_{~r} + R^t_{~t}R^t_{~t}$
and $R^{\mu\nu\rho\sigma}R_{\mu\nu\rho\sigma}= R^{m_1 m_2}_{~~ n_1 n_2} R^{n_1 n_2}_{~~ m_1 m_2} + 2^2 R^{r m}_{~~ r n} R^{r n}_{~~ r m} + 2^2 R^{t m}_{~~ t n} R^{t n}_{~~ t m} + 2^2 R^{r t}_{~~ r t} R^{r t}_{~~ r t}$.

\end{CJK}

\bibliography{000}

\end{document}